\documentclass[conference]{IEEEtran}
\usepackage{cite}
\usepackage{amsmath,amssymb,amsfonts}
\usepackage{subfigure}
\usepackage{algorithmic}
\usepackage{graphicx}
\usepackage{textcomp}
\usepackage[table]{xcolor}
\usepackage{pifont}
\usepackage{makecell}
\def\BibTeX{{\rm B\kern-.05em{\sc i\kern-.025em b}\kern-.08em
    T\kern-.1667em\lower.7ex\hbox{E}\kern-.125emX}}
\AtBeginDocument{\definecolor{ojcolor}{cmyk}{0.93,0.59,0.15,0.02}}

\renewcommand{\baselinestretch}{0.93}

\begin{document}

\title{Reliability of Wi-Fi, LTE, and 5G-Based UAV RC Links in ISM Bands: Uplink Interference Asymmetry Analysis and HARQ Design 
\thanks{\textcolor{red}{This work is supported in part by the NSF awards CNS-1939334 and CNS-2332834}}
}



\author{Donggu Lee$^1$, Sung Joon Maeng$^2$, Ozgur Ozdemir$^1$, Mani Bharathi Pandian$^3$, and Ismail Guvenc$^1$
\\ $^1$Department of Electrical and Computer Engineering, North Carolina State University, Raleigh, NC, USA 
\\ $^2$Department of Electrical and Electronic Engineering, Hanyang University, Ansan, South Korea
\\ $^3$Skydio, Inc., San Mateo, CA, USA 
\\ E-mail: \{dlee42, oozdemi, iguvenc\}@ncsu.edu,
sjmaeng@hanyang.ac.kr, mani.pandian@skydio.com
}


\maketitle

\begin{abstract}
Command and control of uncrewed aerial vehicles (UAVs) is often realized through air-to-ground (A2G) remote control (RC) links that operate in ISM bands. While wireless fidelity (Wi-Fi) technology is commonly used for UAV RC links, ISM-based long-term evolution (LTE) and fifth-generation (5G) technologies have also been recently considered for the same purpose. A major problem for UAV RC links in the ISM bands is that other types of interference sources, such as legacy Wi-Fi and Bluetooth transmissions, may degrade the link quality. Such interference problems are a higher concern for the UAV in the air than the RC unit on the ground due to the UAV being in line-of-sight (LoS) with a larger number of interference sources. To obtain empirical evidence of the asymmetric interference conditions in downlink (DL, UAV to RC unit) and uplink (UL, RC unit to UAV), we first conducted a measurement campaign using a helikite platform in urban and rural areas at NC State University. The results from this measurement campaign show that the aggregate interference can be up to $16.66$~dB at higher altitudes up to $170$~meters, compared with the interference observed at a ground receiver. As a result of this asymmetric UL interference, lost hybrid automatic repeat request (HARQ) indicators (ACK/NACK) in the UL may degrade the DL throughput. To investigate this, we study various HARQ mechanisms, including HARQ Type-I with no combining, HARQ Type-I with chase combining, HARQ Type-III with incremental redundancy, and burst transmission with chase combining. To evaluate the impact of asymmetric UL interference on throughput performance, we consider three steps of evaluation process: 1) standalone physical DL shared channel (PDSCH) throughput evaluation with perfect ACK/NACK assumption; 2) standalone physical UL control channel (PUCCH) decoding reliability evaluation; and 3) PDSCH DL throughput evaluation with asymmetric UL ACK/NACK transmission. Our numerical results show severe performance degradation for larger UL/DL interference asymmetry. Specifically, when the UL signal-to-interference-plus-noise ratio (SINR) is $15$~dB lower than the DL SINR, the average throughput ratio within the operating SINR region decreases by approximately $44$~\%p compared to the perfect ACK/NACK case.
\end{abstract}


\section{INTRODUCTION}
\begin{figure*}[t!]
    \centering
    \includegraphics[width=1.9\columnwidth]{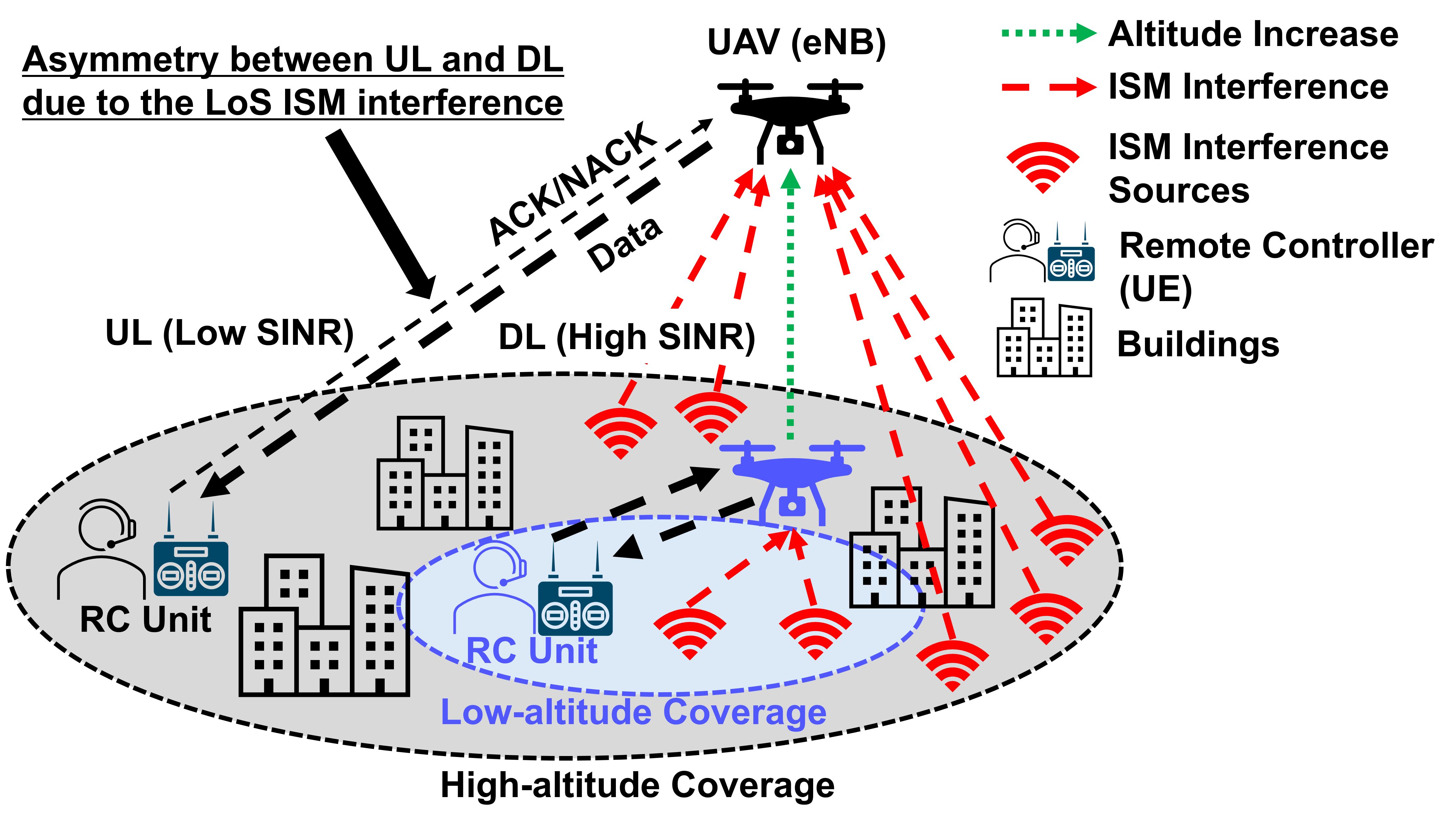}
    \caption{Air-to-ground network model with interference asymmetry in UL and DL. In this scenario, the UAV operates as an eNB (see e.g.~\cite{DJI_paper}), and the RC on the ground operates as a UE. The UL to the UAV gets a higher line-of-sight (LoS) ISM interference as the altitude increases.}
    \label{fig:A2G_model}
\end{figure*}

Ensuring reliable and high-throughput wireless connectivity is critical for uncrewed aerial vehicles (UAVs) remote control (RC) links, especially in dynamic environments with frequent interference. To achieve this, wireless fidelity (Wi-Fi)~\cite{7986413}, and recently, long-term-evolution (LTE)~\cite{LTE_U_UAV_1} technologies have been commonly adopted for air-to-ground (A2G) RC links that operate in the industrial, scientific, and medical (ISM) bands such as the $2.4$~GHz and $5$~GHz spectrum. While LTE-Unlicensed (LTE-U)~\cite{LTE_U}, fifth-generation (5G) New Radio-Unlicensed (NR-U)~\cite{NR_U_UAV_1, NR_U_UAV_2}, and license-assisted access (LAA) \cite{LAA} can be used in ISM bands, some popular UAV vendors also use RC links based on LTE with proprietary modifications~\cite{DJI_paper}. 

Wi-Fi provides certain advantages for UAV RC links in the ISM bands. However, Wi-Fi is not designed to maintain long-range coverage and interference resilience for outdoor connectivity. On the other hand, LTE~\cite{LTE_UAV_magazine} and 5G~\cite{8741719} can serve as promising alternatives for long-range RC link coverage, data rate, and interference resilience. As LTE infrastructure is deployed more pervasively, and due to limited studies evaluating it for A2G RC links despite the future interest in using it for such scenarios over 5G~\cite{FCC_UAV_RuleMaking}, we primarily focus on an LTE-based scenario in this paper. Additionally, we also include a 5G scenario to investigate underlying factors contributing to performance improvement.

Interference is a critical factor affecting the range and stability of RC links in the ISM band. For instance, Wi-Fi access points in the UAV operating area transmit beacon signals every $102.4$ ms, which are also received at the UAV. Therefore, the cumulative traffic of beacon signals can result in high interference at the UAV even when no Wi-Fi data traffic is present. This becomes a challenging problem as the RC-to-UAV link (uplink, UL) observes increasingly higher interference levels at higher altitudes, making continued connectivity difficult, resulting in a low signal-to-interference-plus-noise ratio (SINR), especially at the long link distances. On the other hand, the RC unit on the ground is reasonably isolated from such interference and hence observes better link quality in the downlink (DL), as illustrated in Figure~\ref{fig:A2G_model}. Here, the link qualities of the UL and DL are represented using the thickness of the arrows. Note that the UL in the high-altitude coverage is illustrated with thinner arrows compared to the DL, while the link quality of the UL and DL for low-altitude coverage is the same.

As such, in this work, we consider an UL (RC to UAV) interference-asymmetric scenario to take into account interference sources in the ISM band that more severely affect the signal reception at a UAV. While UAVs can also use cellular bands for RC, the ISM band is commonly used commercially due to the flexibility and cost advantages in the unlicensed spectrum. Moreover, it also allows for the customization of wireless operation in the physical layer. For example, DJI, a drone manufacturer, employs customized RC links using similar waveforms, protocols, and parameters as in LTE, e.g., turbo coding and rate matching \cite{DJI_paper}. 

To our best knowledge, throughput performance evaluation with different hybrid automatic repeat request (HARQ) types and unique features of UAV RC links, such as interference-asymmetric UL conditions in Figure~\ref{fig:A2G_model}, is not available in the literature. Addressing this gap is the main goal of this paper. The main contributions of this paper can be summarized as follows.

\textbf{Evaluation of HARQ Mechanisms for Asymmetric DL/UL Scenarios.}
The HARQ mechanism used both in LTE and 5G is one of the underlying technologies to achieve higher link reliability requirements over noisy and interference-asymmetric UAV channels~\cite{HARQ_survey}. The HARQ process triggers retransmission based on the retransmission request feedback to achieve robust data transmission in lossy channels. In this paper, we adopt 1) HARQ Type-I with no combining, 2) HARQ Type-I with chase combining (CC), 3) HARQ Type-III with incremental redundancy (IR), and 4) (proposed) burst transmission with CC scenarios for the long-range physical DL shared channel (PDSCH) throughput performance evaluation, which is a data-carrying channel in the DL of LTE and 5G. Since the physical UL control channel (PUCCH) carries the HARQ indicator (ACK/NACK) is deteriorated by the asymmetric condition, we set the performance evaluation in three stages: 1) evaluating PDSCH DL standalone throughput under perfect ACK/NACK assumption, 2) assessing decoding reliability of PUCCH independently, and 3) PDSCH DL throughput evaluation with asymmetric and lossy PUCCH, which is a combined analysis of 1) and 2). Our numerical results show the distinctive performance differences when using different HARQ mechanisms in the first stage. Moreover, severe throughput performance degradation is evident when the channel condition on the UL gets worse, which is the last stage of the aforementioned performance evaluation.

\textbf{Measurement Campaign in the ISM Band for Assessing UAV Altitude Effects.}
To validate our UL asymmetric UAV link scenarios, we performed a measurement campaign to capture the aggregate signal power in the ISM bands, which can act as interference in the context of unlicensed spectrum usage, using a helikite platform at different altitudes. The measurement campaign was conducted in $2024$ at the Main Campus of NC State University (during the   Packapalooza Festival) and Lake Wheeler Field Lab at NC State University, for urban and rural scenarios, respectively. Our experiment results show that the received signal strength tends to be higher at a higher altitude due to the higher chance of line-of-sight (LoS) signal reception, as in our A2G network model with asymmetric link SINR in Figure~\ref{fig:A2G_model}.


\textbf{Performance Evaluation of RC Links for Wi-Fi, LTE, 5G.}
To better understand performance trade-offs when different wireless communication standards are used, we describe the distinctive features of LTE, 5G, and Wi-Fi. We comparatively analyze the decoding reliability performance of data and HARQ indicator channels. Our numerical results show that LTE and 5G outperform Wi-Fi in reliability, when considering UL asymmetric A2G network scenarios, due to being optimized to serve at longer link distances when compared with Wi-Fi.


\begin{table*}[t!]
    \centering    
    \caption{Literature review on performance evaluation of A2G UAV communications considering HARQ, UL asymmetry, and experimental verification.}
    \begin{tabular}{|c|c|c|c|c|c|c|c|}
        \hline  \textbf{Ref.} & \textbf{Analysis objectives} & \makecell{ \textbf{UL} \\ \textbf{asymmetry}} & \makecell{\textbf{UAV} \\ \textbf{channel}} & \makecell{\textbf{Various} \\ \textbf{HARQ types}} & \makecell{\textbf{Cellular} \\ \textbf{links}} & \makecell{ \textbf{ISM} \\ \textbf{bands}} & \textbf{Measurements} \\ \hline
        
        \cite{DJI_paper} & Communication protocol analysis of UAV-RC links & \cellcolor{red!25} \ding{55} & \cellcolor{green!25} \ding{51} & \cellcolor{red!25} \ding{55} & \cellcolor{green!25} \ding{51} & \cellcolor{green!25} \ding{51} & \cellcolor{red!25} \ding{55} \\ \hline  

        \cite{rural_coverage_LTE} & Rural coverage analysis for LTE & \cellcolor{red!25} \ding{55} & \cellcolor{green!25} \ding{51} & \cellcolor{red!25} \ding{55} & \cellcolor{green!25} \ding{51} & \cellcolor{red!25} \ding{55} & \cellcolor{green!25} \ding{51} \\ \hline 
        
        \cite{PLE_table_asymmetry} & UAV channel characterization & \cellcolor{green!25} \ding{51} & \cellcolor{green!25} \ding{51} & \cellcolor{red!25} \ding{55} & \cellcolor{red!25} \ding{55} & \cellcolor{green!25} \ding{51} & \cellcolor{green!25} \ding{51} \\ \hline
        
        \cite{asymmetry_UAV} & Asymmetric UAV link analysis & \cellcolor{green!25} \ding{51} & \cellcolor{green!25} \ding{51} & \cellcolor{red!25} \ding{55} & \cellcolor{red!25} \ding{55} & \cellcolor{green!25} \ding{51} & \cellcolor{green!25} \ding{51} \\ \hline
        
        \cite{asymmetry_uav_journal} & Asymmetric UAV link analysis with mobility & \cellcolor{green!25} \ding{51} & \cellcolor{green!25} \ding{51} & \cellcolor{red!25} \ding{55} & \cellcolor{red!25} \ding{55} & \cellcolor{green!25} \ding{51} & \cellcolor{green!25} \ding{51} \\ \hline  

        \cite{uav_channel_model_urban} & UAV channel modeling with measurements & \cellcolor{red!25} \ding{55} & \cellcolor{green!25} \ding{51} & \cellcolor{red!25} \ding{55} & \cellcolor{red!25} \ding{55} & \cellcolor{red!25} \ding{55} & \cellcolor{green!25} \ding{51} \\ \hline
        
        \cite{CC_HARQ_Evaluation} & Packet error rate and throughput & \cellcolor{red!25} \ding{55} & \cellcolor{red!25} \ding{55} & \cellcolor{green!25} \ding{51} &  \cellcolor{green!25} \ding{51} & \cellcolor{red!25} \ding{55} & \cellcolor{red!25} \ding{55} \\ \hline
        
        \cite{CC_HARQ_Evaluation_2} & Throughput, delay, and error rate & \cellcolor{red!25} \ding{55} & \cellcolor{red!25} \ding{55} & \cellcolor{green!25} \ding{51} &  \cellcolor{green!25} \ding{51} & \cellcolor{red!25} \ding{55} & \cellcolor{red!25} \ding{55} \\ \hline

        \cite{LTE_performance_eval_matlab} & Throughput and error rate evaluation & \cellcolor{red!25} \ding{55} & \cellcolor{red!25} \ding{55} & \cellcolor{green!25} \ding{51} & \cellcolor{green!25} \ding{51} & \cellcolor{red!25} \ding{55} & \cellcolor{red!25} \ding{55} \\ \hline
        
        \cite{LTE_based_UAV} & Measurement of LTE-based throughput & \cellcolor{red!25} \ding{55} & \cellcolor{green!25} \ding{51} & \cellcolor{red!25} \ding{55} & \cellcolor{green!25} \ding{51} & \cellcolor{red!25} \ding{55} & \cellcolor{green!25} \ding{51}  \\ \hline
        
        \cite{UAV_adhoc} & Throughput, coverage, and energy consumption & \cellcolor{red!25} \ding{55} & \cellcolor{green!25} \ding{51} & \cellcolor{red!25} \ding{55} &  \cellcolor{red!25} \ding{55} & \cellcolor{red!25} \ding{55} & \cellcolor{red!25} \ding{55} \\ \hline 

        \cite{cognitive_UAV} & Throughput and latency & \cellcolor{red!25} \ding{55} & \cellcolor{green!25} \ding{51} & \cellcolor{red!25} \ding{55} & \cellcolor{green!25} \ding{51} & \cellcolor{red!25} \ding{55} & \cellcolor{red!25} \ding{55} \\ \hline
        
        \cite{UAV_multiple}, \cite{UAV_disaster} & Throughput with multiple UAVs & \cellcolor{red!25} \ding{55} & \cellcolor{green!25} \ding{51} & \cellcolor{red!25} \ding{55} & \cellcolor{red!25} \ding{55} & \cellcolor{red!25} \ding{55} & \cellcolor{red!25} \ding{55} \\ \hline
        
        \cite{UAV_eval_letter} & Throughput and latency & \cellcolor{red!25} \ding{55} & \cellcolor{green!25} \ding{51} & \cellcolor{red!25} \ding{55} & \cellcolor{green!25} \ding{51} & \cellcolor{red!25} \ding{55} & \cellcolor{green!25} \ding{51} \\ \hline
        
        \cite{endtoend_measure} & Measurement of 5G-based throughput & \cellcolor{red!25} \ding{55} & \cellcolor{green!25} \ding{51} & \cellcolor{red!25} \ding{55} &  \cellcolor{green!25} \ding{51} & \cellcolor{red!25} \ding{55} & \cellcolor{green!25} \ding{51} \\ \hline     
        
        \textbf{This Work} & Throughput, UL asymmetry, and latency & \cellcolor{green!25} \ding{51} & \cellcolor{green!25} \ding{51} & \cellcolor{green!25} \ding{51} &  \cellcolor{green!25} \ding{51} & \cellcolor{green!25} \ding{51} & \cellcolor{green!25} \ding{51} \\ \hline 
    \end{tabular}
    \label{tab:literature_review}
\end{table*}

The rest of the paper is organized as follows. Relevant works from the existing literature are discussed in Section~\ref{ch:related_works}. The experiment settings and results of the measurement campaign are presented in Section~\ref{ch:measurement}. The description of the system model, overview of HARQ types, and performance evaluation settings are given in Section~\ref{CH:sys}. The details of HARQ types and UL asymmetric scenarios for evaluation are described in Sections~\ref{ch:HARQs} and~\ref{Sec:performance_evaluation_method}. Simulation results of each evaluation scenario are provided in Section~\ref{ch:results}. Finally, the last section concludes the paper.

\section{RELATED WORKS}\label{ch:related_works}
There are limited studies evaluating different HARQ options for interference-asymmetric channels, such as those used in A2G RC links with UAVs. Our literature review with representative works related to performance evaluation for A2G UAV communications, and the differences with the present work are summarized
in Table \ref{tab:literature_review}. 

\subsection{UL Asymmetry in UAV Channels}
In~\cite{PLE_table_asymmetry}, the characterization of asymmetric links in aerial networks is discussed by using path loss exponent measurement over ground-to-air (G2A) and A2G channels. The path loss exponent over G2A and A2G is given by $2.51$ and $2.32$, respectively, emphasizing higher path loss in the UL side. This study shows that assuming symmetric UL and DL in UAV networks can cause a gap with real-world performance.

The asymmetry of UL and DL in UAV-assisted wireless sensor networks is investigated in~\cite{asymmetry_UAV}. The DL quality is consistently higher than the UL, which the authors attribute to interference at the UAV and asymmetric path loss. Performance metrics of packet reception rate and received signal strength indicator are used to demonstrate asymmetric characteristics in UAV networks, which can cause adverse impacts on data collection efficiency and network stability in wireless sensor networks. A similar study can be found in~\cite{asymmetry_uav_journal}. Here, the asymmetry of UL and DL in UAV networks is analyzed from the perspective of the mobility of the UAV. The mobility of the UAV introduces rapid fluctuations in UL signal quality compared to the DL due to dynamic path loss variation caused by changing UAV position and propagation conditions.

\subsection{Various HARQ Types}
In \cite{CC_HARQ_Evaluation}, packet error rate and throughput performance with CC and different types of HARQ for an LTE orthogonal frequency-division multiple access (OFDMA) system are investigated. The HARQ Type-II yields the best performance in both packet error rate and throughput, while it has an additional cost of memory to operate. However, the HARQ Type-III provides a good trade-off compared to the Type-II cases, which is also reported in \cite{CC_HARQ_Evaluation_2}. 

Performance evaluation of the LTE physical layer using a link-level simulator in MATLAB is conducted in~\cite{LTE_performance_eval_matlab}. Among the key findings, the study examines the impact of HARQ retransmissions on error rate and throughput. Rather than exploring various types of HARQ mechanisms in the aforementioned studies, this work primarily evaluates the performance difference between HARQ-enabled and disabled cases. The numerical results indicate that the HARQ can significantly improve the error rate in low SNR ranges.



\subsection{Cellular Links for UAV Communications}
A cognitive UAV-assisted cellular network has been studied in~\cite{cognitive_UAV}. This study models cellular environments by optimizing power allocation to maintain a balance between energy efficiency and latency constraint to support ultra-reliable and low-latency communications (URLLCs) and massive machine-type communication (mMTC) services in 5G. A Lagrangian method with Karush–Kuhn–Tucker conditions is used for formulating the joint optimization problem. The numerical results show that energy efficiency is improved while meeting latency constraints.

In~\cite{endtoend_measure}, an end-to-end performance evaluation of UAV communication systems over a 5G network is presented. The performance evaluation focuses on the command and control and video transmission with real-world measurements, including throughput, signal quality, and network slicing performance. The results highlight that temporal performance degradation is observed at a high altitude due to the frequent handovers. 

\subsection{ISM Bands for UAV Communications}
In~\cite{DJI_paper}, the security vulnerabilities of the DJI DroneID protocol are investigated, which operates in the ISM bands. The study reveals a lack of encryption, leading to eavesdropping and location spoofing. The authors also conduct reverse engineering and fuzz testing to analyze the UAV communication protocol in the ISM bands. This research highlights critical privacy and security risks associated with UAV communication in the ISM bands.

In~\cite{PLE_table_asymmetry}, link characterization in aerial wireless sensor networks for low-power ISM bands with ZigBee links. The authors emphasize that the standard free space propagation assumption does not capture realistic environmental factors such as interference and multi-path fading. The measurement results show that ISM band interference and multi-path fading significantly impact communication reliability.

The characterization of low-power UAV-assisted wireless sensor networks operating in ISM bands is studied in~\cite{asymmetry_uav_journal}. Here, the impact of UAV dynamics and interference on link reliability and UL asymmetry is shown by the experimental results. It is highlighted that deploying UAVs in ISM bands causes unpredictable fluctuation due to the presence of interference in ISM bands and the mobility of UAVs. Similar results are also reported in~\cite{asymmetry_UAV}.

\subsection{Measurements for UAV Communications}
In~\cite{UAV_eval_letter}, a cellular-connected UAV testbed for real-time video streaming and control is presented. This study evaluates end-to-end performance using metrics such as delay, throughput, and reliability with the different settings of UAV flight conditions, video resolutions, and control signal transmission frequencies. The experiment results show that the frequent transmission of the control signals degrades the video transmission quality and control reliability due to buffer overflows.

LTE-based throughput performance in a suburban environment is analyzed in \cite{LTE_based_UAV}. Real measurements of received signal strength in a suburban area in Malaysia are used not only for performance evaluation but also for considering the coexistence of terrestrial and aerial users. The results show that the throughput performance is degraded when the aerial users reach a higher altitude between $40$~m and $60$~m.

In~\cite{uav_channel_model_urban}, a measurement-based channel model is developed for A2G and air-to-air (A2A) urban scenarios. A wide-band channel sounder is used for the measurement campaign to analyze path loss, shadowing, and multi-path components. A path loss model with a simplified Saleh-Valenzuela model is derived to characterize multi-path clustering and scattering in urban scenarios.   

In~\cite{rural_coverage_LTE}, the capability of using LTE networks for UAV command and control links in rural areas is investigated. This study is based on real measurements and numerical results, focusing on the impact of LTE coverage on the altitude of the UAV. The key findings of this paper highlight that interference is the primary limiting factor in LTE coverage.

\begin{figure}[t!]
    \centering
    \subfigure[Urban, 2024 Packapalooza.]{\includegraphics[width=0.97\columnwidth]{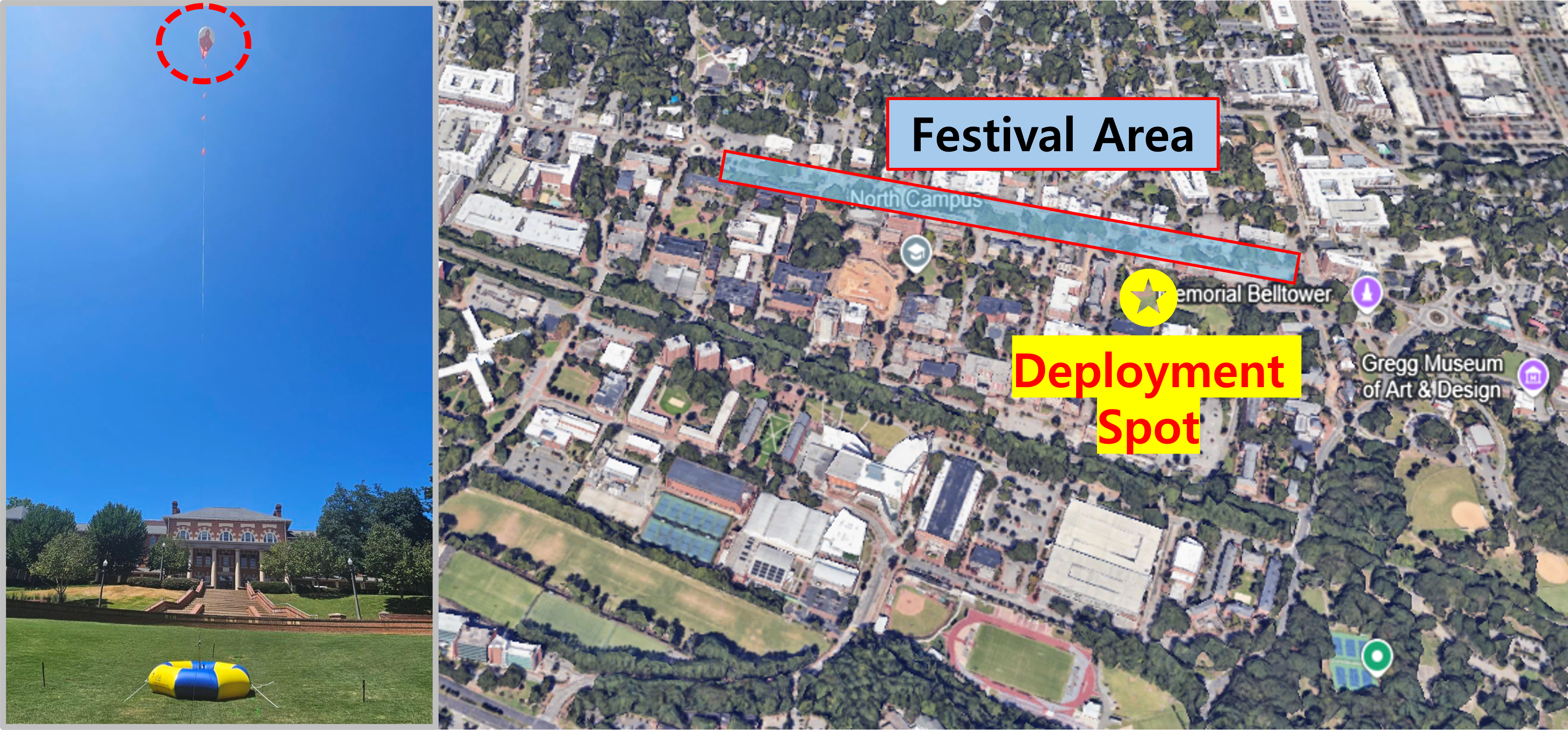}
    \label{fig:measurement_campaign_2024_packapalooza}}
    \subfigure[Rural, 2024 Lake Wheeler Field Lab.]{\includegraphics[width=0.97\columnwidth]{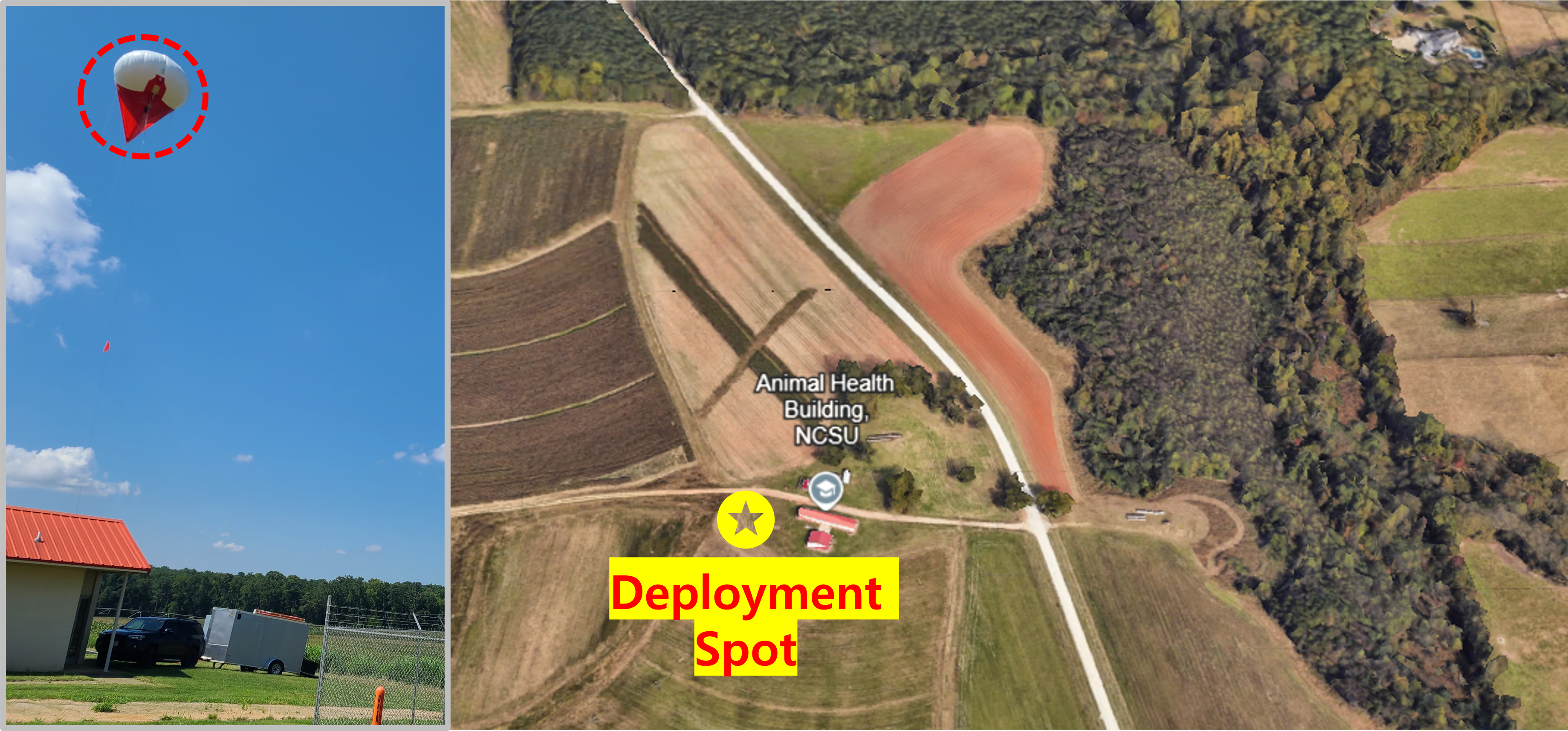}
    \label{fig:measurement_campaign_2024_LW}}
    \caption{Helikite measurement campaign in 2024 Packapalooza Festival and Lake Wheeler Field Lab.}
    \label{fig:measurement_campaign_pictures}
\end{figure}

\begin{figure}[t!]
    \centering
    \subfigure[Urban, 3D trajectory.]{\includegraphics[trim={0.6cm, 0, 1.3cm, 0.35cm},clip, width=0.47\columnwidth]{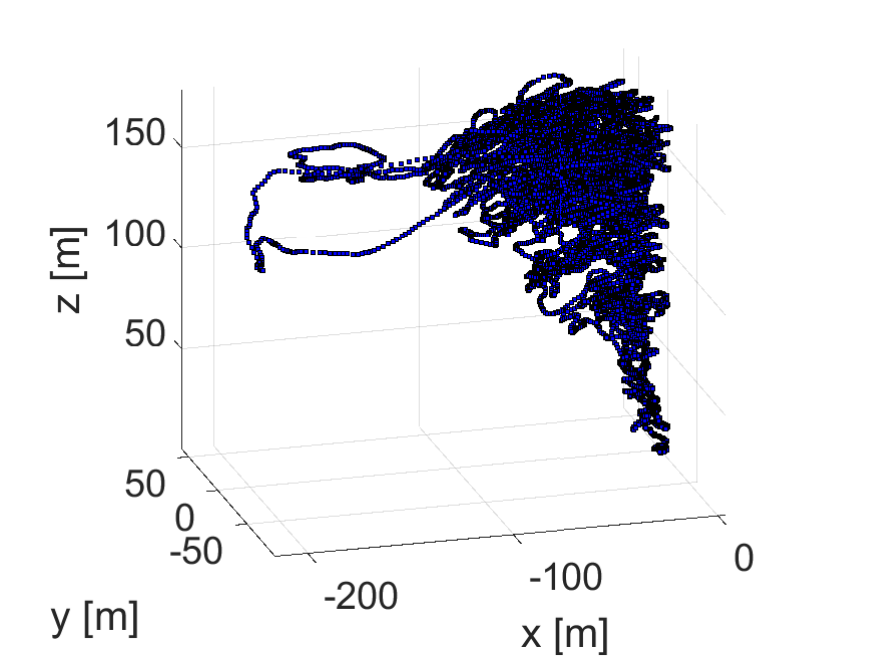}
    \label{fig:3D_trajectory_2024_packapalooza}}
    \subfigure[Urban, altitude over time.]{\includegraphics[trim={0.3cm, 0, 1.3cm, 0.35cm},clip, width=0.47\columnwidth]{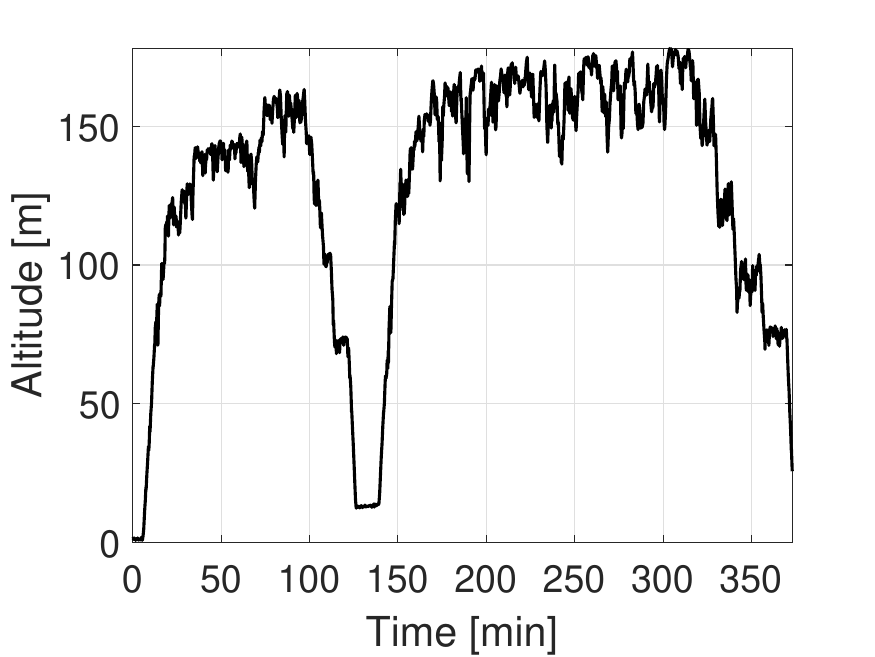}
    \label{fig:altitude_time_2024_packapalooza}}
    \subfigure[Rural, 3D trajectory.]{\includegraphics[trim={0.6cm, 0, 1.3cm, 0.35cm},clip, width=0.47\columnwidth]{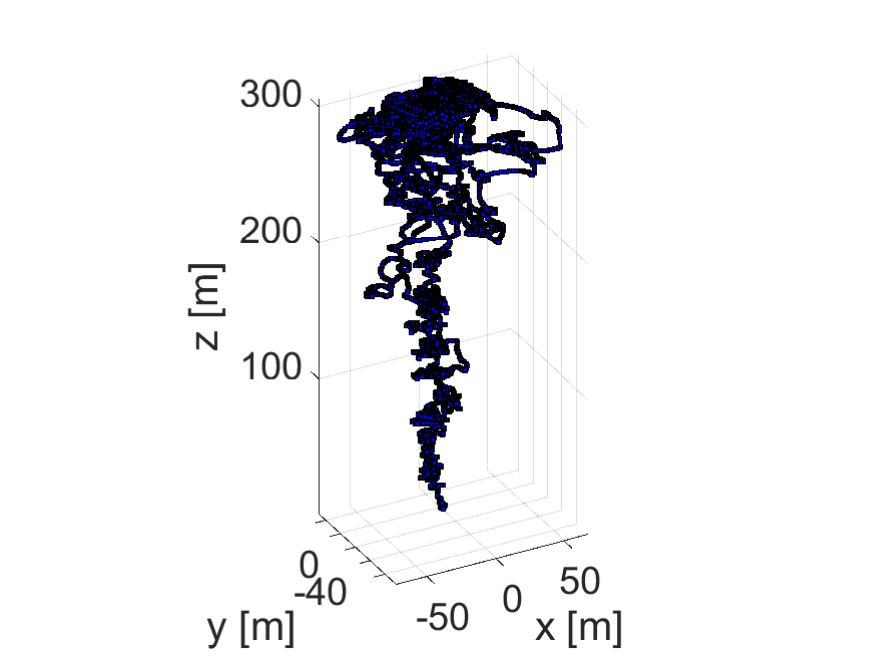}
    \label{fig:3D_trajectory_2024_LW}}
    \subfigure[Rural, altitude over time.]{\includegraphics[trim={0.2cm, 0, 0.9cm, 0.35cm},clip, width=0.47\columnwidth]{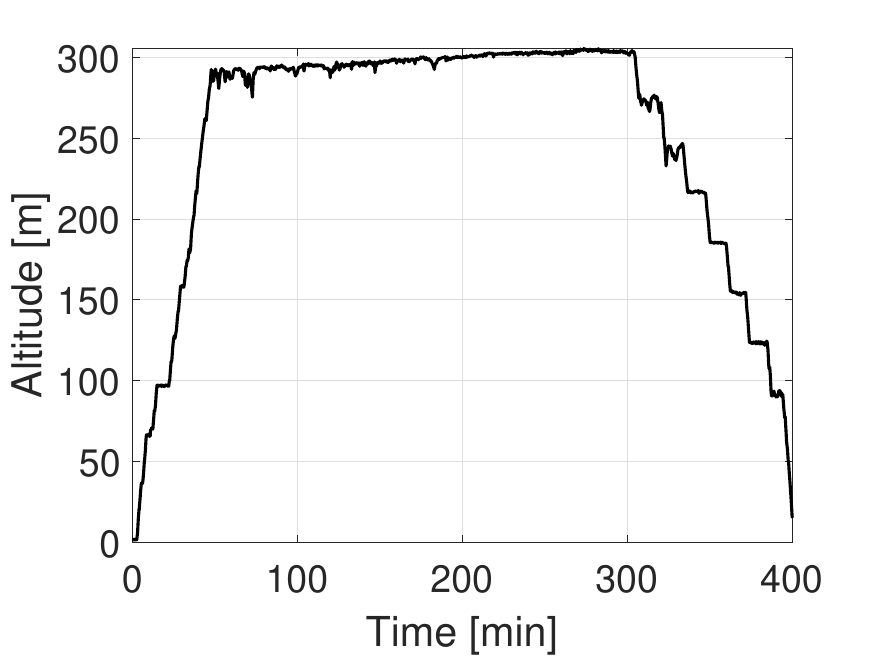}
    \label{fig:altitude_time_2024_LW}}
    \caption{3D helikite trajectory and altitude over time for urban 2024 Packapalooza in (a) and (b), and rural Lake Wheeler in (c) and (d).}
    \label{fig:experiement_setups_helikite}
\end{figure}

\section{HELIKITE MEASUREMENT CAMPAIGN}\label{ch:measurement}
Understanding the distribution of UL ISM band signals is crucial for identifying UL asymmetry in RC-UAV links. To obtain empirical evidence of the UL asymmetry, we conducted a measurement campaign using the helikite platform in both urban and rural environments~\cite{maeng_2024_Packapalooza, Maeng_2025}. By capturing received signal power variations at varying altitudes, the experiments provide physical insight into the interference affecting RC-UAV links. 

The helikite experiment for the urban areas was conducted at the Main Campus of NC State University during the 2024 Packapalooza Festival in August $2024$. For the rural areas, a similar experiment was conducted at the Lake Wheeler Field Lab at NC State University in August 2024. The locations of the deployment spot over the experiment areas are shown in Figure~\ref{fig:measurement_campaign_pictures}. Helikite was floated without external power for up to $400$~minutes, so it is possible to capture a large amount of received signal strength data within ISM bands at different altitudes, which we consider to represent the interference signals over RC-UAV links as demonstrated in Figure~\ref{fig:A2G_model}. The 3D trajectories and altitude over time of the helikite for each location are shown in Figure~\ref{fig:experiement_setups_helikite}. Note that the rural experiment was conducted in controlled environments exclusively designed for the measurement campaign. On the other hand, the helikite in the urban area was also used by the festival organizer for safety monitoring purposes during the festival, hence, it was not possible to control the altitude of the helikite during the experiments.

\subsection{Helikite A2G Measurement Setup}

\begin{figure}[t!]
    \centering
    \subfigure[Spectrum sweep procedure.]{\includegraphics[trim={0.2cm, 0.9cm, 0.2cm, 0.2cm},width=0.7\columnwidth]{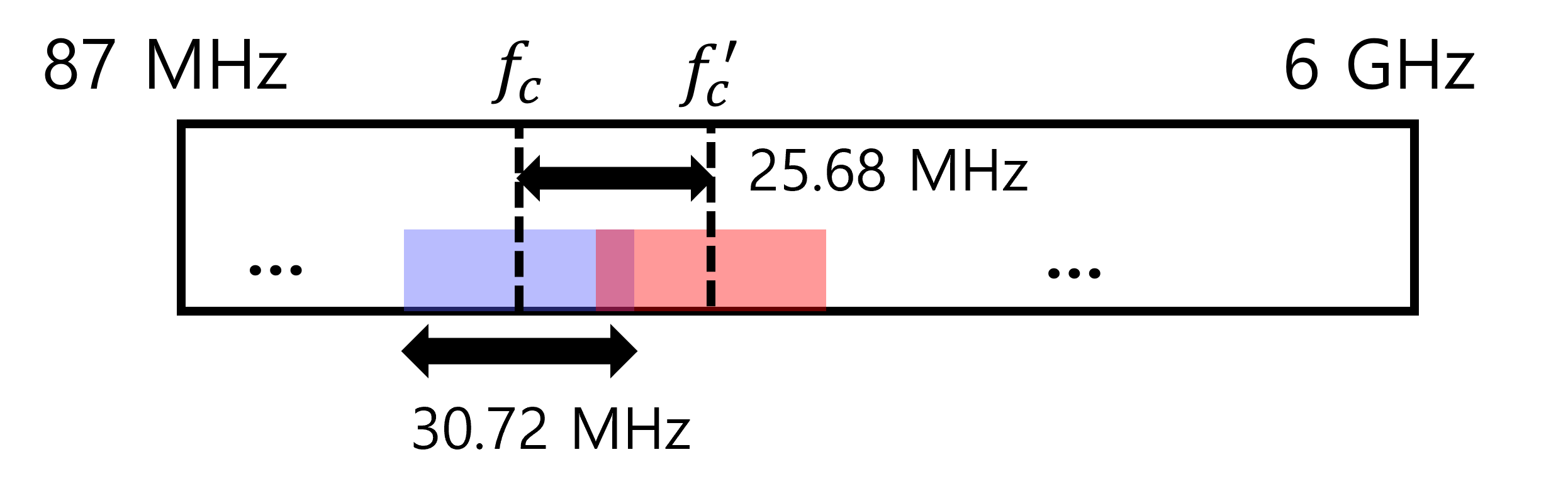}
    \label{fig:spectrum_sweep}
    }
    \subfigure[Non-overlapping channels in $2.4$~GHz ISM band.]{\includegraphics[trim={0.2cm, 0.9cm, 0.2cm, 0.2cm},width=0.75\columnwidth]{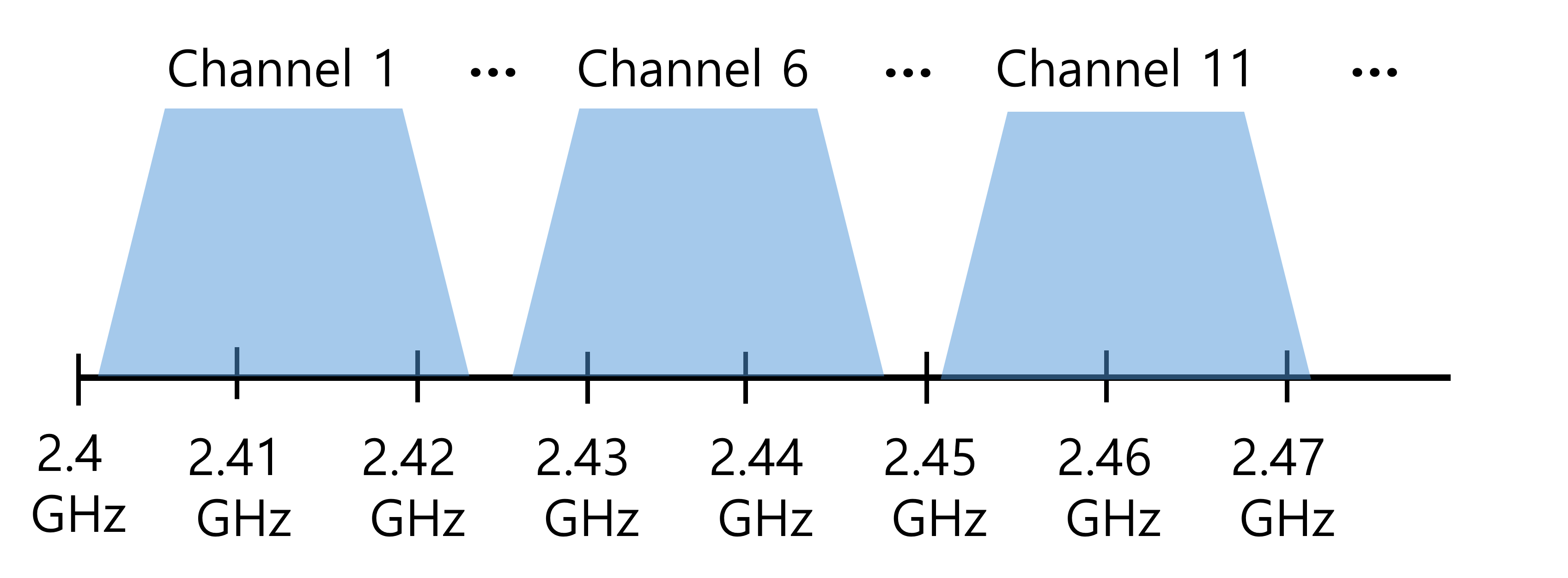}
    \label{fig:non_overlapping_channels}}
    \caption{Spectrum sweep procedure and non-overlapping channels in $2.4$~GHz ISM band.}    
    \label{fig:spectrum_and_channels}
\end{figure}


\begin{figure*}[t!]
    \centering
    \subfigure[Channel 1.]{\includegraphics[trim={0.4cm, 0, 1.3cm, 0.6cm},clip, width=0.65\columnwidth]{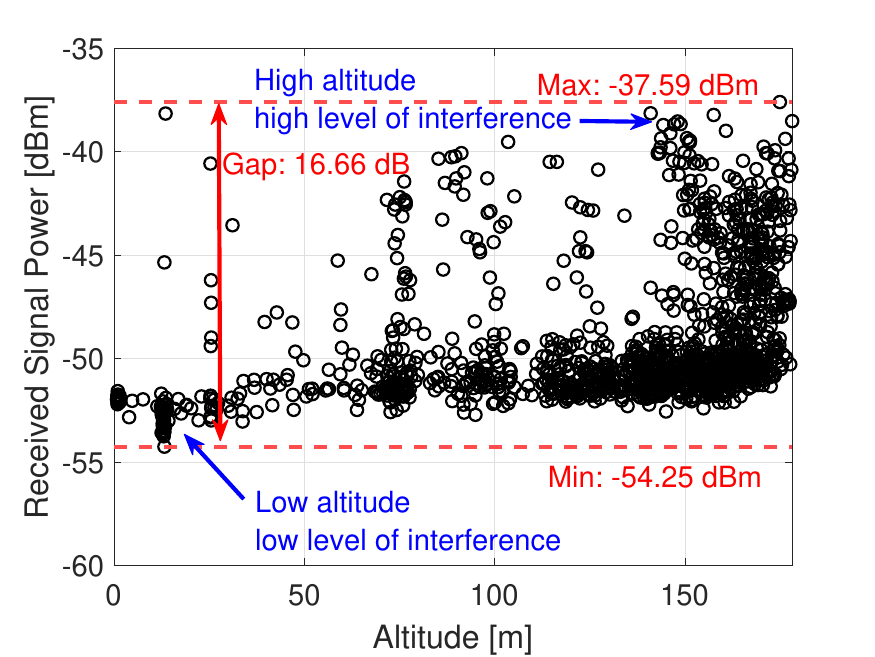}
    \label{fig:rss_altitude_2024_Packapalooza_channel_1}}
    \subfigure[Channel 6.]{\includegraphics[trim={0.4cm, 0, 1.3cm, 0.6cm},clip, width=0.65\columnwidth]{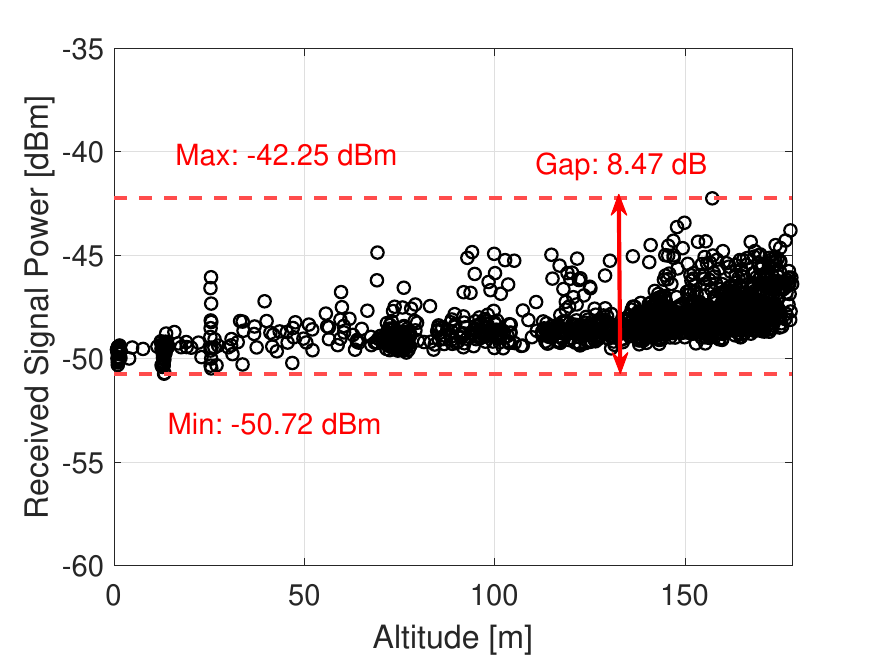}
    \label{fig:rss_altitude_2024_Packapalooza_channel_6}}
    \subfigure[Channel 11.]{\includegraphics[trim={0.4cm, 0, 1.3cm, 0.6cm},clip, width=0.65\columnwidth]{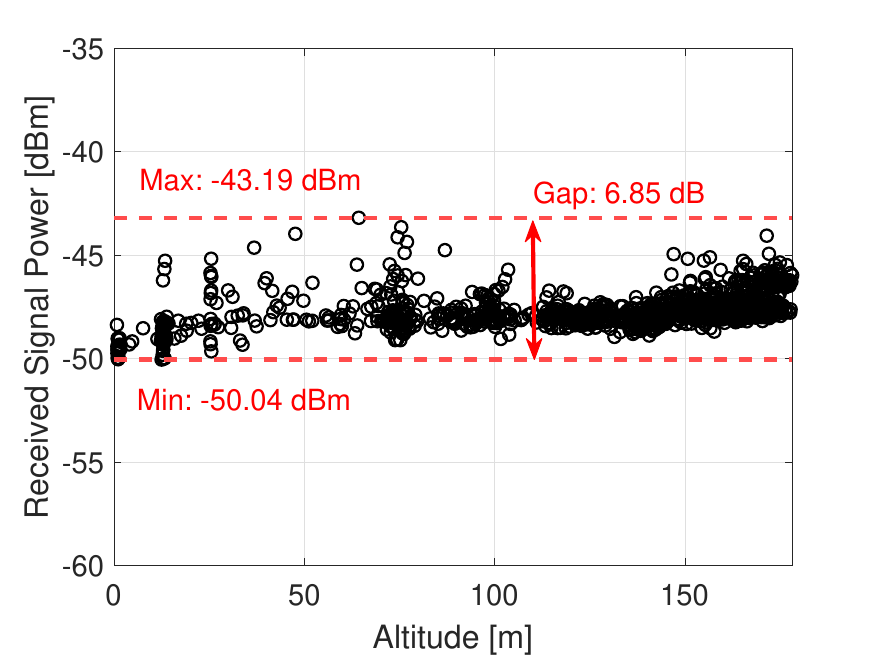}
    \label{fig:rss_altitude_2024_Packapalooza_channel_11}}
    \subfigure[Channel 1.]{\includegraphics[trim={0.4cm, 0, 0cm, 0.6cm},clip, width=0.65\columnwidth]{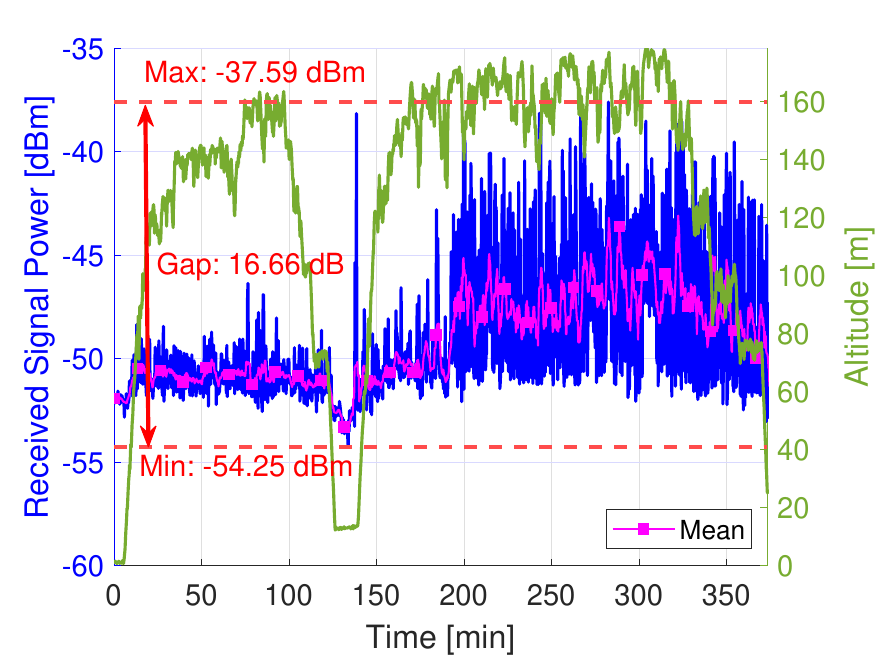}    \label{fig:rss_time_altitude_2024_Packapalooza_channel_1}}
    \subfigure[Channel 6.]{\includegraphics[trim={0.4cm, 0, 0cm, 0.6cm},clip, width=0.65\columnwidth]{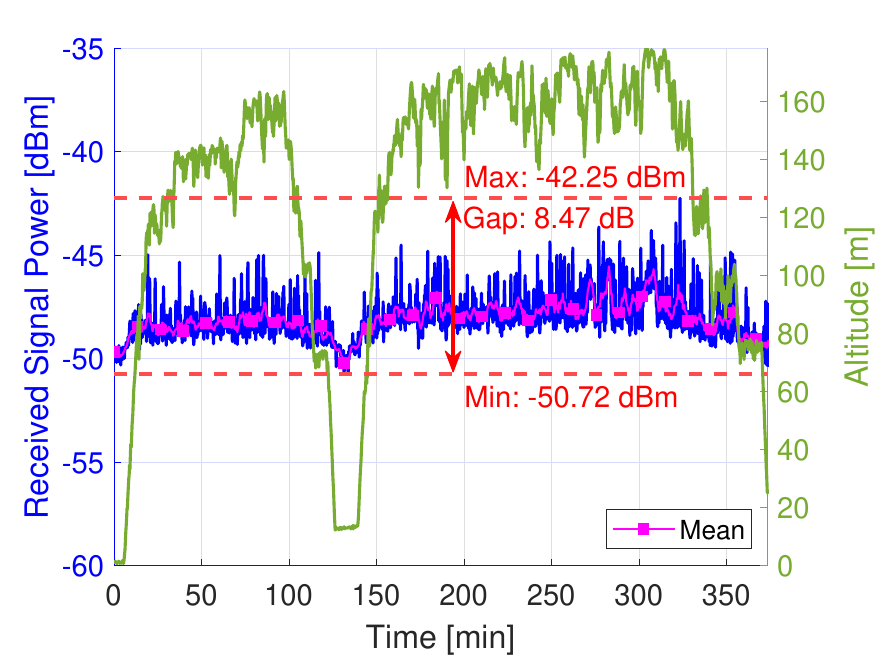}    \label{fig:rss_time_altitude_2024_Packapalooza_channel_6}}
    \subfigure[Channel 11.]{\includegraphics[trim={0.4cm, 0, 0cm, 0.6cm},clip, width=0.65\columnwidth]{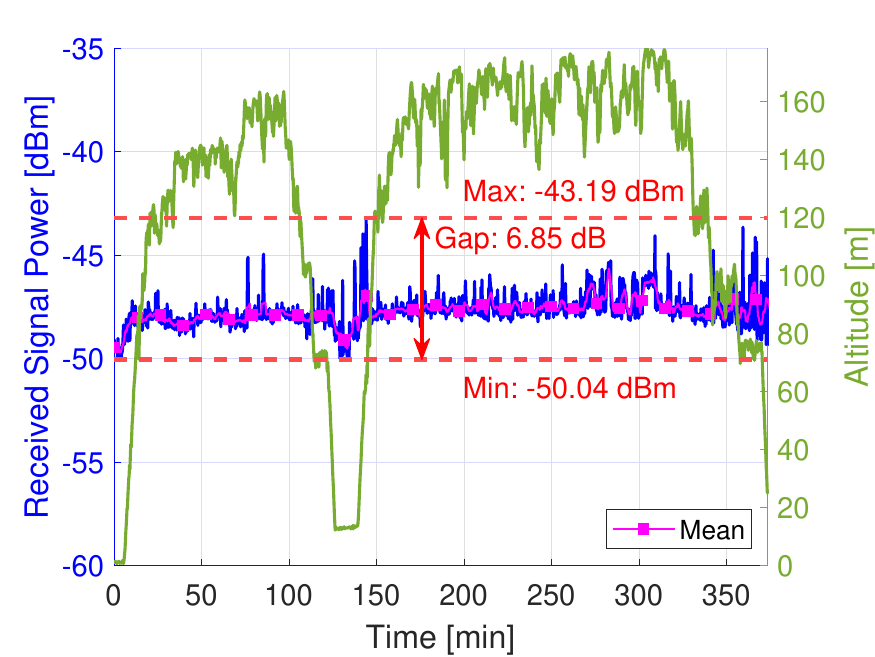}    \label{fig:rss_time_altitude_2024_Packapalooza_channel_11}}
    \subfigure[Channel 1.]{\includegraphics[trim={0.4cm, 0, 0.8cm, 0.6cm},clip, width=0.65\columnwidth]{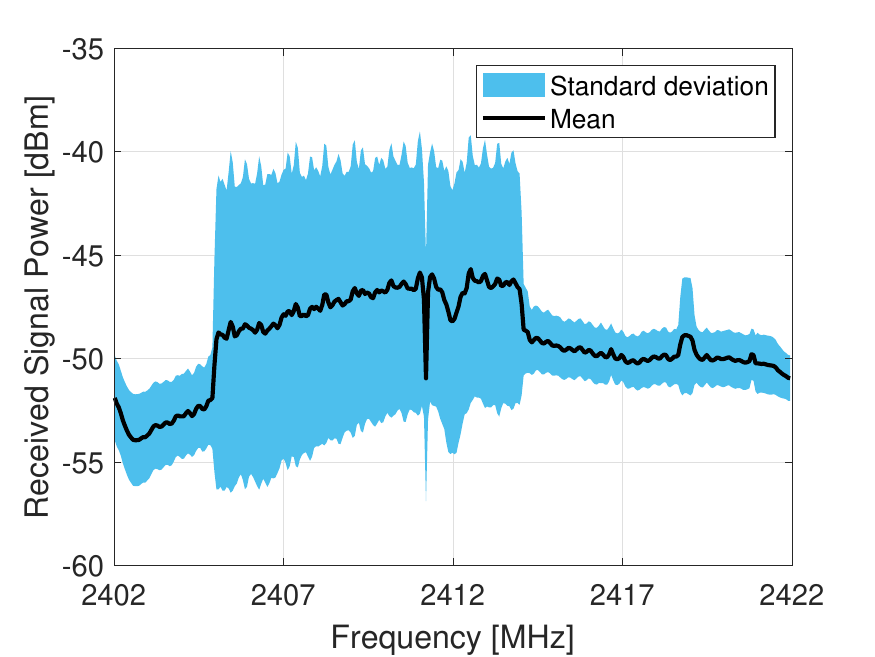}
    \label{fig:rss_freq_2024_Packapalooza_channel_1}}
    \subfigure[Channel 6.]{\includegraphics[trim={0.4cm, 0, 0.8cm, 0.6cm},clip, width=0.65\columnwidth]{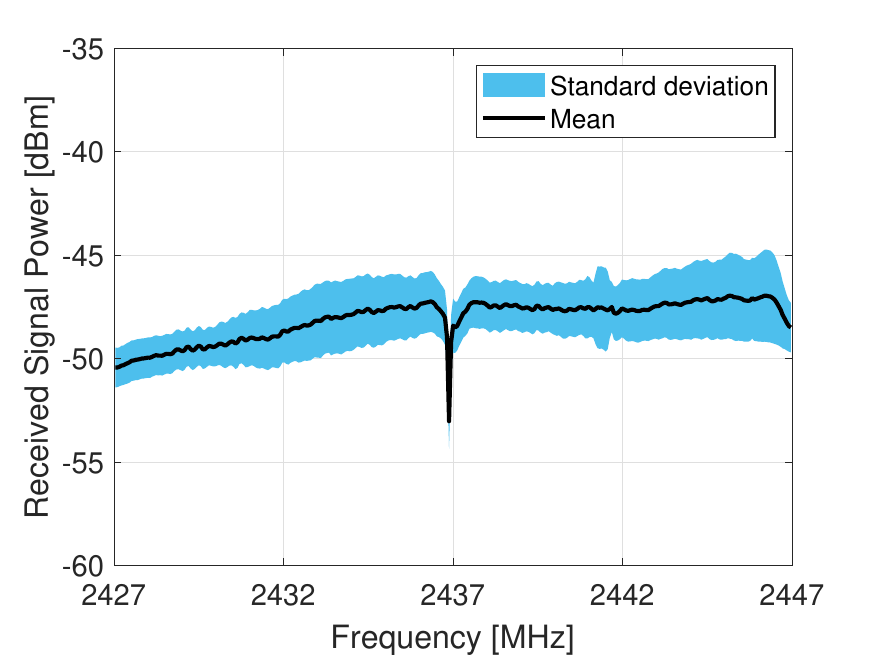}
    \label{fig:rss_freq_2024_Packapalooza_channel_6}}
    \subfigure[Channel 11.]{\includegraphics[trim={0.4cm, 0, 0.8cm, 0.6cm},clip, width=0.65\columnwidth]{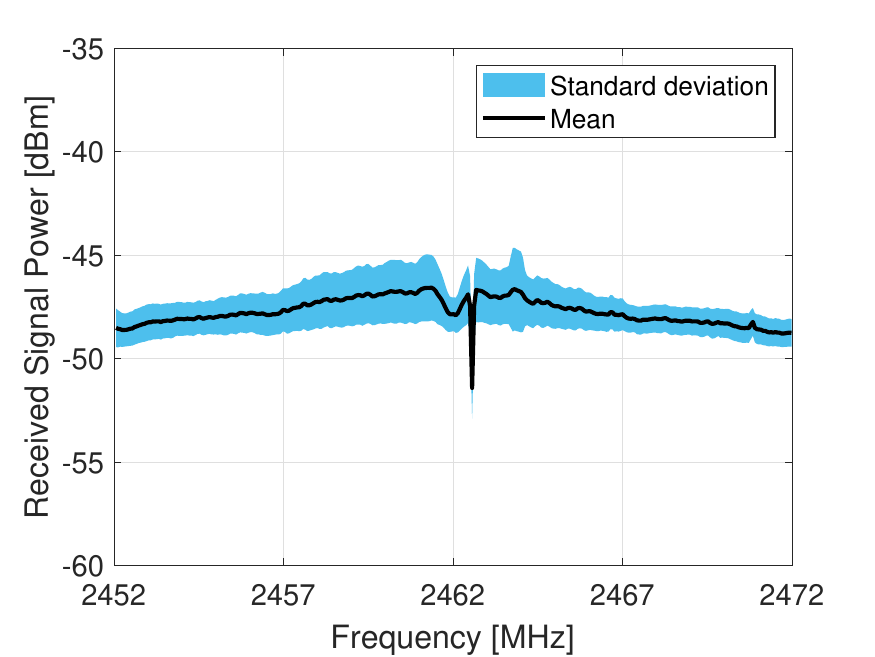}
    \label{fig:rss_freq_2024_Packapalooza_channel_11}}    \vspace{-0.1cm}
    \caption{Received signal power measurements in the urban areas of the 2024 Packapalooza Festival. (a), (b), and (c) show measurements of received signal power over altitude. (d), (e), and (f) show received signal power over time along with the altitude for analysis purposes. Lastly, (g), (h), and (i) show the received signal power over the frequency domains.}
    \label{fig:rss_altitude_freq_2024_Packapalooza}
\end{figure*}

\begin{figure*}[t!]
    \centering
    \subfigure[Channel 1.]{\includegraphics[trim={0.4cm, 0, 1.2cm, 0.65cm},clip, width=0.65\columnwidth]{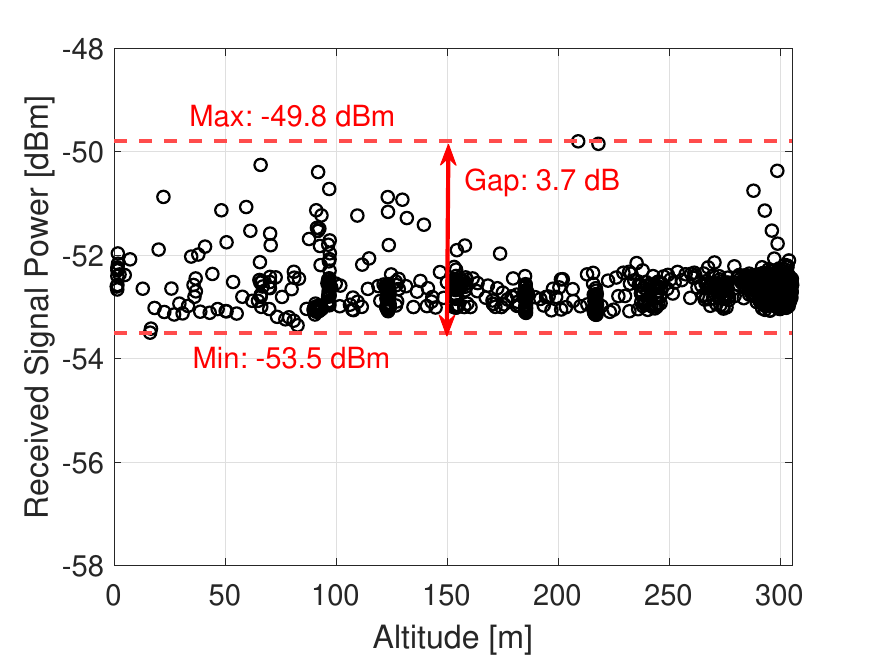}
    \label{fig:rss_altitude_2024_LW_channel_1}}
    \subfigure[Channel 6.]{\includegraphics[trim={0.4cm, 0, 1.2cm, 0.65cm},clip, width=0.65\columnwidth]{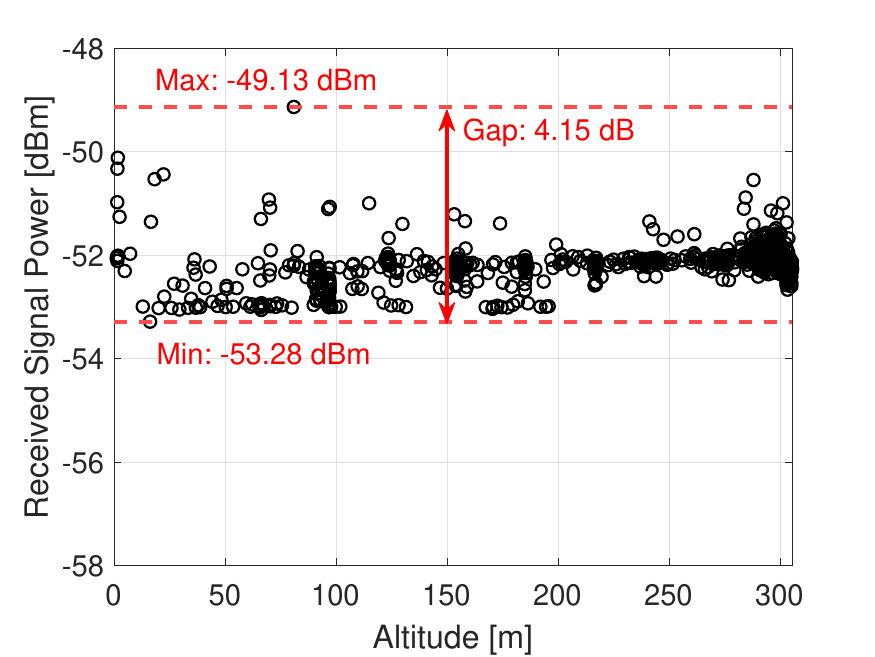}
    \label{fig:rss_altitude_2024_LW_channel_6}}
    \subfigure[Channel 11.]{\includegraphics[trim={0.4cm, 0, 1.2cm, 0.65cm},clip, width=0.65\columnwidth]{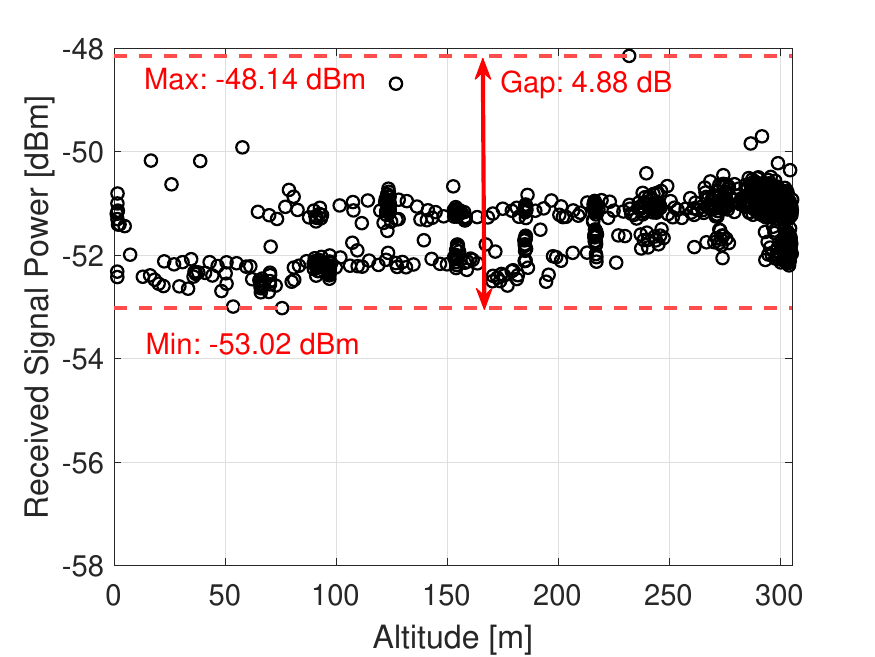}
    \label{fig:rss_altitude_2024_LW_channel_11}}
    \subfigure[Channel 1.]{\includegraphics[trim={0.4cm, 0, 0cm, 0.6cm},clip, width=0.65\columnwidth]{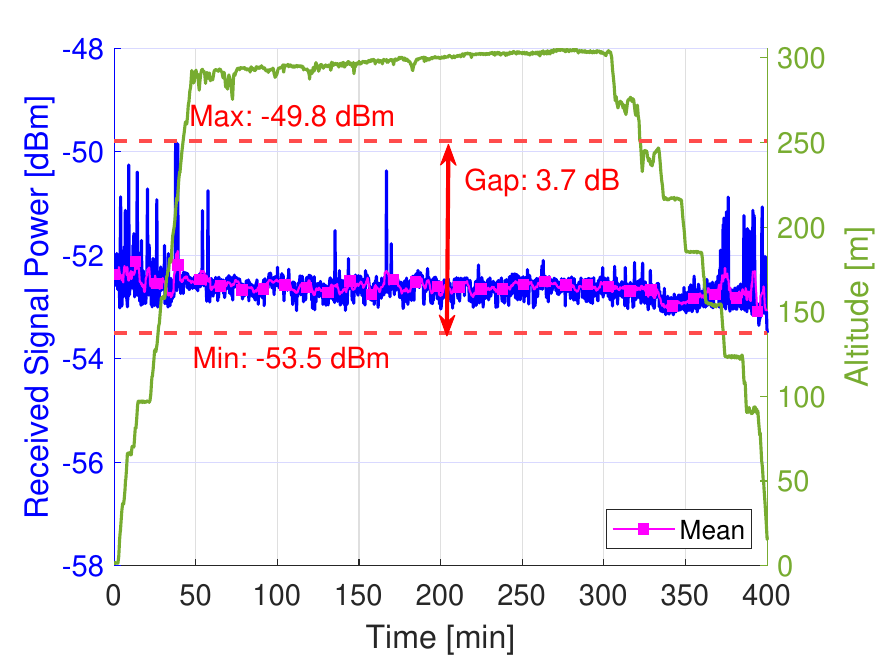}
    \label{fig:rss_time_altitude_2024_LW_channel_1}}
    \subfigure[Channel 6.]{\includegraphics[trim={0.4cm, 0, 0cm, 0.6cm},clip, width=0.65\columnwidth]{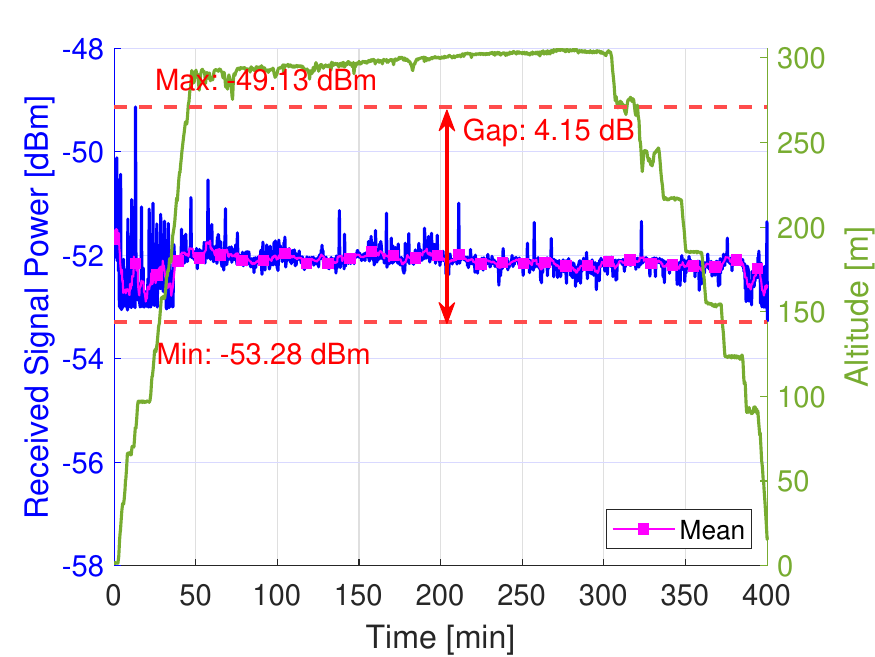}
    \label{fig:rss_time_altitude_2024_LW_channel_6}}
    \subfigure[Channel 11.]{\includegraphics[trim={0.4cm, 0, 0cm, 0.6cm},clip, width=0.65\columnwidth]{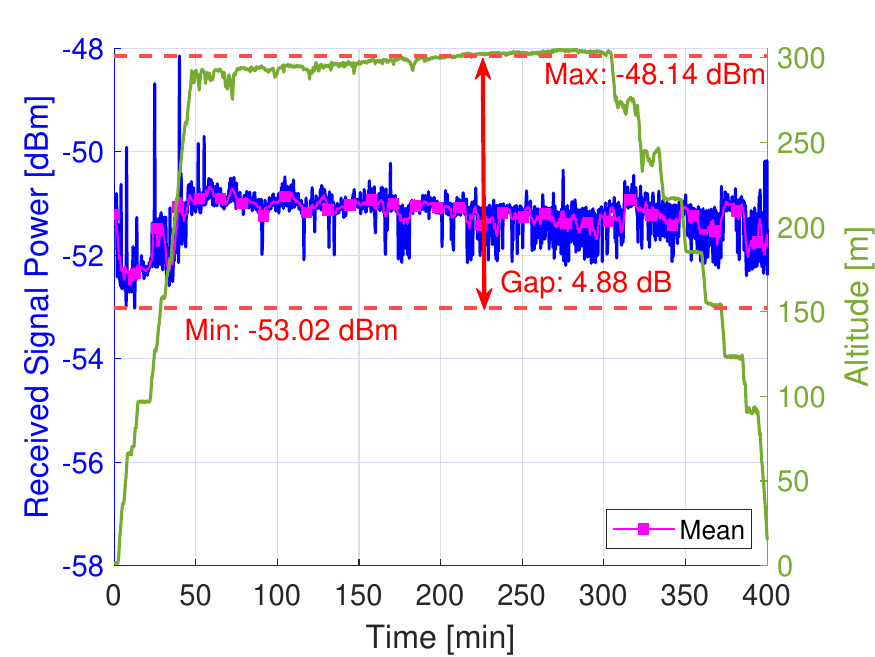}
    \label{fig:rss_time_altitude_2024_LW_channel_11}}
    \subfigure[Channel 1.]{\includegraphics[trim={0.4cm, 0, 0.8cm, 0.65cm},clip, width=0.65\columnwidth]{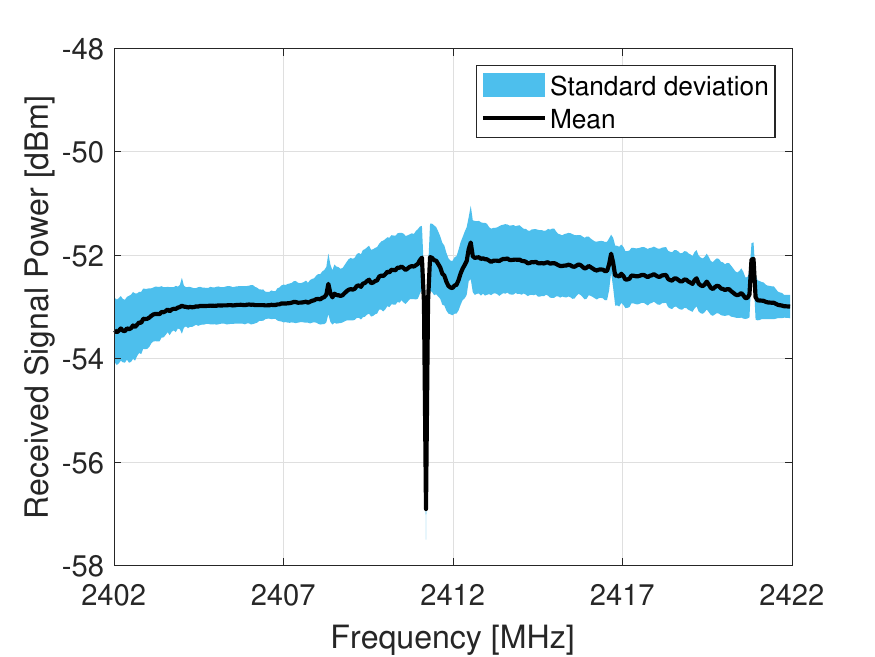}
    \label{fig:rss_freq_2024_LW_channel_1}}
    \subfigure[Channel 6.]{\includegraphics[trim={0.4cm, 0, 0.8cm, 0.65cm},clip, width=0.65\columnwidth]{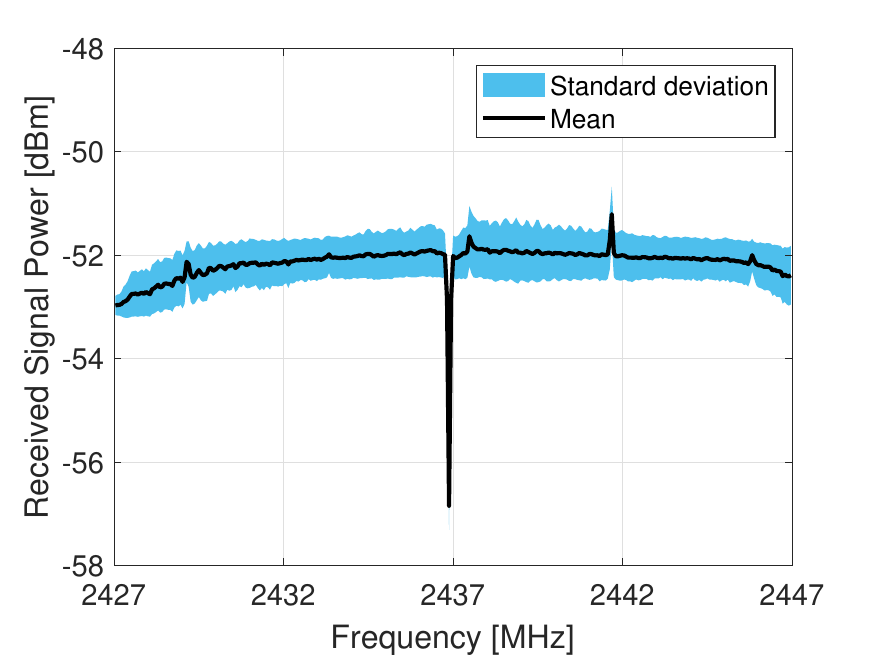}
    \label{fig:rss_freq_2024_LW_channel_6}}
    \subfigure[Channel 11.]{\includegraphics[trim={0.4cm, 0, 0.8cm, 0.65cm},clip, width=0.65\columnwidth]{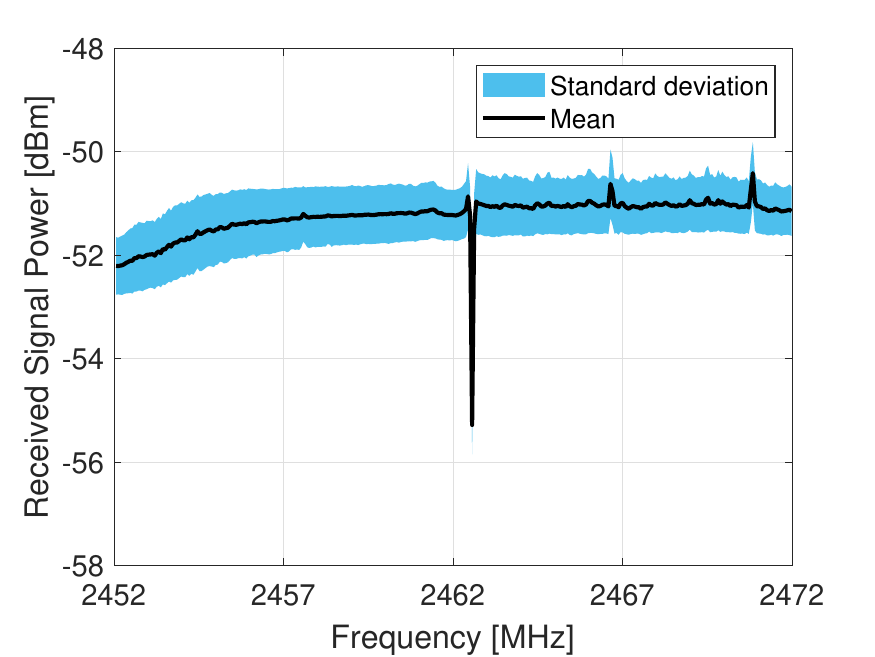}
    \label{fig:rss_freq_2024_LW_channel_11}}
    \vspace{-0.1cm}
    \caption{Received signal power measurements in the rural areas of the 2024 Lake Wheeler Field Lab. (a), (b), and (c) show measurements of received signal power over altitude. (d), (e), and (f) show received signal power over time along with the altitude for analysis purposes. Lastly, (g), (h), and (i) show the received signal power over the frequency domains.}
    \label{fig:rss_altitude_freq_2024_LW}
\end{figure*}

A software-defined radio (SDR) and a global positioning system (GPS) receiver were mounted on the helikite. The SDR module continuously captures signal power over the spectrum from $87$~MHz to $6$~GHz. The procedure of spectrum sweep for the experiments is depicted in Figure~\ref{fig:spectrum_sweep}~\cite{maeng_2024_Packapalooza}, where the center frequency shift is $25.68$~MHz and the sampling rate is $30.72$~MHz, respectively. Both urban and rural experiments were conducted for over six hours starting from noon, where each $6$~GHz sweep takes around $15$ seconds. Over the whole spectrum sweeping measurement, we isolated the $2.4$~GHz ISM band, i.e., from $2.4$~GHz to $2.5$~GHz. Moreover, we consider $20$~MHz bandwidth of non-overlapping channels in the $2.4$~GHz ISM band: Channels $1$, $6$, and $11$ with center frequencies of $2.412, 2.437,$ and $2.462$~GHz, respectively, which are depicted in Figure~\ref{fig:non_overlapping_channels}. The dataset used in this experiment is publicly available at~\cite{aerpawWebsite}.

The continuous GPS data from the GPS receiver mounted on the helikite facilitates tracking the altitude and location of the helikite, which are shown in Figure~\ref{fig:experiement_setups_helikite}, for the urban and rural areas. The altitude of the helikite was controlled by the length of the tether tied to the helikite. Both urban and rural cases in Figure~\ref{fig:3D_trajectory_2024_packapalooza} and~\ref{fig:3D_trajectory_2024_LW} show that the location of the helikite varied, which came from weather conditions such as wind speed and direction. In Figure~\ref{fig:altitude_time_2024_packapalooza}, the helikite in urban areas gradually ascended to $160$~m and then descended to around $20$~m in altitude. After this descent, the helikite ascended again to the $160$~m altitude range and stayed there for around $3$~hours before gradually descending to the ground. In addition, Figure~\ref{fig:altitude_time_2024_LW} shows the altitude over time for rural areas. The helikite gradually ascended to the $300$~m range and stayed there for around $4$~hours for measurements. After this, the helikite gradually descended to the ground over around $2$~hours duration.



\subsection{Helikite Measurement Results}
The helikite measurement results of the received signal power for the urban 2024 Packapalooza Festival and Lake Wheeler Field Lab are shown in Figures~\ref{fig:rss_altitude_freq_2024_Packapalooza} and~\ref{fig:rss_altitude_freq_2024_LW}, respectively. The measurement results are organized as follows: 1) received signal power over altitude in (a)-(c), 2) received signal power over time with altitude information in (d)-(f), and 3) received signal power over frequency domain in (g)-(i). The measurement results are shown separately for each channel. 

The received signal power measurements over altitude are shown in Figures~\ref{fig:rss_altitude_2024_Packapalooza_channel_1}-\ref{fig:rss_altitude_2024_Packapalooza_channel_11}. It is observed that channel $1$ has the strongest signals among all these channels. This is consistent with the fact that channel $1$ is assigned as a default channel for ISM band operating equipment. Received signal power in channel $1$ has the largest range of variation, which is observed in the $16.66$~dB range between minimum and maximum measurements. In contrast, the range of deviation with altitude is around $8$~dB in channels $6$ and $11$. Moreover, the received signal power tends to be higher at higher altitudes due to the high chance of LoS ISM signal sources. This implies higher interference at UAVs flying at higher altitudes.

The received signal power of the measurements over time along with the altitude information are shown in Figures~\ref{fig:rss_time_altitude_2024_Packapalooza_channel_1}-\ref{fig:rss_time_altitude_2024_Packapalooza_channel_11}, where blue curves show the aggregate received signal power with the y-axis on the left, magenta curve with square markers indicate the moving average aggregate signal power over the time with the window size of $10$ time stamps, which is equivalent to approximately $2.5$~minutes, and green curves show the altitude over time with the y-axis on the right side. The received signal power in channel $1$ is the strongest and deviates most with altitude, which has a range of around $16$~dB. An altitude-dependent pattern is also seen at a low altitude, specifically during the periods from $0$ to $10$~minutes and from $130$ to $140$~minutes. Within these intervals, the received signal power shows subtle fluctuations of an amount of $3$~dB, closely following the changes of altitude. In Figure~\ref{fig:rss_time_altitude_2024_Packapalooza_channel_1}, the variation range of the received signal power during the first ascending phase (from $20$ to $100$~minutes) is smaller than the second ascending phase (from $170$ to $320$~minutes). This increased fluctuation in the later phase is likely due to the higher density of people in the festival area later in the day. 

In Figures~\ref{fig:rss_freq_2024_Packapalooza_channel_1}-\ref{fig:rss_freq_2024_Packapalooza_channel_11}, the received signal power over frequency domain measurements is shown. In all cases, sharp drops are observed, caused by the null subcarrier in orthogonal frequency division multiplexing (OFDM). Similar to the previous results, channel $1$ shows the largest range of standard deviation, which tends to be severe at the $2405$-$2414$~MHz. 

The helikite experiment results for rural areas are shown in Figure~\ref{fig:rss_altitude_freq_2024_LW}, with the same structure as the urban cases. Compared to the urban cases, the rural measurements show a relatively lower signal power variation with altitude for all channels, which are shown in Figures~\ref{fig:rss_altitude_2024_LW_channel_1}-\ref{fig:rss_altitude_2024_LW_channel_11}. This trend is connected to the characteristics of the rural area in which our measurement campaign was conducted. There is a limited number of Wi-Fi sources within several kilometers of the helikite location, and it is an open field in general, which leads to a lower chance of receiving interference in the ISM band compared to the urban cases. The received signal power over time is close to constant in all channels as seen in Figure~\ref{fig:rss_altitude_freq_2024_Packapalooza}. A similar pattern can also be observed in the frequency domain, except for the null subcarrier in the center of each channel.


\section{SYSTEM MODEL}\label{CH:sys}

\subsection{A2G Network Model}
In our A2G network model, a UAV acts as the cellular base station, as in other commercial drones (see e.g.,~\cite{DJI_paper}), and the RC unit on the ground is the UE, as shown in Figure~\ref{fig:A2G_model}. We consider that this cellular link operates in the ISM bands, where other unlicensed communication links exist, such as Wi-Fi and Bluetooth. Therefore, the UAV observes a lower SINR in the UL, which results in DL/UL SINR asymmetry due to the LoS interference from other signal sources on the ground. It is expected that the SINR asymmetry will be exacerbated with a higher altitude of the UAV due to the higher probability of LoS with the interference sources~\cite{10200994}. 

\subsection{HARQ Types}
There are three commonly used versions of HARQ~\cite{HARQ_survey}. In HARQ Type-I, erroneous packets are discarded, and a retransmission request is sent. The transmitter sends an identical packet while the receiver successfully decodes the data, or the maximum number of retransmissions is reached. HARQ Type-I can be extended with CC to get a combining gain. HARQ Type-II is known as full IR. The receiver combines the retransmitted packet with the stored packet from the previous attempts. The IR gradually decreases the effective code rate of the packet in every retransmission attempt. Compared to the full IR in HARQ Type-II, HARQ Type-III employs partial IR, including the punctured data version. With this method, a receiver can extract the information bits independently in each retransmission, which is known as self-decodability \cite{HARQ_survey}.

\subsection{HARQ Indicators in LTE}
\begin{figure*}[t!]
    \centering
    \includegraphics[width=1.85\columnwidth]{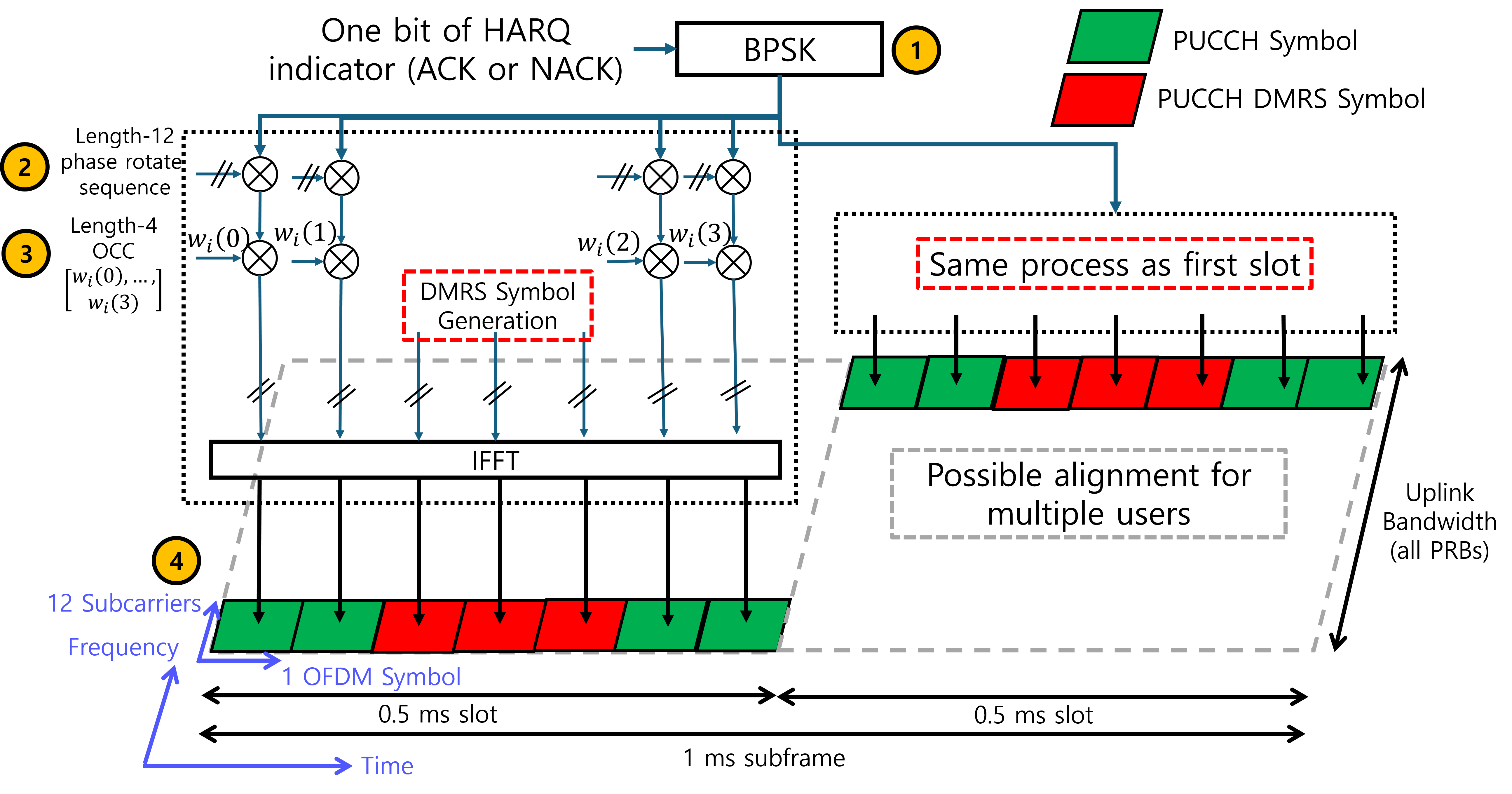}
    \caption{Structure and procedure for PUCCH format 1a in LTE (modified from~\cite{DAHLMAN2016}).}
    \label{fig:PUCCH_format1a}
\end{figure*}


In LTE, the HARQ indicator for ACK or NACK is carried by PUCCH. In this paper, we limit our scope of the PUCCH to the HARQ indicator in format 1a, which only carries a single bit of ACK or NACK. The structure and procedure for PUCCH format 1a in LTE are shown in Figure~\ref{fig:PUCCH_format1a}~\cite{DAHLMAN2016}, where the green blocks are PUCCH symbols and the red blocks are demodulation reference signal (DMRS) symbols for channel estimation purposes. The procedure can be summarized as follows. 
\begin{itemize}
    \item \textbf{Step-1.} A single bit of the HARQ indicator for ACK/NACK is modulated by the binary phase shift keying (BPSK) modulation method.
    
    \item \textbf{Step-2.} The modulated signal is spread into four possible locations and phase-rotated with the length-$12$ phase rotation sequence, which can be expressed as
    \begin{equation}
        x_n' = x_n \cdot e^{j \frac{m \pi}{6}n},
    \end{equation}
    where $x_n'$ is the phase-rotated output, $x_n$ indicates the input BPSK modulated signal, $n$ is the subcarrier index, and $m$ is the phase rotation index from $0$ to $11$, respectively.
    
    \item \textbf{Step-3.} The length-$4$ of the orthogonal cover code (OCC) with an OCC index $I_{\mathrm{occ}}$ is applied to the phase-rotated output for four possible locations, which is listed in Table~\ref{tab:occ_table}~\cite{chen2021fundamentals}. The four elements of the OCC are selected with $I_{\mathrm{occ}}$, e.g., OCC with $I_{\mathrm{occ}}=2$ is selected as $[w_{2}(0), w_{2}(1), w_{2}(2), w_{2}(3)]=[1, -1, 1, -1]$. Each element of the sequence is separately applied to the corresponding streams as illustrated in Figure~\ref{fig:PUCCH_format1a}.

    \item \textbf{Step-4.} The resulting sequence is mapped to the left and right edges of the uplink spectrum granted for the UE.
    
\end{itemize}
A total of $8$ OFDM symbols are allocated for the HARQ indicator, spanning $12$ subcarriers in the frequency domain. This results in $8 \times 12 = 96$ HARQ indicator modulated symbols distributed across the resource grid's time and frequency domains. An alternative PUCCH placement at the edges of the UL bandwidth can provide additional frequency diversity gain by exposing control signaling to different channel conditions across frequency resources~\cite{DAHLMAN2016}.

\begin{table}[t!]
    \centering
    \caption{OCC used for PUCCH process in LTE and 5G~\cite{chen2021fundamentals}.}
    \begin{tabular}{|c|c|c|c|c|} \hline
        \multicolumn{5}{|c|}{\textbf{Length-4 OCC}} \\ \hline
        $I_{\mathrm{occ}}$ & $w_{I_{\mathrm{occ}}}(0)$ & $w_{I_{\mathrm{occ}}}(1)$ & $w_{I_{\mathrm{occ}}}(2)$ & $w_{I_{\mathrm{occ}}}(3)$ \\ \hline
        $1$ & $1$ & $1$ & $1$ & $1$ \\ \hline
        $2$ & $1$ & $-1$ & $1$ & $-1$ \\ \hline
        $3$ & $1$ & $1$ & $-1$ & $-1$ \\ \hline
        $4$ & $1$ & $-1$ & $-1$ & $1$ \\ \hline
        \multicolumn{5}{|c|}{\textbf{Length-3 OCC}} \\ \hline
        $I_{\mathrm{occ}}$ & $w_{I_{\mathrm{occ}}}(0)$ & $w_{I_{\mathrm{occ}}}(1)$ & $w_{I_{\mathrm{occ}}}(2)$ & \\ \hline
        $1$ & $1$ & $1$ & $1$ & \\ \hline
        $2$ & $1$ & $e^{j\frac{2\pi}{3}}$ & $e^{j\frac{4\pi}{3}}$ & \\ \hline
        $3$ & $1$ & $e^{j\frac{4\pi}{3}}$ & $e^{j\frac{2\pi}{3}}$ & \\ \hline
    \end{tabular}
    \label{tab:occ_table}
\end{table}

In the decoding stage of PUCCH, the decoder restores the phase rotation in the received PUCCH symbols and calculates the normalized error metric in the BPSK domain. It can be expressed as~\cite{matlab_LTE}
\begin{equation}\label{eq:error_metric_PUCCH_decode}
    e_{b}= \left| \frac{\sum_{i=1}^{N_r}(y_{i}-s_{b,i})s_{b, i}^* }{\sum_{i=1}^{N_r}|s_{b,i}|^2}\right|^2 ,
\end{equation}
where $e_{b}$ is the error metric for bit $b\in{0, 1}$ for BPSK, $y_i$ is the $i$-th received PUCCH symbol out of $N_r=8 \times 12=96$, $s_{b, i}$ is the desired PUCCH symbol for the corresponding bit of $b$, and $s_{b,i}^*$ is the complex conjugate of $s_{b, i}$ for phase rotation restoration purposes, respectively. As a result of~(\ref{eq:error_metric_PUCCH_decode}), $e_1$ and $e_0$, which show the error metric for BPSK $1$ and $0$, represent normalized and aggregated error metrics over all received PUCCH symbols. After the error metric calculation, the decoder selects the decoded bit that has the minimum error, which can be expressed as
\begin{equation}\label{eq:decoder}
    \hat{b} =
    \begin{cases}
    \displaystyle \arg\min_{b \in \{0,1\}} e_b, & \text{if } \min(e_b) < \tau_{\mathrm{LTE}}, \\
    \varnothing, & \text{otherwise},
    \end{cases}
\end{equation}
where $\tau_{\mathrm{LTE}}$ is a threshold, which is empirically chosen, that controls the sensitivity of the decoding process. The threshold-based selection determines whether the decoder should make a decision based on the error metric in~(\ref{eq:error_metric_PUCCH_decode}) or discard the result as unreliable output. To summarize, LTE PUCCH format 1a leverages $96$ PUCCH symbols to carry a single bit of HARQ indicator. This redundancy and threshold-based decoding process contributes to improving reliability in challenging channel conditions.

\subsection{HARQ Indicators in 5G}

\begin{figure*}[t!]
    \centering
    \includegraphics[width=1.85\columnwidth]{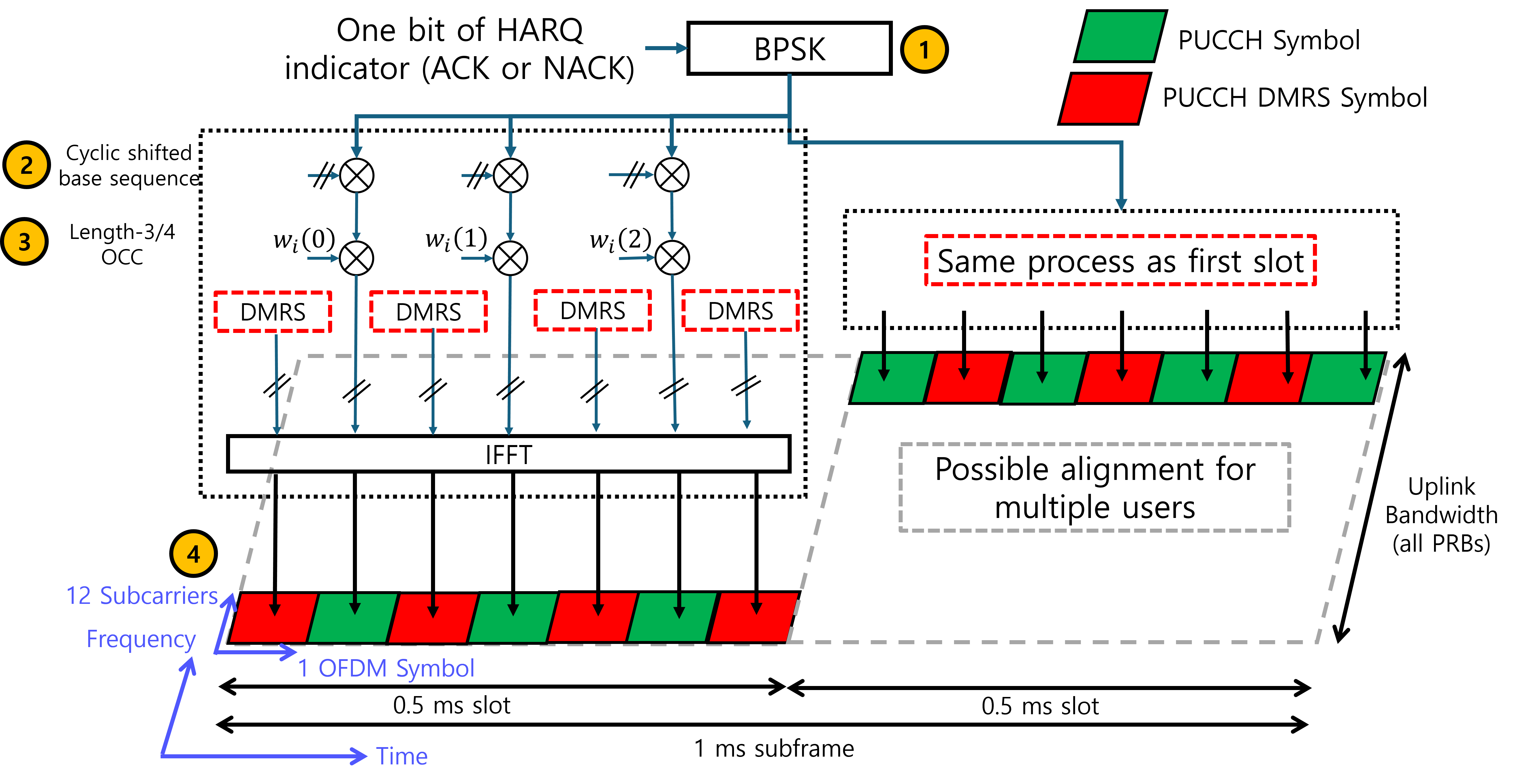}
    \caption{Structure and procedure for PUCCH format 1 in 5G. While 5G PUCCH format 1 allows a flexible number of allocated OFDM symbols, the illustration shows the case where the number of OFDM symbols assigned is the same as LTE for direct comparison. Moreover, the hopping option is also assumed to get a consistent location for PUCCH as in LTE.}
    \label{fig:PUCCH_format1_5G}
\end{figure*}

Similar to LTE, the HARQ indicator is processed by PUCCH in 5G. We consider PUCCH format 1 in 5G to facilitate comparison with LTE. It is known that PUCCH format 1 in 5G supports a flexible number of OFDM symbols, i.e., from $4$ to $14$ OFDM symbols, to carry a single bit of ACK or NACK. Note that we consider $14$ OFDM symbols and intra-slot hopping settings to facilitate direct comparison with the LTE. The structure and procedure for PUCCH format 1 in 5G are shown in Figure~\ref{fig:PUCCH_format1_5G}. The procedure can be summarized as follows.
\begin{itemize}
    \item \textbf{Step-1.} Similar to PUCCH format 1a in LTE, a single bit of HARQ indicator (ACK/NACK) is BPSK modulated.
    \item \textbf{Step-2.} A base sequence is first generated, which can be expressed as~\cite{chen2021fundamentals}
    \begin{equation}\label{eq:base_sequence_5G}
        \bar{r}_{u,v}(n)=e^{j\varphi(n)\pi/4},
    \end{equation}
    where $u$ is the Zadoff-Chu sequence group number, $v$ denotes the sequence index number, $n$ is the subcarrier index, and $\varphi(n)$ is determined from Table 9.4 in~\cite{chen2021fundamentals}. Then, a pseudo-random cyclic shift is applied to the base sequence. It can be written as
    \begin{equation}
        r_{u,v}(n)=e^{j\alpha n} \cdot \bar{r}_{u,v}(n),
    \end{equation}
    where $\alpha$ is the pseudo-random phase rotation sequence, which can be derived from the procedure in Section 6.3.2.2.2 in~\cite{TS_138_211}. The BPSK modulated signal is multiplied by this pseudo-random phase-rotated base sequence, which can be written as
    \begin{equation}\label{eq:final_5G}
        x_n'= x_n \cdot r_{u,v}(n),
    \end{equation}
    where $x_n$ is the BPSK modulated symbol.
    \item \textbf{Step-3.} The OCC of length 3 or 4, derived from Table~\ref{tab:occ_table}, is applied according to the number of PUCCH symbols within each slot. Specifically, length-3 OCC is applied in the first slot and length-4 OCC in the following slot, as illustrated in Figure~\ref{fig:PUCCH_format1_5G}.    
    \item \textbf{Step-4.} The resulting sequences are mapped into allocated locations in the resource grid.
\end{itemize}
Compared to LTE, where a total of $96$ PUCCH symbols are used for a single HARQ indicator, a total of $7 \times 12 = 84$ symbols are assigned for PUCCH format 1 in 5G with the $14$ OFDM symbol allocation. Moreover, the alternative DMRS placement in 5G implies a $50$~\% DMRS overhead in PUCCH format 1, which is close to an optimal overhead ratio through unfavorable channel conditions~\cite{chen2021fundamentals}. 

In the decoding stage, the decoder generates a conjugated version of the desired PUCCH symbols to calculate the normalized correlation coefficient, which can be written as~\cite{matlab_5G}
\begin{equation}\label{eq:5G_PUCCH_decode_correlation}
    c=\frac{|\sum_{i=1}^{N_r}y_i \cdot s_{1,i}^*|}{\sqrt{(\sum_{i=1}^{N_r}|y_i|^2)(\sum_{i=1}^{N_r}|s_{1,i}|^2)}},
\end{equation}
where all of the variables are same from~(\ref{eq:error_metric_PUCCH_decode}) and~(\ref{eq:decoder}), except that $N_r=7 \times 12 = 84$ for the received PUCCH symbol $y_i$, and $s_{1, i}^*$ is the complex conjugate of the PUCCH symbol generated with the input bit fixed to $1$, following the procedure described from~(\ref{eq:base_sequence_5G}) to~(\ref{eq:final_5G}). Here, since the actual transmitted bit $b\in{0,1}$ is unknown at the decoder, the reference (desired) symbol $s_{b, i}^*$ is always generated using bit $1$ as input for detection purposes.

Based on the normalized correlation coefficient $c$ defined in~(\ref{eq:5G_PUCCH_decode_correlation}), the decoder determines whether it is a valid PUCCH transmission. If $c$ exceeds a predefined detection threshold $\tau_{\mathrm{5G}}$, the decoder proceeds to recover the phase rotation and performs a hard decision for BPSK demodulation. Otherwise, the decoder considers the reception to be a discontinuous transmission (DTX) with an empty result, which can be written as 
\begin{equation}\label{eq:decoder_5G}
    \hat{b} =
        \begin{cases}
        \text{sign}\left( \Re\left\{ \frac{1}{N_r} \sum_{i=1}^{N_r} y_i \cdot s_{1,i}^* \right\} \right), & \text{if } c \geq \tau_{\mathrm{5G}}, \\
        \varnothing, & \text{otherwise},
        \end{cases}
\end{equation}
where $\Re\{\cdot\}$ denotes the real part and the function sign$(\cdot)$ maps positive values to bit $1$ and negative values to bit $0$. Here, the expression inside the sign function represents the average matched filter output, which restores the transmitted BPSK symbol.

\subsection{HARQ Indicators in WI-FI}

\begin{figure}[t!]
    \centering
    \includegraphics[width=0.97\linewidth]{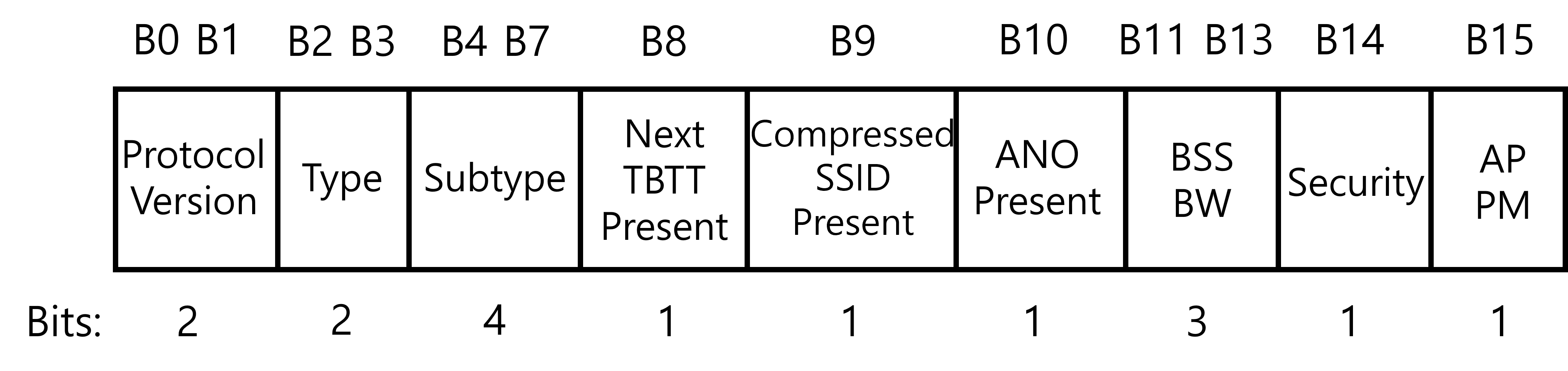}
    \caption{Structure of Wi-Fi control frame (reused from~\cite{wifi_standard}). Bit indices (e.g., B0, B1, etc.) and the number of bits for each field are indicated.}
    \label{fig:wifi_control_frame}
\end{figure}

Unlike in LTE/5G, Wi-Fi does not support the HARQ mechanism. Alternatively, Wi-Fi relies on explicit ACK frames in the medium access control (MAC) layer. It is known that the ACK frame is identified by the ACK type (bit value of $1011$ from B4 to B7~\cite{wifi_standard}) in the frame control field, which is illustrated in Figure~\ref{fig:wifi_control_frame}. In this study, we solely focus on ACK frames specified by the Subtype field. Hence, we do not cover other fields and their abbreviations or usages. When a receiver successfully decodes a data frame, the ACK frame is transmitted after a short interframe space (SIFS) to ensure the ACK frame is prioritized. If the sender does not receive an ACK frame within a specified time duration, it assumes the transmitted packet was lost and initiates retransmission. Moreover, in contrast to LTE/5G, Wi-Fi does not use a NACK indicator for retransmission strategies. These fundamental differences highlight that the HARQ indicator in LTE/5G is more robust than Wi-Fi's ACK mechanism in an asymmetric environment. Specifically, the HARQ indicator in LTE is spread across $96$ symbols, providing enhanced reliability from the structured and time-aligned feedback. Meanwhile, Wi-Fi delivers an ACK in a single standalone ACK frame.


\section{HARQ and Burst Transmission for RC Links}\label{ch:HARQs}
\begin{figure*}[t!]
    \centering
    \subfigure[HARQ Type-I with no combining.]{\includegraphics[width=0.97\columnwidth]{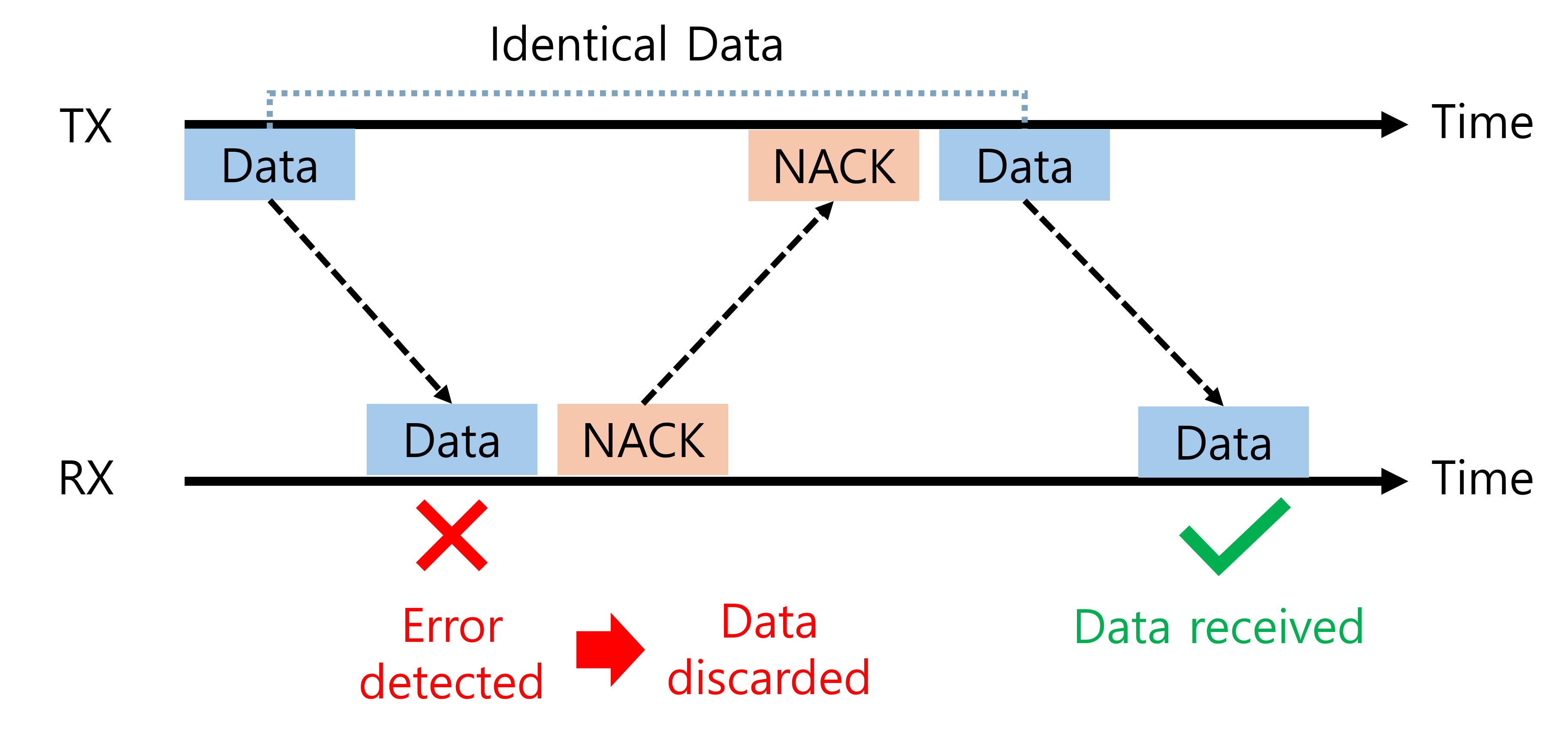}    \label{fig:description_HARQ_type_I_with_no_combining}}
    \subfigure[HARQ Type-I with CC.]{\includegraphics[width=0.97\columnwidth]{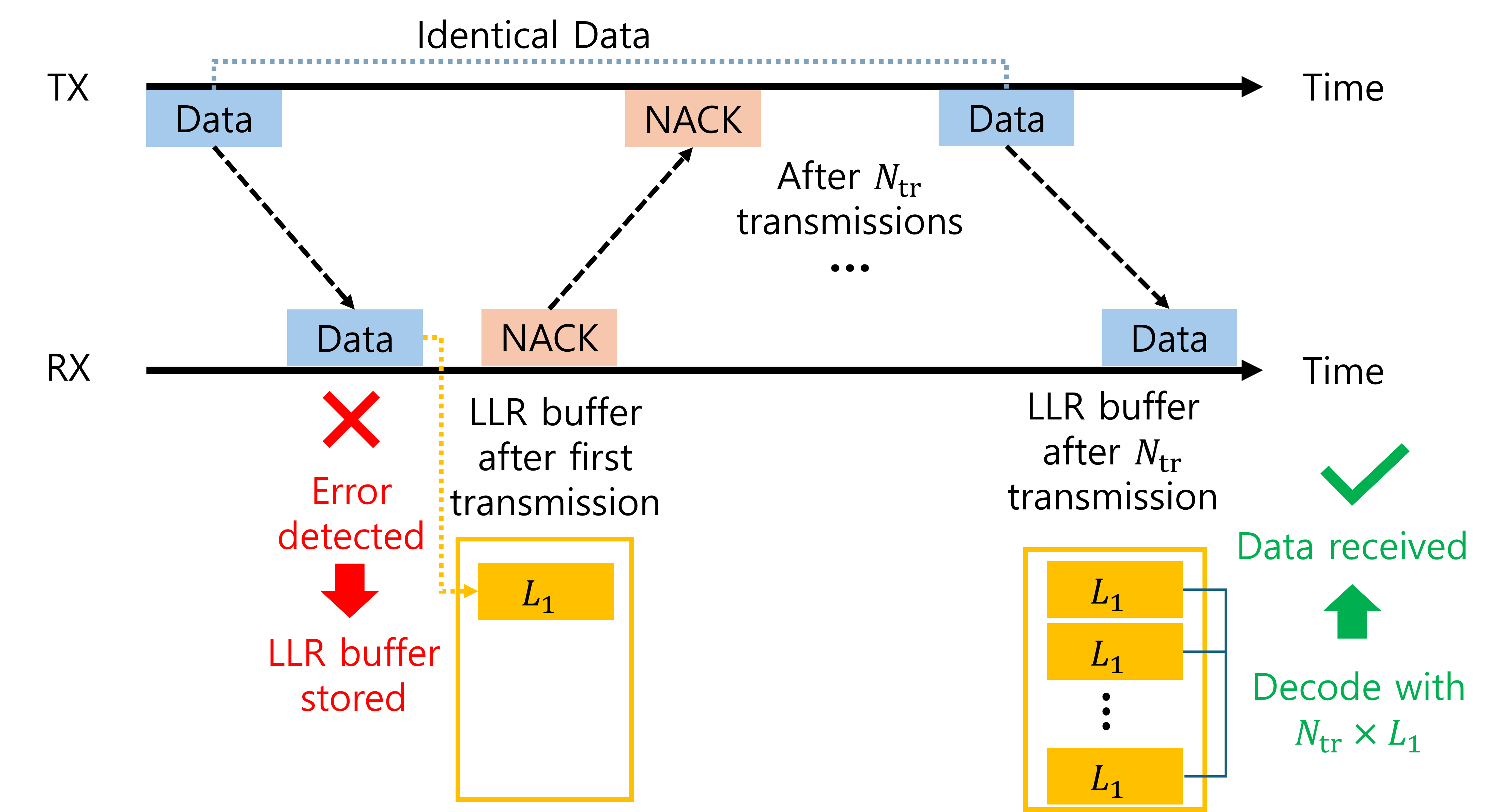}
    \label{fig:description_HARQ_type_I_with_CC}}
    \subfigure[HARQ Type-III with IR.]{\includegraphics[width=0.97\columnwidth]{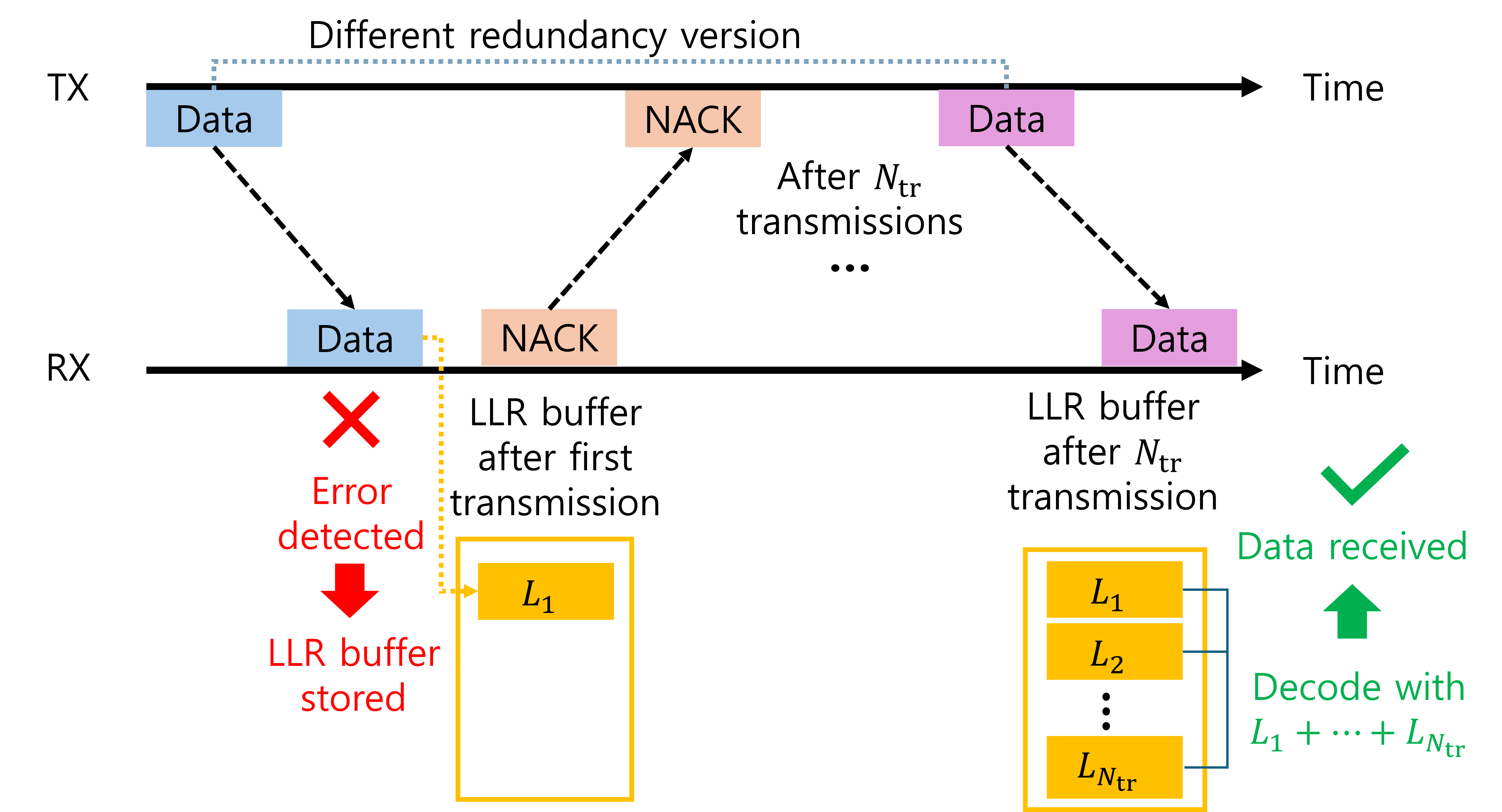}
    \label{fig:description_HARQ_type_III_with_IR}}
    \subfigure[Burst transmission with CC.]{\includegraphics[width=0.97\columnwidth]{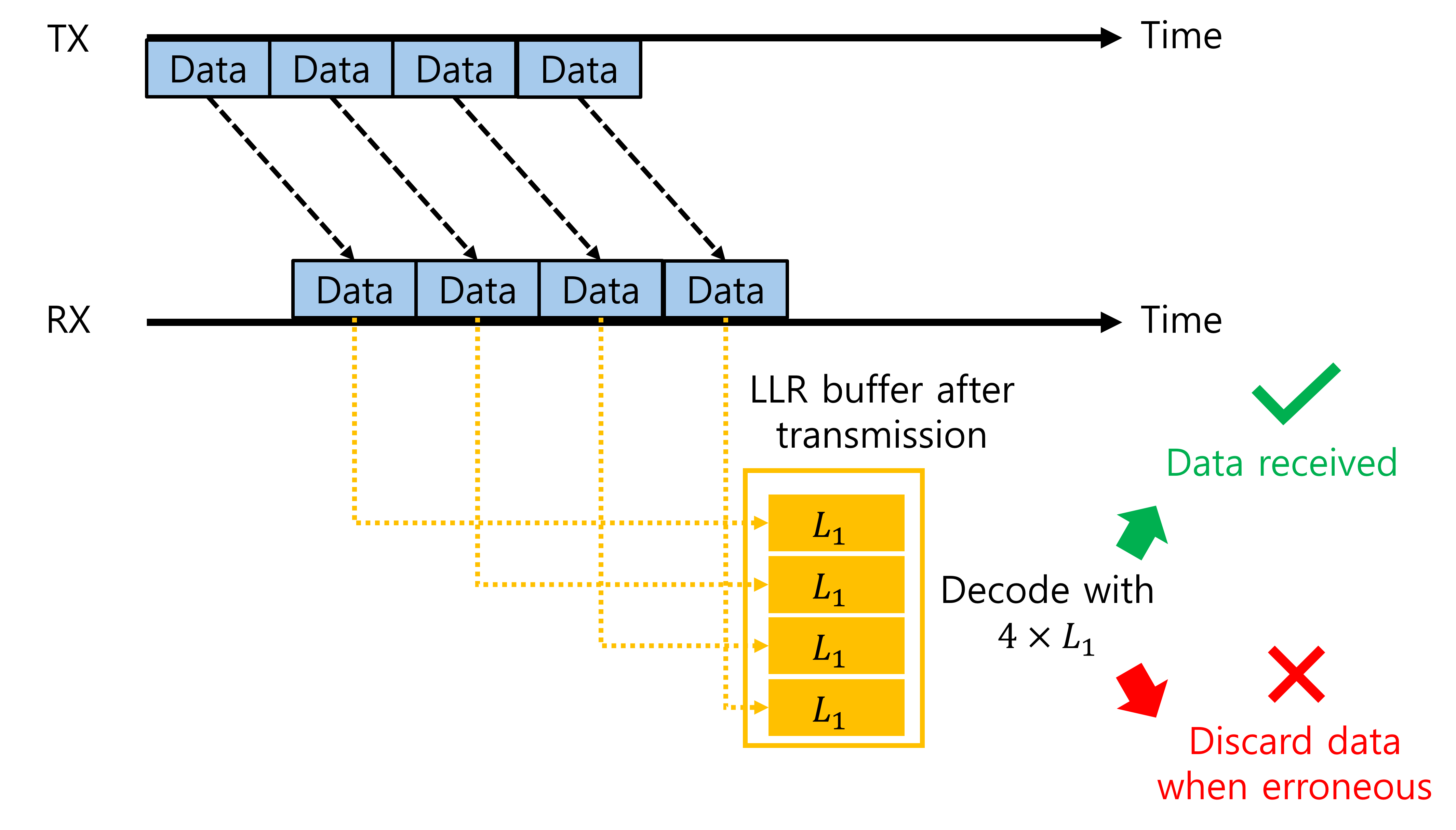}
    \label{fig:description_burst_transmission_with_CC}}
    \caption{HARQ types and combining process.}
    \label{fig:description_HARQs}
\end{figure*}

In this section, we describe the details of the retransmission schemes with different types of HARQ that we will evaluate for A2G UAV RC links~\cite{Matlab_HARQ}, where the HARQ and combining processes are illustrated in Figure~\ref{fig:description_HARQs}.

\subsection{HARQ Type-I with No Combining}
\begin{table}[t!]
    \begin{tabular}{l}\hline
        \textbf{Algorithm 1} Throughput evaluation with HARQ Type-I with no \\ combining.\\ \hline        
        1: Initialize simulation parameters\\ 
        2: \textbf{for} $\gamma$ = $\gamma_{\mathrm{min}}:\gamma_{\mathrm{max}}$ dB \textbf{do}\\ 
        3: ~~~~\textbf{for} $n_{\mathrm{SF}}$ = $1: N_{\rm SF}$ \textbf{do} \\ 
        4: ~~~~~~~~ \textbf{if} $B_{\rm d}$ is empty \textbf{then} Generate data \\ 
        5: ~~~~~~~~ \textbf{else then} Load data from $B_{\rm d}$ \\ 
        6: ~~~~~~~~ \textbf{end if} \\ 
        7: ~~~~~~~~ Modulate transmit waveform \\ 
        8: ~~~~~~~~ Propagate through channel \\ 
        9: ~~~~~~~~ Demodulate and decode \\ 
        10: ~~~~~~~ \textbf{if} $I_{\rm err} == 0$ \textbf{then} $TH \leftarrow TH + \rm{size}(data)$ \\
        11: ~~~~~~~ \textbf{else then} Store data in $B_{\rm d}$ for repetition \\ 
        12: ~~~~~~~~~~~ $N_{\rm tr} \leftarrow N_{\rm tr} + 1$ \\
        13: ~~~~~~~ \textbf{end if} \\
        14: ~~~~~~~ \textbf{if} $N_{\rm tr} == N_{\rm tr, max}$ \textbf{or} $I_{\rm err} == 0$ \textbf{then} \\
        15: ~~~~~~~~~~~ Empty $B_{\rm d}$ and set $N_{\rm tr} = 0$ \\
        16: ~~~~~~~ \textbf{end if} \\
        17: ~~~ \textbf{end for} \\ 
        18: ~~~ $TH \leftarrow TH/(N_{\rm SF} \times 10^{-3})$ \\
        19: \textbf{end for} \\ \hline        
    \end{tabular}
    \label{tab:sim_alg_HARQ_I_no_comb}
\end{table}

In this setting shown in Figure~\ref{fig:description_HARQ_type_I_with_no_combining}, the erroneous packets are discarded at the receiver side. Then, a repeat request is sent to the transmitter. The throughput evaluation procedure is summarized in Algorithm 1, where $\gamma$ is the SINR defined within range of $\gamma_{\mathrm{min}} \leq \gamma \leq \gamma_{\mathrm{max}}$, $n_{\mathrm{SF}}$ is the index of the subframe, $N_{\rm SF}$ indicates the maximum number of subframe index of transmission, $B_{\rm d}$ is the data buffer to facilitate the retransmission procedure, $I_{\rm err}$ is the cyclic redundancy check (CRC) error indicator,  $TH$ is the throughput, $N_{\rm tr}$ indicates the transmission number counter, and $N_{\rm tr, max}$ is the maximum number of transmission, respectively. The number of parallel HARQ streams ($N_{\rm HARQ}$) is set to $1$. Moreover, the soft buffer (LLR buffer), $B_{\rm s}$, is excluded to implement the packet discard process in the retransmission. 

After parameter initialization, the transmitter prepares the data to be sent by checking the data buffer. Then, the modulated data is propagated through the channel. The receiver demodulates and decodes the received signal to check whether the received data packets are erroneous or not based on the CRC error indicator. If the CRC error indicator is off, the size of the received packet is accumulated for throughput calculation. The data is stored in the data buffer, and the received packet is discarded for the throughput calculation when the packet is erroneous. Then, the transmitter sends identical data packets with the receiver's repeat request until the successful reception of the receiver or the maximum number of transmission attempts is reached. After the inner loop of subframe transmission is ended in Line~$17$ of Algorithm~$1$, the accumulated size of successfully decoded packets is divided by the total elapsed time duration. 

\begin{table}[t!]
    \begin{tabular}{l}\hline
        \textbf{Algorithm 2} Throughput evaluation with HARQ Type-I with CC.\\ \hline        
        1: Initialize simulation parameters\\ 
        2: \textbf{for} $\gamma$ = $\gamma_{\mathrm{min}}:\gamma_{\mathrm{max}}$ dB \textbf{do}\\ 
        3: ~~~~\textbf{for} $n_{\mathrm{SF}}$ = $1: N_{\rm SF}$ \textbf{do} \\ 
        4: ~~~~~~~~ \textbf{if} $B_{\rm d}$ is empty \textbf{then} Generate data \\ 
        5: ~~~~~~~~ \textbf{else then} Load data from $B_{\rm d}$ \\ 
        6: ~~~~~~~~ \textbf{end if} \\ 
        7: ~~~~~~~~ Modulate transmit waveform \\ 
        8: ~~~~~~~~ Propagate through channel \\ 
        9: ~~~~~~~~ Demodulate and decode with $B_{\rm s}$ \\ 
        10: ~~~~~~~ \textbf{if} $I_{\rm err} == 0$ \textbf{then} $TH \leftarrow TH + \rm{size}(data)$ \\
        11: ~~~~~~~ \textbf{else then} Combine as in (\ref{eq:LLR_CC}) and store in $B_{\rm s}$ \\ 
        12: ~~~~~~~~~~~ Store data in $B_{\rm d}$  \\ 
        13: ~~~~~~~~~~~ $N_{\rm tr} \leftarrow N_{\rm tr} + 1$ \\        
        14: ~~~~~~~ \textbf{end if} \\
        15: ~~~~~~~ \textbf{if} $N_{\rm tr} == N_{\rm tr, max}$ \textbf{or} $I_{\rm err} == 0$ \textbf{then} \\
        16: ~~~~~~~~~~~ Empty $B_{\rm d}, B_{\rm s},$ and set $N_{\rm tr} = 0$ \\
        17: ~~~~~~~ \textbf{end if} \\
        18: ~~~ \textbf{end for} \\ 
        19: ~~~ $TH \leftarrow TH/(N_{\rm SF} \times 10^{-3})$ \\
        20: \textbf{end for} \\ \hline        
    \end{tabular}
    \label{tab:sim_alg_HARQ_I_CC}
\end{table}

\subsection{HARQ Type-I with CC}
The simulation procedure for HARQ Type-I with CC is described in Algorithm~$2$ and the overall flow is shown in Figure~\ref{fig:description_HARQ_type_I_with_CC}. Most of the procedure is similar to Algorithm~$1$. However, in the HARQ Type-I with CC scenario, the soft buffer, $B_{\rm s}$, remains to store and combine soft log-likelihood ratio (LLR) data over repetitive receptions. The LLR of the received symbol over an AWGN channel can be expressed as
\begin{equation}
    L = \log{\frac{P(y|x=+1)}{P(y|x=-1)}} = \frac{2y}{\sigma^2},
\end{equation}
where $y$ is the received symbol, $x$ is the BPSK modulated symbol, and $\sigma^2$ is the variance of AWGN, respectively. In this paper, we adopt QPSK modulation for throughput evaluation purposes, which can be represented as two independent BPSK signals on the in-phase and quadrature axes. However, it is worthwhile to note that we express LLR with BPSK for simplicity and clarity in subsequent descriptions of combining techniques in different HARQ types.  

The number of parallel HARQ streams is set to 1, similar to the HARQ Type-I with no combining settings. The LLR in CC is accumulated over $N_{\mathrm{tr}}$ repetitive receptions, where $N_{\mathrm{tr}}$ ranges from $1$ to the maximum number of allowable retransmissions $N_{\mathrm{tr, max}}$. This can be expressed as
\begin{equation}\label{eq:LLR_CC}
    L_{\mathrm{CC}}=\sum_{i=1}^{N_{\mathrm{tr}}} L_i = N_{\mathrm{tr}} \times L_1, 
\end{equation}
where $L_i$ represents the obtained LLR from the $i$-th transmission and $N_{\mathrm{tr}}$ is the number of retransmissions. Since HARQ Type-I with CC transmits identical symbols in each retransmission, the combined LLR after $N_{\mathrm{tr, max}}$ transmissions can reach $N_{\mathrm{tr, max}} \times L_1$. The CC with repetitive reception results in an effective signal power increase~\cite{LTE_book}, and hence improves the throughput at the low-SINR region, as will be demonstrated in the numerical results section. 

\subsection{HARQ Type-III with IR}
\begin{table}[t!]
    \begin{tabular}{l}\hline
        \textbf{Algorithm 3} Throughput evaluation with HARQ Type-III with IR. \\ \hline        
        1: Initialize simulation parameters\\ 
        2: \textbf{for} $\gamma$ = $\gamma_{\mathrm{min}}:\gamma_{\mathrm{max}}$ dB \textbf{do}\\ 
        3: ~~~~\textbf{for} $n_{\mathrm{SF}}$ = $1: N_{\rm SF}$ \textbf{do} \\ 
        4: ~~~~~~~~ $I_{\rm HARQ} = {\rm mod}(Subframe, N_{\rm HARQ})+1$ \\ 
        5: ~~~~~~~~ \textbf{if} $B_{\rm d} [I_{\rm HARQ}]$ is empty \textbf{then} Generate data \\ 
        6: ~~~~~~~~ \textbf{else then} Load data from $B_{\rm d} [I_{\rm HARQ}]$ \\ 
        7: ~~~~~~~~ \textbf{end if} \\ 
        8: ~~~~~~~~ Modulate transmit waveform \\ 
        9: ~~~~~~~~ Propagate through channel \\ 
        10: ~~~~~~~ Demodulate and decode with $B_{\rm s}[I_{\rm HARQ}]$ \\ 
        11: ~~~~~~~ \textbf{if} $I_{\rm err} == 0$ \textbf{then} $TH \leftarrow TH + \rm{size}(data)$ \\
        12: ~~~~~~~ \textbf{else then} Combine as in (\ref{eq:LLR_IR}) and store in $B_{\rm s}[I_{\rm HARQ}]$ \\ 
        13: ~~~~~~~~~~~ $N_{\rm tr}[I_{\rm HARQ}] \leftarrow N_{\rm tr}[I_{\rm HARQ}] + 1$ \\
        14: ~~~~~~~~~~~ Store data in $B_{\rm d}$ \\
        15: ~~~~~~~ \textbf{end if} \\
        16: ~~~~~~~ \textbf{if} $N_{\rm tr}[I_{\rm HARQ}] == N_{\rm tr, max}$ \textbf{or} $I_{\rm err} == 0$ \textbf{then} \\
        17: ~~~~~~~~~~~ Empty $B_{\rm s}[I_{\rm HARQ}], B_{\rm d}$, and set $N_{\rm tr}[I_{\rm HARQ}] = 0$ \\
        18: ~~~~~~~ \textbf{end if} \\
        19: ~~~ \textbf{end for} \\ 
        20: ~~~ $TH \leftarrow TH/(N_{\rm SF} \times 10^{-3})$ \\
        21: \textbf{end for} \\ \hline        
    \end{tabular}
    \label{tab:sim_alg_HARQ_III}
\end{table}

In the IR scheme, different redundancy versions of the data packet are transmitted to obtain additional reliability gain by decreasing the effective coding rate over repetitive reception \cite{LTE_book}, as summarized in Figure~\ref{fig:description_HARQ_type_III_with_IR}. The receiver can decode the information bits from each redundancy version of data packets, which is known as self-decodability. The LLR in the IR scheme is accumulated over the $N_{\mathrm{tr}}$ receptions, where $N_{\mathrm{tr}}$ ranges from $1$ to $N_{\mathrm{tr, max}}$. This can be expressed as
\begin{equation}\label{eq:LLR_IR}  
    L_{\mathrm{IR}} = \sum_{i=1}^{N_{\mathrm{tr}}} L_i = L_1 + L_2 + \cdots + L_{N_{\mathrm{tr}}}.
\end{equation}
Unlike CC, which transmits identical symbols over each transmission, the IR transmits different coded bits in each transmission. This fundamental difference provides additional coding diversity, allowing the receiver to lower the effective coding rate and improve decoding reliability.

In the HARQ Type-III with the IR scheme, $8$ parallel HARQ processes are employed. The soft-combining of LLR data is also adopted for this scheme. The simulation procedure with HARQ Type-III is described in Algorithm~$3$, where $I_{\rm HARQ}$ is the HARQ index. After the initialization process, the soft buffer of the corresponding HARQ index, which can be derived by the modulo operation of the subframe index and the number of parallel HARQ processes, is checked to see if the soft buffer is empty. Data to be transmitted is generated if the corresponding index number of the data buffer is empty. Then, physical layer processing is conducted from Lines $8$ to $10$ in Algorithm~$3$. After the decoding process at the receiver side, the CRC error indicator, $I_{\rm err}$, is checked to decide the next procedure between the throughput calculation and the combining process when the CRC error indicator is on and off, respectively. If the receiver successfully decodes the data or the number of retransmissions has been exceeded, the corresponding data and soft buffers are emptied. After the subframe loop ends, throughput is divided by the total elapsed time. 

\begin{table}[t!]
    \begin{tabular}{l}\hline
        \textbf{Algorithm 4} Throughput evaluation with burst transmission and CC.\\ \hline        
        1: Initialize simulation parameters\\ 
        2: \textbf{for} $\gamma$ = $\gamma_{\mathrm{min}}: \gamma_{\mathrm{max}}$ dB \textbf{do}\\ 
        3: ~~~~ \textbf{for} $n_{\mathrm{SF}}$ = $1: N_{\rm SF}$ \textbf{do} \\ 
        4: ~~~~ Generate data \\
        5: ~~~~~~~~ \textbf{for} $N_{\rm tr} = 1: 4$ \textbf{do} \\
        6: ~~~~~~~~~~~~ Modulate transmit waveform \\ 
        7: ~~~~~~~~~~~~ Propagate through channel \\ 
        8: ~~~~~~~~~~~~ Demodulate \\
        9: ~~~~~~~~~~~~ \textbf{if} $N_{tr} == 4$ \textbf{then} Decode with $B_{\rm s}$ \\
        10: ~~~~~~~~~~~ \textbf{else then} Decode and combine as in (\ref{eq:LLR_CC}) and store in $B_{\rm s}$ \\
        11: ~~~~~~~~~~~ \textbf{end if} \\         
        12: ~~~~~~~ \textbf{end for} \\        
        13: ~~~~~~~ \textbf{if} $I_{\rm err} == 0$ \textbf{then} $TH \leftarrow TH + \rm{size}(data)$ \\
        14: ~~~~~~~ \textbf{end if} \\
        15: ~~~ \textbf{end for} \\ 
        16: ~~~ $TH \leftarrow TH/(4 \times N_{\rm SF} \times 10^{-3})$ \\
        17: \textbf{end for} \\ \hline        
    \end{tabular}
    \label{tab:sim_alg_burst}
\end{table}

\subsection{Burst Transmission with CC}
In addition to other conventional HARQ mechanisms, we also consider a burst transmission scheme with no retransmissions, high cell-edge reliability, and hence to improve cell range and latency, at the cost of reduced throughput. This can be critical for scenarios where the high UL interference can significantly reduce SINR and hence reliability. In the burst transmission settings as summarized in Figure~\ref{fig:description_burst_transmission_with_CC}, we assume the data packet is transmitted four times in four consecutive LTE subframes (transmit time intervals (TTIs)), which can be found in Lines 5 to 10 of Algorithm 4. After four back-to-back packet receptions in consecutive subframes, the received packets are decoded based on accumulated soft LLR data in the soft buffer. Thus, the total elapsed time for throughput calculation should be four times, i.e., $TH = TH/(4 \times N_{\rm SF} \times 10^{-3})$. New data packets are transmitted after every back-to-back transmission.

\section{PERFORMANCE EVALUATION AND METHODOLOGY}\label{Sec:performance_evaluation_method}

\subsection{Performance Evaluation Scenarios}
\begin{table}[t!]
    \centering    
    \caption{Simulation parameters for throughput evaluation.}
    \begin{tabular}{|c|c|c|}
        \hline  \textbf{Parameter}   & \textbf{Value} & \textbf{Description} \\ \hline
       $N_{\rm SF}$  & $500$ & Maximum number of subframe index \\ \hline
       $N_{\rm HARQ}$ & $8$ & Number of parallel HARQ streams \\ \hline
       $N_{\rm tr,max}$ & $4$ & Maximum number of transmission \\ \hline
       $N_{\rm RB, DL}$ & $50$ & Number of resource blocks for DL \\ \hline
       $N_{\rm RB,UL}$ & $6$ & Number of resource blocks for UL \\ \hline
       TM & TM1 (SISO) & Transmission mode \\ \hline
       Modulation & QPSK & Modulation scheme for MCSs 1 - 3 \\ \hline 
       $c_1$ & 0.25 & Coding rate for MCS1 \\ \hline
       $c_2$ & 0.5 & Coding rate for MCS2 \\ \hline
       $c_3$ & 0.75 & Coding rate for MCS3 \\ \hline
       $W_{\rm DL}$ & $10$ MHz & Bandwidth for DL \\ \hline
       $W_{\rm UL}$ & $1.4$ MHz & Bandwidth for UL \\ \hline 
       Duplex Mode & FDD & Frequency division duplex \\ \hline
       $\tau_{\mathrm{LTE}}$ & $0.83$ & Decoding threshold in~(\ref{eq:decoder}) \\ \hline
       $\tau_{\mathrm{5G}}$ & $0.22$ & Decoding threshold in~(\ref{eq:decoder_5G}) \\ \hline
       $I_{\mathrm{occ}}$ & $1$ & OCC index for LTE/5G \\ \hline
    \end{tabular}
    \label{tab:sim_params}
\end{table}

We consider a point-to-point communication scenario between the UAV and the remote controller on the ground. MATLAB's LTE~\cite{Matlab_HARQ}, 5G~\cite{matlab_5G}, and WLAN~\cite{matlab_WLAN} Toolboxes are used for carrying out the corresponding simulations. MATLAB scripts for this work are publicly available at~\cite{github_repo}. We have modified the default HARQ Type-III with IR simulation settings available in the LTE toolbox to realize other retransmission mechanisms as described earlier in Section~\ref{ch:HARQs}, in Algorithms~1-4.

The following assumptions are made for performance evaluation: 1) the maximum number of transmission subframes is set to $500$; 2) the number of parallel HARQ streams of HARQ Type-III is $8$; 3) the maximum number of transmission attempts is set to $4$; 4) transmission mode $1$ of single-input single-output (SISO) antenna settings with $50$ resource blocks for DL and $6$ resource blocks for UL are assumed; 5) quadrature phase shift keying (QPSK) with three different coding rates is adopted (in particular we denote, the lowest modulation and coding schemes (MCS) of MCS1, MCS2, and MCS3 in the increasing order of coding rate); 6) perfect ACK/NACK indicator exchange is assumed except for the UL/DL SINR asymmetry scenario; 7) all overheads in the resource grid are taken into account for data transmission and throughput evaluation; 8) an extended vehicular A model (EVA) with the maximum Doppler frequency of $5$~Hz is used for Rayleigh fading channel model; and 9) other simulation parameters are selected to match the LTE and 5G PUCCH structure as closely as possible, except the inherently different DMRS placement, which is illustrated in Figures~\ref{fig:PUCCH_format1a} and~\ref{fig:PUCCH_format1_5G}. The key parameters for throughput evaluation simulations are shown in Table~\ref{tab:sim_params}.

\subsection{Throughput Evaluation with UL/DL SINR Asymmetry}\label{Sec:asymmetry}
\begin{figure}[t!]
    \centering
    \includegraphics[width=0.97\columnwidth]{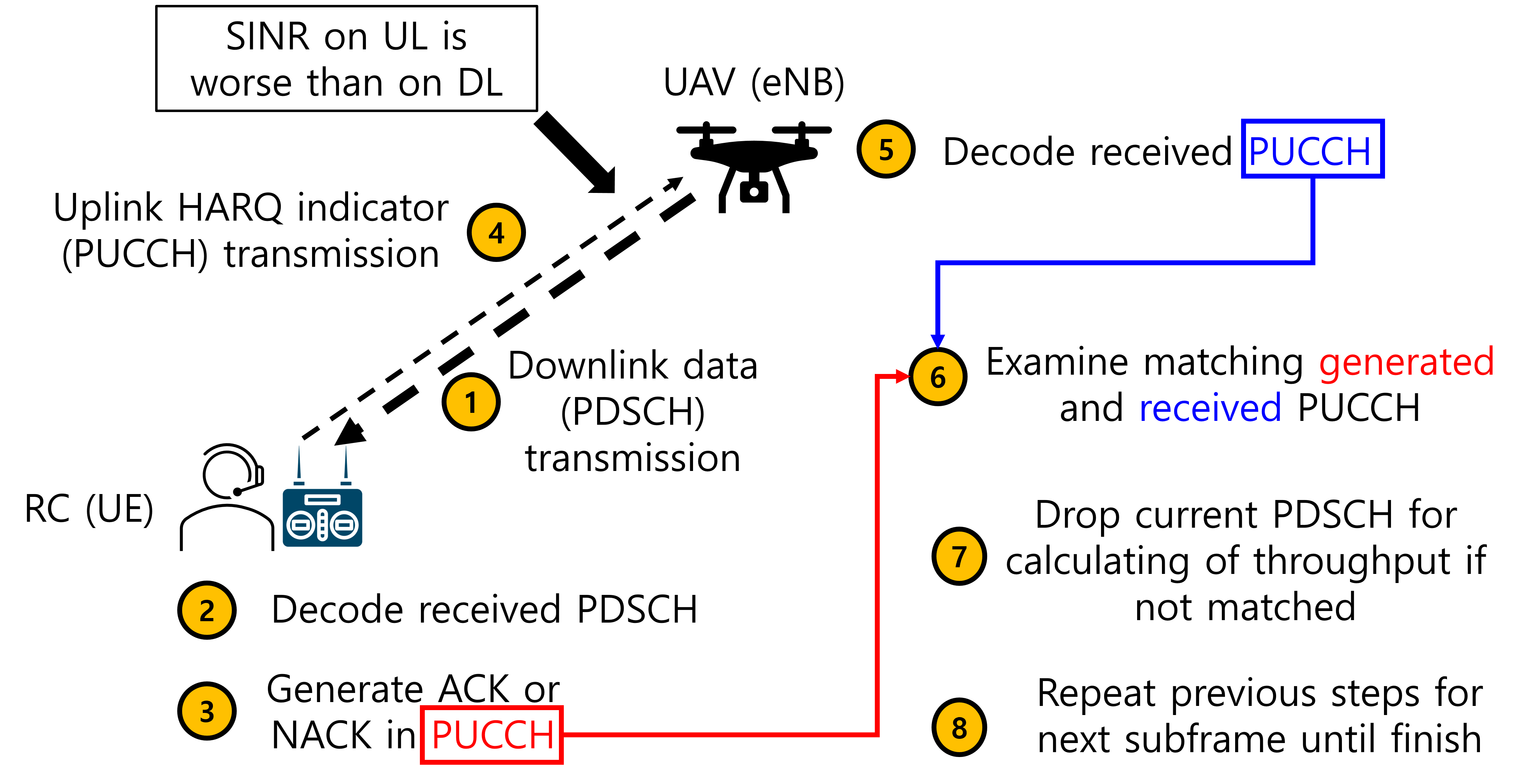}
    \caption{UL/DL SINR asymmetry analysis procedure.}
    \label{fig:uplink_asymmetry_procedure}
\end{figure}

The procedure for performance evaluation in the UL/DL SINR asymmetry scenario is illustrated in Figure~\ref{fig:uplink_asymmetry_procedure}. The procedure can be summarized as follows.
\begin{itemize}
    \item \textbf{Step-1.} UAV transmits DL data through the PDSCH.
    \item \textbf{Step-2.} RC unit decodes the received PDSCH subframe and conducts a CRC error check.
    \item \textbf{Step-3.} RC unit generates the HARQ indicator using PUCCH format $1$a.
    \item \textbf{Step-4.} RC unit transmits the UL HARQ indicator via the PUCCH. The SINR of the UL is worse than the DL due to the interference, which leads to a negative bias on the SINR on the DL side.
    \item \textbf{Step-5.} UAV decodes the received HARQ indicator in PUCCH.
    \item \textbf{Step-6.} UAV examines the matching of generated PUCCH in Step-4 and decoded PUCCH symbols in Step-5.
    \item \textbf{Step-7.} UAV drops the current PDSCH for calculating throughput if the generated and decoded PUCCH symbols are different.
    \item \textbf{Step-8.} Repeat the previous steps for all subframe transmissions.
\end{itemize}
This evaluation procedure emphasizes how the UL SINR degradation affects the throughput performance, considering the accuracy of HARQ indicator signaling. Even successfully received DL data subframes may be discarded due to the UL HARQ indicator loss, which results in throughput degradation in the asymmetric SINR scenarios. 

Additionally, standalone decoding error performance analysis is conducted for the LTE/5G PUCCH and Wi-Fi ACK frame independently from the data transmission. In these analyses, the error performance of the LTE/5G PDSCH and Wi-Fi data frames is also included as reference cases.


\begin{table}[t!]
    \centering    
    \caption{Definition and duration for latency evaluation.}
    \begin{tabular}{|c|c|c|}
        \hline  \textbf{Parameters}   & \textbf{Values} & \textbf{Description} \\ \hline
       $T_{\rm L1/L2}$ & $1$~ms (1 TTI) & L1/L2 processing delay \\ \hline
       $T_{\rm Align}$ & $1$~ms (1 TTI) & Data alignment delay \\ \hline
       $T_{\rm Tx}$ & $1$~ms (1 TTI) & Transmission time \\ \hline
       $T_{\rm Proc}$ & $3$~ms (3 TTI) & Data processing delay \\ \hline
    \end{tabular}
    \label{tab:sim_params_latency}
\end{table}

\subsection{Latency Evaluation with HARQ}
\begin{figure}[t!]
    \centering
    \subfigure[Latency with HARQ.]{    \includegraphics[width=0.85\columnwidth]{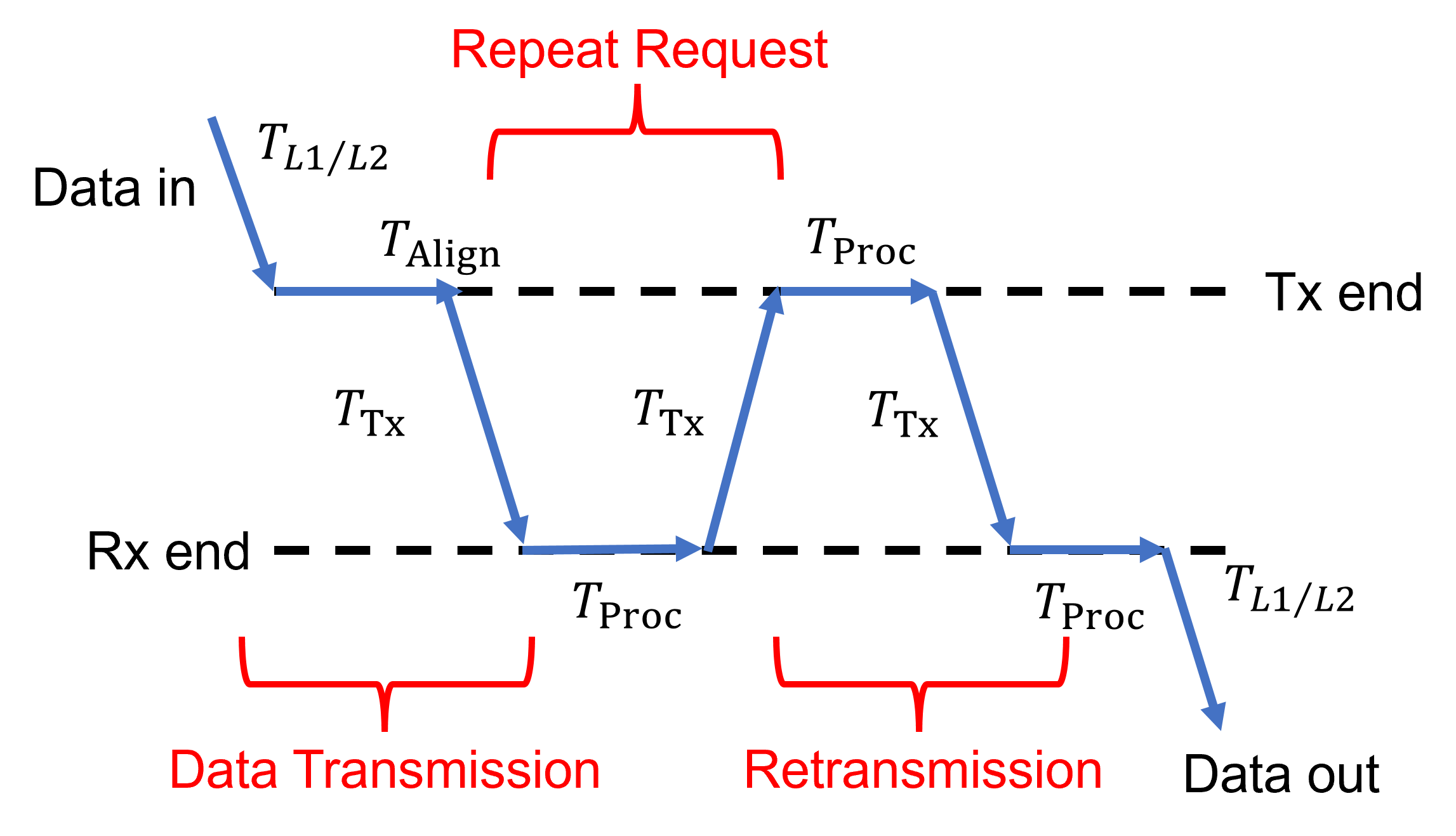}\label{fig:latency_HARQs}
    }
    \subfigure[Latency with burst transmission.]{    \includegraphics[width=0.85\columnwidth]{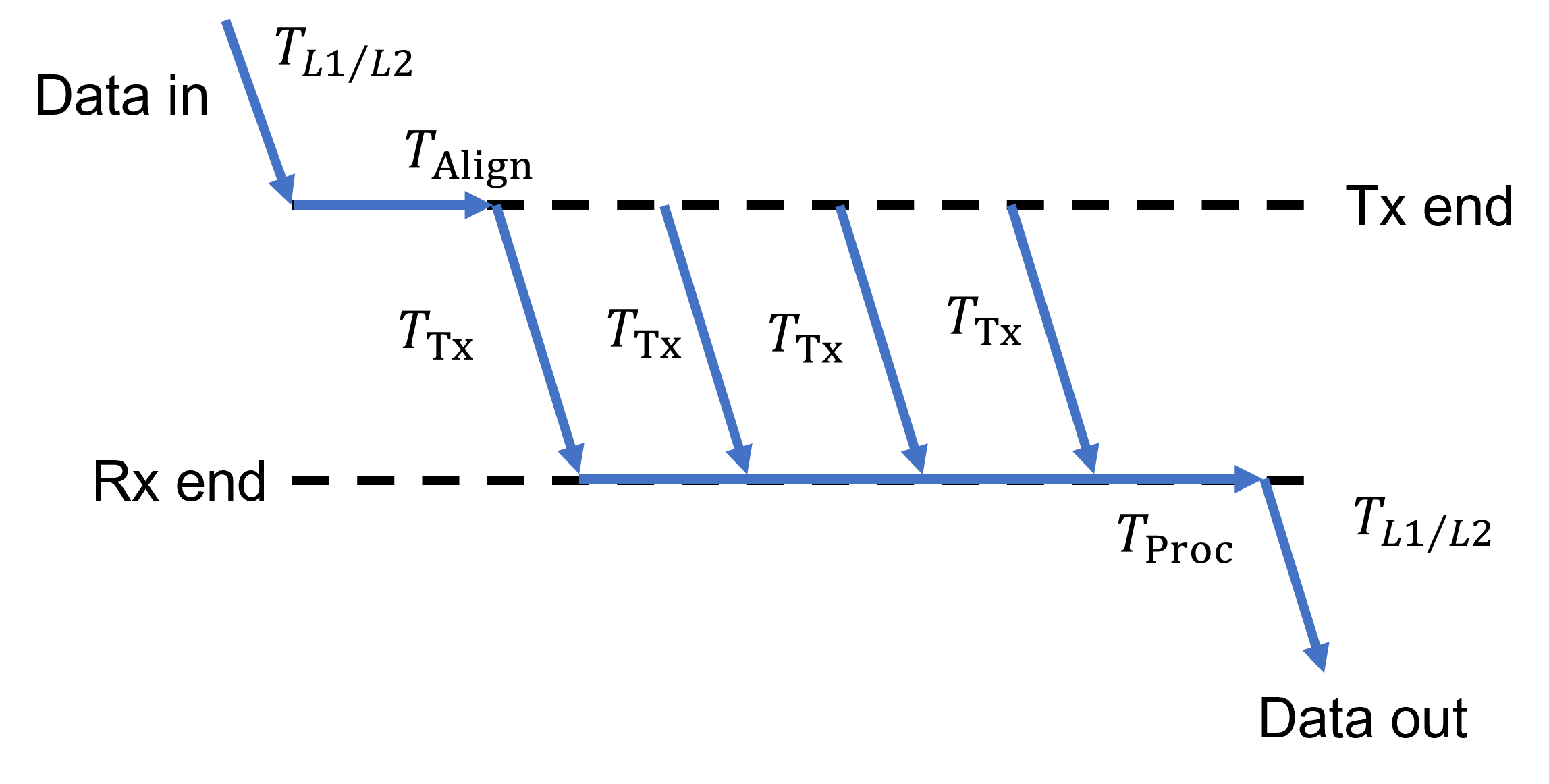}\label{fig:latency_burst}
    }
    \caption{Illustration of latency components for the HARQ-based approaches~\cite{latency_components_paper} and burst transmission.}
    \label{fig:latency_components}
\end{figure}

\begin{figure*}[t!]
    \centering
    \subfigure[MCS1.]{    \includegraphics[trim={0.4cm, 0, 1.1cm, 0.35cm},clip, width=0.65\columnwidth]{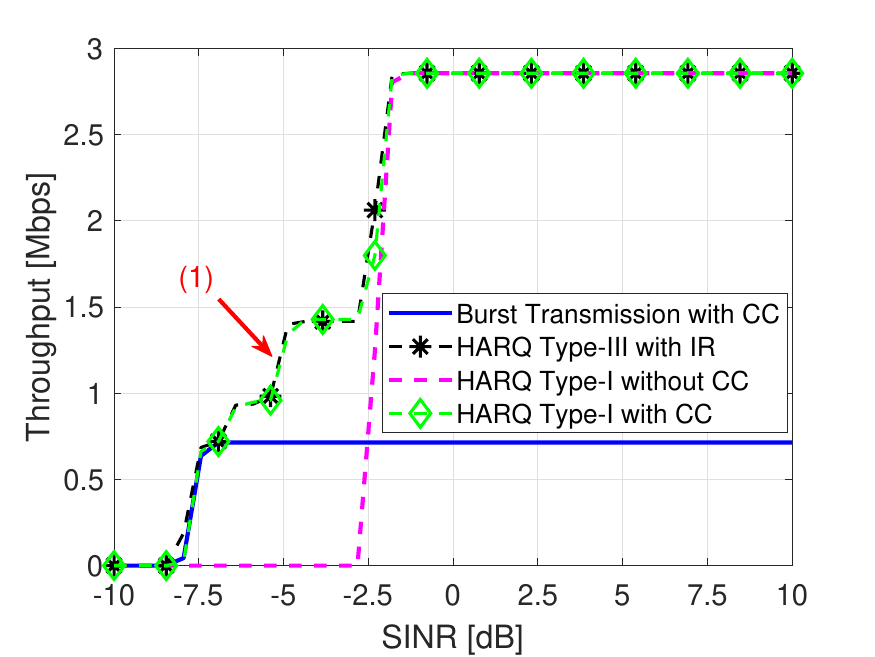}\label{fig:throughput_AWGN_MCS1}}
    \subfigure[MCS2.]{    \includegraphics[trim={0.4cm, 0, 1.1cm, 0.35cm},clip, width=0.65\columnwidth]{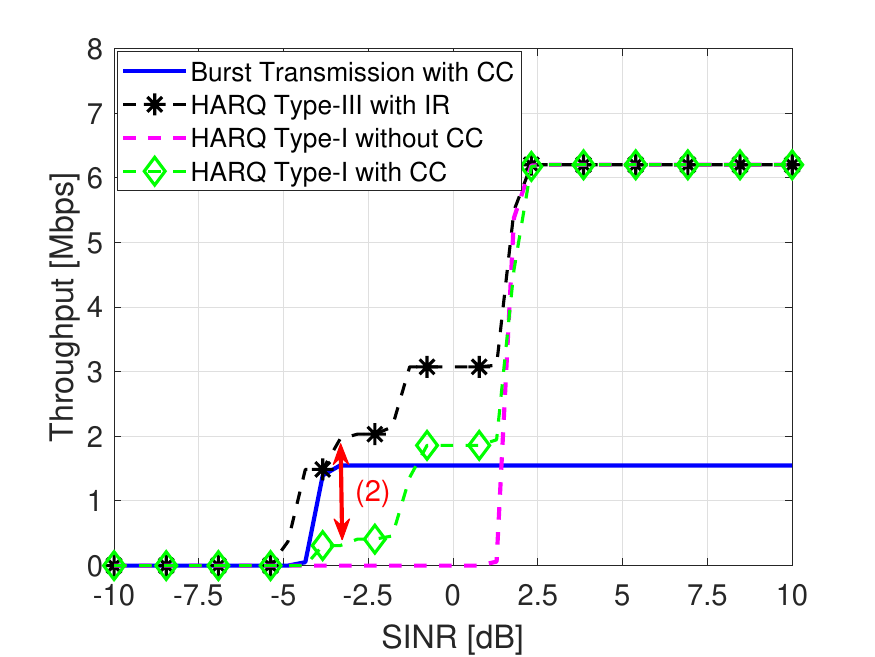}\label{fig:throughput_AWGN_MCS2}}
    \subfigure[MCS3.]{    \includegraphics[trim={0.4cm, 0, 1.1cm, 0.35cm},clip, width=0.65\columnwidth]{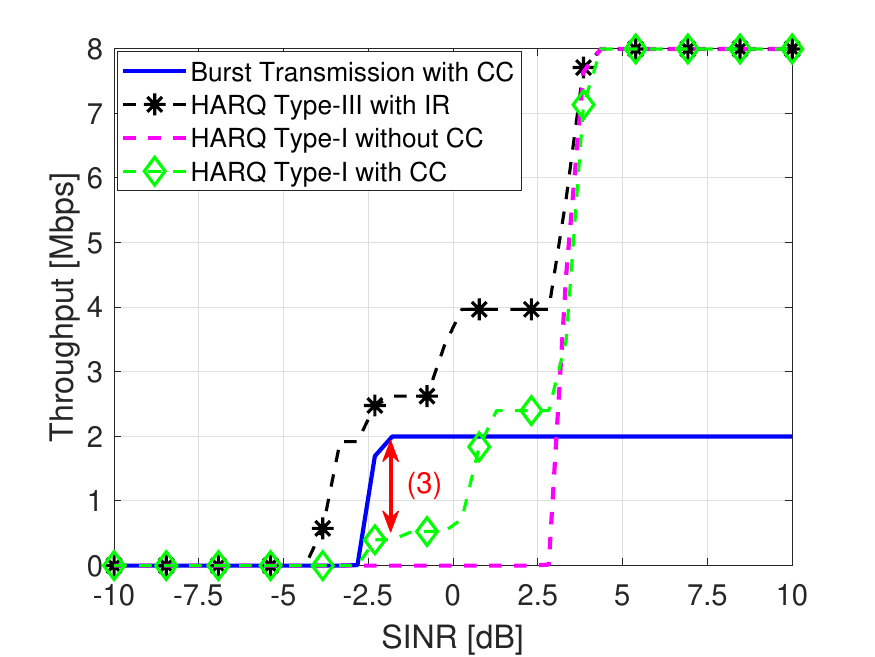}\label{fig:throughput_AWGN_MCS3}}
    \caption{DL throughput evaluation over AWGN channel with different retransmission schemes. Perfect ACK/NACK is assumed for all cases.}
    \label{fig:throughput_AWGN}
\end{figure*}

\begin{figure*}[t!]
    \centering
    \subfigure[MCS1.]{    \includegraphics[trim={0.4cm, 0, 1.1cm, 0.35cm},clip, width=0.65\columnwidth]{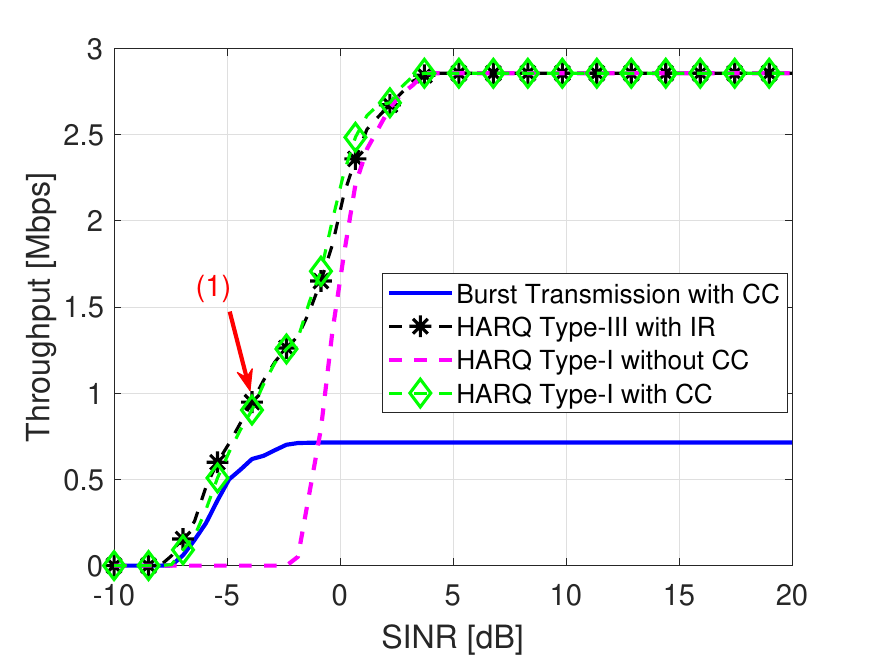}}
    \subfigure[MCS2.]{    \includegraphics[trim={0.4cm, 0, 1.1cm, 0.35cm},clip, width=0.65\columnwidth]{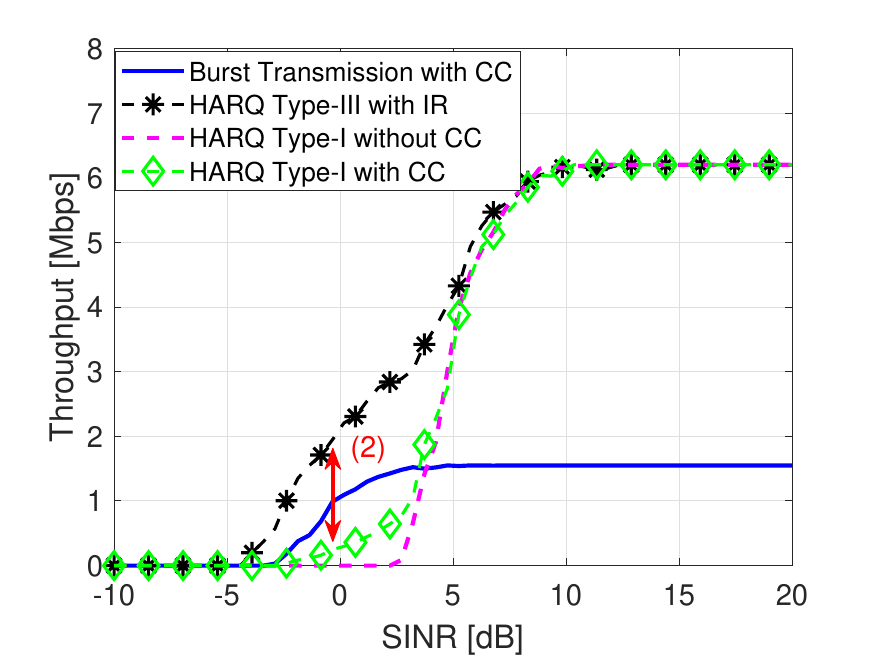}}
    \subfigure[MCS3.]{    \includegraphics[trim={0.4cm, 0, 1.1cm, 0.35cm},clip, width=0.65\columnwidth]{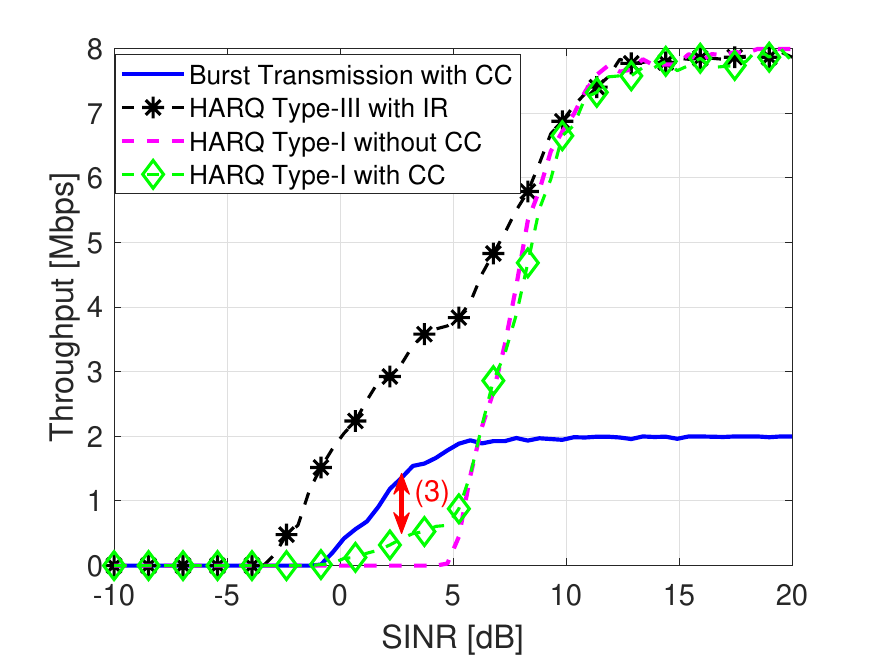}}
    \caption{DL throughput evaluation over Rayleigh fading channel with different retransmission schemes. Perfect ACK/NACK is assumed for all cases.}
    \label{fig:throughput_Rayleigh}
\end{figure*}

In Figure~\ref{fig:latency_components}, the latency components of HARQ-based approaches \cite{latency_components_paper} and burst transmission are depicted. The definition and time duration of each latency component can be summarized in Table~\ref{tab:sim_params_latency}. 
Note that the assumptions of TTI and delay components in Table~\ref{tab:sim_params_latency} are adopted from \cite{latency_components_paper}. The total latency of the HARQ scheme with $k$ retransmissions can be calculated as~\cite{latency_components_paper} (see also Figure~\ref{fig:latency_HARQs}):
\begin{equation}
    T = 2T_{\rm L1/L2} + T_{\rm Align} + 2(k+1)T_{\rm Proc} + 2(k+1)T_{\rm Tx}.
\end{equation}
While calculating the latency, we also take into account the ACK/NACK processing time at the transmitter, except for burst transmissions. On the other hand, for the burst transmission case with four transmissions,  the total latency is given as (see also Figure~\ref{fig:latency_burst}):
\begin{equation}
    T = 2T_{\rm L1/L2} + T_{\rm Align} + 4T_{\rm Tx} + T_{\rm Proc}.\label{Eq:burstTX}
\end{equation}
Note that with burst transmissions, \eqref{Eq:burstTX} requires processing delay to be added only once for the last transmission because the data processing for the earlier transmissions is done in parallel, which is depicted in a blue arrow starting from the first reception in Figure \ref{fig:latency_components}. 

\section{NUMERICAL RESULTS}\label{ch:results}
In this section, we provide numerical results for various scenarios described earlier in  Sections~\ref{ch:HARQs} and~\ref{Sec:performance_evaluation_method} using the modeling and simulation assumptions discussed in Section~\ref{CH:sys}.

\subsection{AWGN Channel Analysis}\label{CH:AWGN}
Simulation results of throughput over AWGN channel with different retransmission strategies are shown in Figure \ref{fig:throughput_AWGN}. Recall that MCS1, MCS2, and MCS3 refer to QPSK modulation with coding rates of $0.25$, $0.5$, and $0.75$, respectively. As seen in Figures \ref{fig:throughput_AWGN_MCS1} - \ref{fig:throughput_AWGN_MCS3}, the HARQ Type-III with IR achieves the best performance, particularly at low SINR and the higher coding rate as in the MCS2 and MCS3 cases. However, at the lowest MCS (MCS1, coding rate of $0.25$), HARQ Type-I with CC closely matches the throughput of HARQ Type-III with IR. This similarity arises because sufficient redundancy at the lowest coding rate leaves little room for IR to provide additional gains compared to the CC, which is highlighted by the annotation ($1$). 

The burst transmission with CC cases consistently has the lowest throughput at high SINR due to the degradation from the total elapsed time of back-to-back transmission of identical packets, as in Line~$16$ in Algorithm~$4$. However, the burst method outperforms the HARQ Type-I with CC at low SINR for MCS2 and MCS3, which is indicated by the annotation ($3$). This performance gain is from $N_{\mathrm{tr}}$ in the Line~$5$ of the Algorithm~$4$, leading to transmission of the same subframe four times back-to-back, allowing for LLR accumulation. Consequently, in SINR regions where HARQ Type-I with CC requires four transmissions to succeed, only one out of four data frames contributes to throughput. In contrast, the burst method rapidly reaches the maximum achievable throughput by using four times back-to-back transmissions at the same region, which requires four transmissions. 

It is noteworthy that the HARQ Type-III with IR shows the throughput improvement from the soft combining gain and reliability improvement from IR, which is highlighted by the annotation ($2$). Meanwhile, the soft combining gain can be observed in HARQ Type-I with CC compared to HARQ Type-I with no combining.   

\subsection{Rayleigh Fading Channel Analysis}\label{CH:Rayleigh}
In the Rayleigh fading channel analysis, the maximum Doppler frequency of $5$ Hz is applied for the simulation. Simulation results of throughput over the Rayleigh fading channel are shown in Figure \ref{fig:throughput_Rayleigh}, where the throughput curves over SINR tend to be smoother than AWGN cases, which is caused by the impact of fading over the propagation channel. In the MCS1 case, similar to the AWGN scenario, the throughput of HARQ Type-I with CC and HARQ Type-III with IR are closely matched, which is shown by the annotation (1). Moreover, performance improvement from the IR can also be observed in HARQ Type-III with IR cases compared to HARQ Type-I with CC cases by leveraging IR and coding diversity over retransmissions, particularly at higher coding rates as in MCS2 and MCS3, which is highlighted by the annotation (2). 

Meanwhile, the HARQ Type-I with CC benefits from the combining gain over the repetitive receptions, showing limited performance improvement compared to the IR scheme. The annotation (3) also shows that the burst transmission with CC scheme outperforms the HARQ Type-I with CC in a low SINR, which is from repetitive back-to-back transmissions of the burst transmission. Similar to the AWGN scenario, the HARQ Type-III with IR has the highest throughput performance, followed by HARQ Type-I with CC and HARQ Type-I with no combining at low SINR, respectively. It can hence be said that the HARQ Type-III with IR provides more reliable cell-edge connectivity for A2G links.

\subsection{UL/DL SINR Asymmetry Analysis}\label{CH:UL}

\begin{figure}[t!]
    \centering
    \includegraphics[trim={0.4cm, 0cm, 0.8cm, 0.4cm},clip, width=0.95\columnwidth]{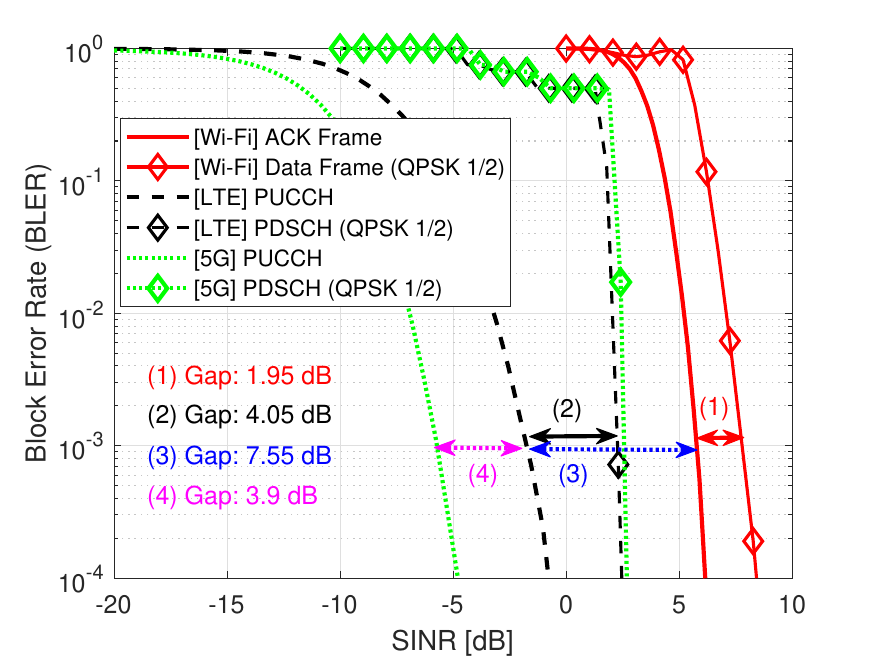}
    \caption{BLER evaluation over AWGN for ACK frame and data frame in Wi-Fi, and PUCCH and PDSCH in LTE/5G. The SINR in the x-axis assumes identical SINR in both UL and DL.}
    \label{fig:BLER_LTE_WiFi}
\end{figure}

\begin{figure}[t!]
    \centering
    \includegraphics[trim={0.4cm, 0cm, 0.8cm, 0.4cm},clip, width=0.95\columnwidth]{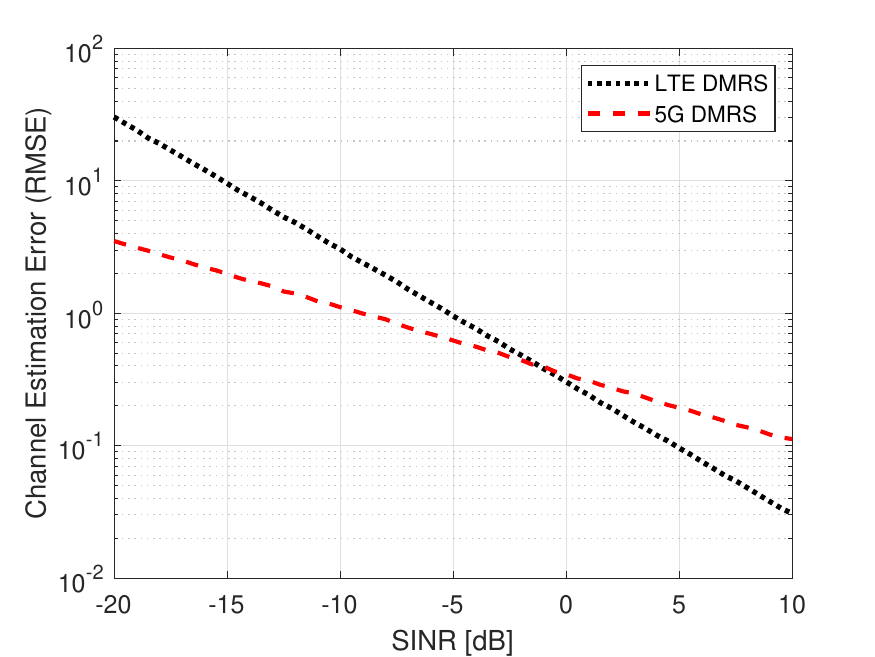}
    \caption{Channel estimation error using DMRS in LTE and 5G.}
    \label{fig:channel_estimation_error}
\end{figure}

\begin{figure}
    \centering
    \subfigure[LTE.]{    \includegraphics[trim={0.4cm, 0, 1.1cm, 0.35cm},clip, width=0.95\columnwidth]{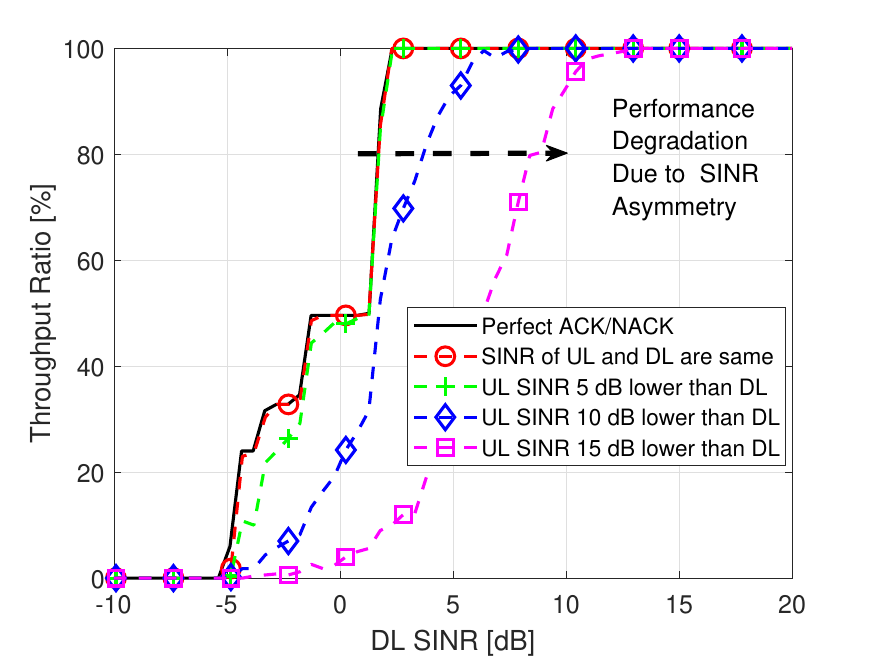}
    \label{fig:UL_AWGN_LTE}}
    \subfigure[5G.]{    \includegraphics[trim={0.4cm, 0, 1.1cm, 0.35cm},clip, width=0.95\columnwidth]{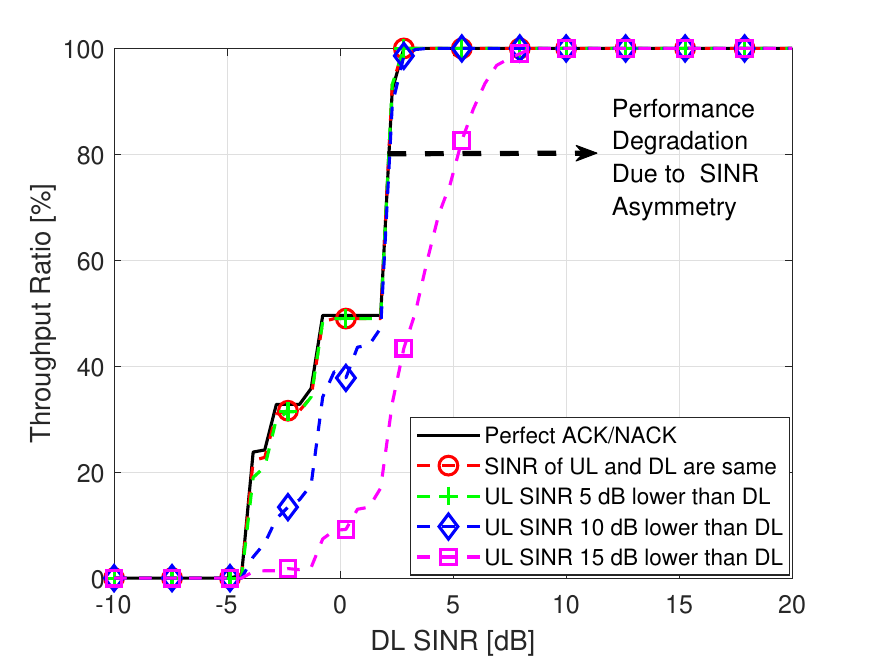}
    \label{fig:UL_AWGN_5G}}
    \caption{LTE and 5G DL throughput evaluation (MCS2) of HARQ Type-III with IR over AWGN with UL/DL SINR asymmetry.}
    \label{fig:UL_AWGN_LTE_5G}
\end{figure}

\begin{figure}
    \centering
    \subfigure[LTE.]{    \includegraphics[trim={0.4cm, 0, 1.1cm, 0.35cm},clip, width=0.95\columnwidth]{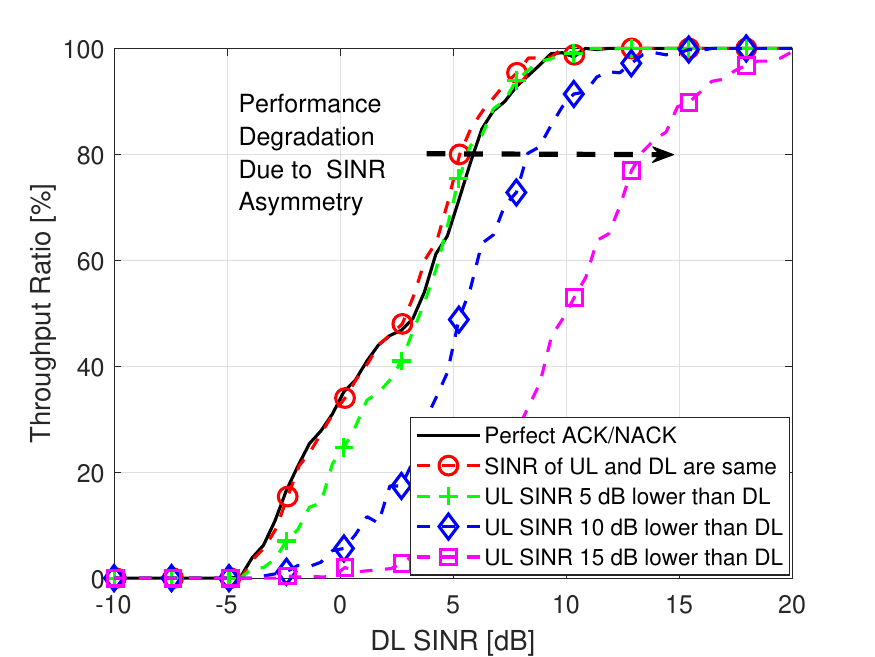}
    \label{fig:UL_Rayleigh_LTE}}
    \subfigure[5G.]{    \includegraphics[trim={0.4cm, 0, 1.1cm, 0.35cm},clip, width=0.95\columnwidth]{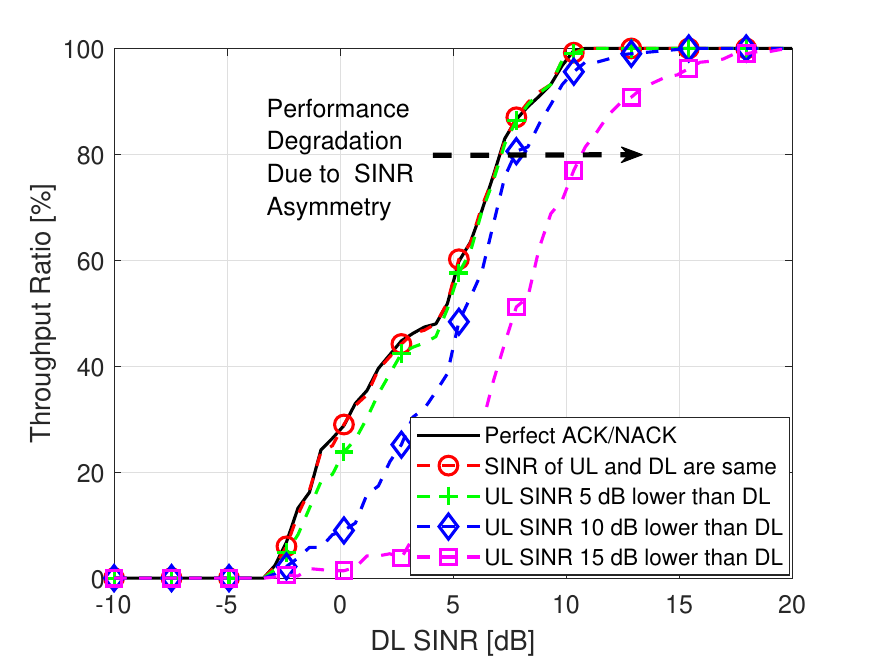}
    \label{fig:UL_Rayleigh_5G}}
    \caption{LTE and 5G DL throughput evaluation (MCS2) of HARQ Type-III with IR over Rayleigh fading channel with UL/DL SINR asymmetry.}
    \label{fig:UL_Rayleigh_LTE_5G}
\end{figure}



In Figure~\ref{fig:BLER_LTE_WiFi}, the block error rate (BLER) evaluations for data and HARQ indicator channels of Wi-Fi, LTE, and 5G over the AWGN channel are shown, where QPSK modulated with a coding rate of $1/2$ is adopted for comparison purposes. It is shown that the HARQ indicator in PUCCH of LTE is $7.55$~dB more robust compared to the ACK frame in Wi-Fi to reach a BLER of $10^{-3}$, as highlighted by the gap (3) in the figure. This robustness stems from the allocation of the PUCCH over $96$ symbols, which provides frequency diversity. On the contrary, the Wi-Fi ACK frame is transmitted as a short burst, making it more susceptible to interference. 

The gaps between data and HARQ indicator channels for the LTE and Wi-Fi are $4.05$~dB (gap (2)) and $1.95$~dB (gap (1)), respectively. It implies that LTE can tolerate up to $4.05$~dB of SINR degradation due to asymmetry in UL and DL while still maintaining a BLER of $10^-3$, whereas Wi-Fi is limited to a $1.95$~dB margin. It is also worthwhile to note that curves for the LTE yield better reliability at low SINRs when compared with Wi-Fi. Two underlying reasons for these differences are as follows. First, LTE employs a turbo coding scheme for channel coding, which provides a significant coding gain compared to the convolutional coding scheme in Wi-Fi. Moreover, LTE adopts a structured frame design and optimized resource allocation scheme over the time and frequency domains to enhance link reliability. These factors collectively enable LTE to achieve a more robust performance at low SINR compared to Wi-Fi.

It is observed that PUCCH in 5G outperforms PUCCH in LTE by $3.9$~dB to reach a BLER of $10^{-3}$. This performance gain is from the different DMRS structures and procedures in LTE and 5G. In LTE, the three DMRSs are allocated at the center of each slot, which is shown in Figure~\ref{fig:PUCCH_format1a}. In contrast, the DMRSs in 5G are assigned to every other symbol over the allocated OFDM symbol duration, which is shown in Figure~\ref{fig:PUCCH_format1_5G}. In the channel estimation of LTE PUCCH, the channel coefficients on the PUCCH symbols are interpolated by using DMRSs. Meanwhile, the channel coefficients on the PUCCH symbols in 5G are averaged by the alternating DMRS. The underlying structure and procedures of DMRS and the impact on channel estimation lead to a performance gain in 5G compared to LTE at low SINRs, which is highlighted in the gap ($4$) in Figure~\ref{fig:BLER_LTE_WiFi}. 

In Figure~\ref{fig:channel_estimation_error}, the simulation results of the channel estimation root mean squared error (RMSE) using DMRS in LTE and 5G are shown, which can be expressed as
\begin{equation}
    E_{\mathrm{RMSE}} = \sqrt{\frac{1}{N}\sum_{i=1}^{N}|\hat{H}_i-H_i|^2},
\end{equation}
where $N$ is the number of samples, $\hat{H}_i$ is the $i$-th estimated channel coefficient, and $H_i$ is the $i$-th channel coefficient, respectively. Here, the number of samples $N$ consists of the total number of subcarriers in $1$ second, i.e., $N=14 \times 12 \times 10^3$ (considering $14$ symbols and $12$ subcarriers in a subframe). Channel estimation error of 5G in the SINR range between $-20$~dB to $-2$~dB outperforms that of LTE, which contributes to the lower BLER of PUCCH in 5G at low SINR. At a higher SINR, channel estimation of LTE outperforms 5G. However, the accurate channel estimation of LTE in this region does not impact BLER because coded subframes have already been successfully decoded at high SINR. 

In Figures~\ref{fig:UL_AWGN_LTE_5G} and~\ref{fig:UL_Rayleigh_LTE_5G}, DL throughput simulation results of HARQ Type-III with IR that consider the SINR asymmetry between UL and DL are shown for AWGN and Rayleigh fading channels, respectively. The x-axis shows the SINR of the DL channel, while the UL channel either has an identical SINR as the DL, a $5$~dB worse SINR, a $10$~dB worse SINR, or a $15$~dB worse SINR that will result in an increasingly higher percentage of lost HARQ indicator (ACK/NACK) as discussed in Section V-\ref{Sec:asymmetry}. For comparison purposes, we also include the scenario where perfect ACK/NACK reception is assumed in the UL. Since the size of the transport block with the given MCS and the number of resource blocks within the same bandwidth are different in LTE and 5G, we introduce the throughput ratio as a normalized metric. The throughput ratio is defined as the throughput divided by the maximum achievable throughput with a given MCS setting. This can be expressed as $TH_{\mathrm{ratio}}=TH/TH_{\mathrm{max}}\times 100$, where $TH_{\mathrm{max}}$ is the maximum achievable throughput with a given MCS.

\begin{table}[t!]
\centering
\caption{Average throughput ratio gap in $-5$~dB to $10$~dB DL SINR range compared to the perfect ACK/NACK scenarios for LTE and 5G in AWGN channels.}
\begin{tabular}{|c|c|}
\hline
\multicolumn{2}{|c|}{\textbf{LTE}} \\
\hline
\textbf{Asymmetry scenarios} & \makecell[c]{\textbf{Average throughput} \\ \textbf{ratio gap [\%p]}} \\
\hline
SINR of UL and DL are the same & $0.4$ \\
\hline
UL SINR $5$~dB lower than DL & $2.61$ \\
\hline
UL SINR $10$~dB lower than DL & $16.19$ \\
\hline
UL SINR $15$~dB lower than DL & $44.33$ \\
\hline
\multicolumn{2}{|c|}{\textbf{5G}} \\
\hline
\textbf{Asymmetry scenarios} & \makecell[c]{\textbf{Average throughput} \\ \textbf{ratio gap [\%p]}} \\
\hline
SINR of UL and DL are the same & $0.39$ \\
\hline
UL SINR $5$~dB lower than DL & $0.61$ \\
\hline
UL SINR $10$~dB lower than DL & $5.53$ \\
\hline
UL SINR $15$~dB lower than DL & $22.79$ \\
\hline
\end{tabular}\label{tab:throughput_gap_AWGN}
\end{table}

In the AWGN channel scenarios in Figure~\ref{fig:UL_AWGN_LTE_5G}, the throughput performance of identical SINR in UL and DL closely approaches the perfect ACK/NACK. This indicates that the HARQ indicator through PUCCH has a minimal impact when the SINR of the UL is identical to the DL. However, the performance degradation can be observed as the asymmetry at the UL side gets worse, especially from the $0$~dB to $10$~dB SINR region in the LTE case when the SINR of the UL is $15$~dB worse than the DL. This highlights how HARQ indicator decoding error in the asymmetric UL, which is described in Step-$7$ in Figure~\ref{fig:uplink_asymmetry_procedure}, substantially limits throughput performance even in the AWGN channel. Table~\ref{tab:throughput_gap_AWGN} presents the average throughput ratio gaps between various UL SINR asymmetry scenarios and the perfect ACK/NACK scenario in the operating SINR range (from $-5$~dB to $10$~dB DL SINR) over the AWGN channels. Note that \%p denotes that the arithmetic difference in throughput ratio, not a relative percentage change. For instance, if the baseline throughput ratio is $80$~\%, and it decreases to $60$~\%, this corresponds to a $20$~\%p drop rather than a $25$~\% drop.  


Since the HARQ indicator performance in 5G outperforms that of LTE, a lower throughput degradation is observed in all asymmetric SINR scenarios compared to LTE. For instance, in the $15$~dB asymmetric SINR scenario, 5G achieves the maximum achievable throughput at a DL SINR of $10$~dB, corresponding to an UL SINR of $-5$~dB. Specifically, as shown in Figure~\ref{fig:BLER_LTE_WiFi}, 5G PUCCH already reaches a BLER of $10^{-4}$ at $-5$~dB. However, in the LTE case, the BLER of PUCCH is still relatively high.

\begin{table}[t!]
\centering
\caption{Average throughput ratio gap in $-5$~dB to $20$~dB DL SINR range compared to the perfect ACK/NACK scenarios for LTE and 5G in Rayleigh fading channels.}
\begin{tabular}{|c|c|}
\hline
\multicolumn{2}{|c|}{\textbf{LTE}} \\
\hline
\textbf{Asymmetry scenarios} & \makecell[c]{\textbf{Average throughput} \\ \textbf{ratio gap [\%p]}} \\
\hline
SINR of UL and DL are the same & $0.96$ \\
\hline
UL SINR $5$~dB lower than DL & $2.33$ \\
\hline
UL SINR $10$~dB lower than DL & $12.67$ \\
\hline
UL SINR $15$~dB lower than DL & $30.43$ \\
\hline
\multicolumn{2}{|c|}{\textbf{5G}} \\
\hline
\textbf{Asymmetry scenarios} & \makecell[c]{\textbf{Average throughput} \\ \textbf{ratio gap [\%p]}} \\
\hline
SINR of UL and DL are the same & $0.07$ \\
\hline
UL SINR $5$~dB lower than DL & $1.28$ \\
\hline
UL SINR $10$~dB lower than DL & $6.57$ \\
\hline
UL SINR $15$~dB lower than DL & $17.78$ \\
\hline
\end{tabular}\label{tab:throughput_gap_Rayleigh}
\end{table}

On the other hand, in the Rayleigh fading channel scenarios in Figure~\ref{fig:UL_Rayleigh_LTE_5G}, a gradual trend of throughput can be observed due to the fading effects. Especially, for the $15$~dB UL asymmetric SINR case, the throughput degradation is observed over the wide range of SINRs, e.g., from $0$ to $15$~dB DL SINR range. Moreover, Table~\ref{tab:throughput_gap_Rayleigh} demonstrates the average throughput ratio gaps between asymmetric SINR and the perfect ACK/NACK scenarios in the operating SINR range (from $-5$~dB to $20$~dB DL SINR) over Rayleigh fading channels. Specifically, as shown in Figure~\ref{fig:rss_altitude_freq_2024_Packapalooza}, our urban scenario helikite measurement demonstrates maximum UL/DL SINR asymmetry reaches more than $16$~dB, and the results presented in Tables~\ref{tab:throughput_gap_AWGN} and~\ref{tab:throughput_gap_Rayleigh} indicate significant throughput degradation due to HARQ indicator loss in the UL. Consequently, ensuring a balanced UL and DL is essential for fully exploiting the achievable capacity of the link in A2G networks. 


\begin{figure}[tb!]
    \centering
    \subfigure[Latency in AWGN channel.]{    \includegraphics[trim={0.1cm, 0, 1.1cm, 0.35cm},clip, width=0.95\columnwidth]{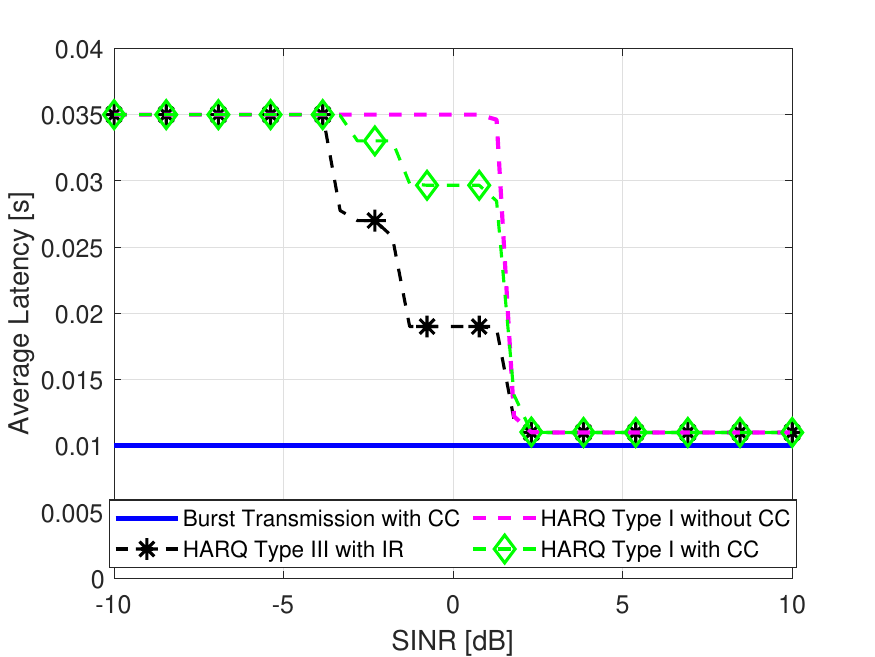}}
    \subfigure[Latency in Rayleigh fading channel.]{    \includegraphics[trim={0.1cm, 0, 1.1cm, 0.35cm},clip, width=0.95\columnwidth]{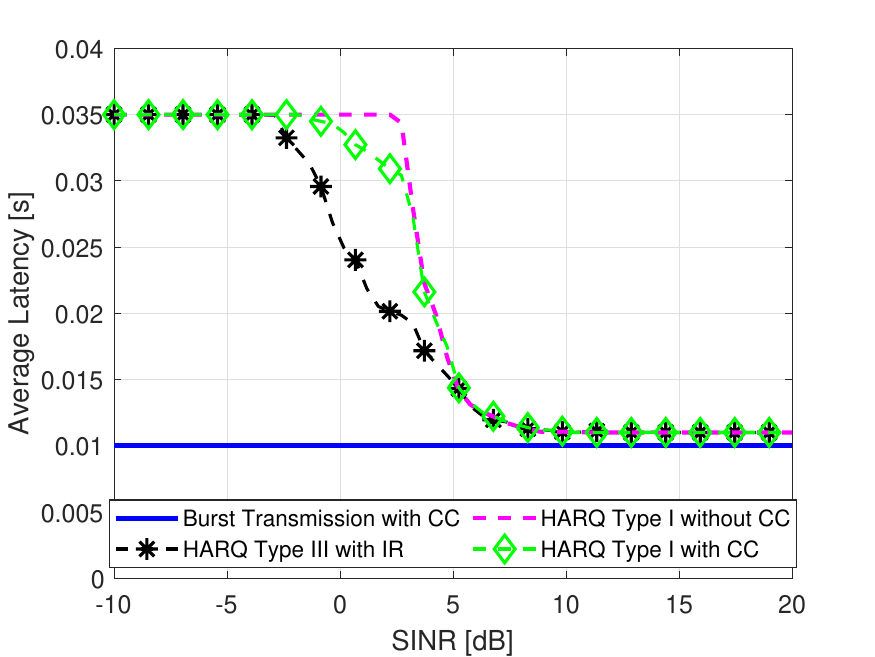}}
    \caption{DL latency evaluation over AWGN and Rayleigh fading channels with different retransmission schemes. Burst transmission provides the lowest latency at the expense of lower throughput, as highlighted in Figures~\ref{fig:throughput_AWGN} and \ref{fig:throughput_Rayleigh}.}
    \label{fig:latency}
\end{figure}

\subsection{Latency Analysis}\label{CH:latency}
Simulation results of average latency with AWGN and Rayleigh fading channels are shown in Figure \ref{fig:latency}. The HARQ-based approaches have a higher average latency than the burst transmission cases in the low SINR region due to the latency incurred by the repeat request. The latency of HARQ-based cases gradually decreases from $-5$~dB to $5$~dB as the number of retransmission requests decreases. Here, the HARQ Type-III with IR significantly outperforms HARQ Type-I with CC in both AWGN and Rayleigh fading channel cases by exploiting the decoding reliability improvement of IR. The average latency for burst transmission is constant over the whole SINR range due to the fixed number of back-to-back transmissions. The HARQ Type-I without CC in both channels shows a rapid decrease in latency from $0$ to $5$~dB. The simulation results highlight that the burst transmission provides the lowest latency, while the HARQ Type-III with IR shows a reasonable balance between latency and throughput performance. 


\section{CONCLUSIONS}
In this paper, considering UL and DL SINR asymmetry, we evaluated the performance of LTE, 5G, and Wi-Fi-based A2G links in terms of throughput and latency for various HARQ options. As a piece of practical evidence supporting our UL asymmetric scenarios, we also conducted a measurement campaign using a helikite platform for urban and rural environments. To evaluate the throughput performance over A2G links, we adopted 1) HARQ Type-I with no combining, 2) HARQ Type-I with CC, 3) HARQ Type-III with IR, and 4) burst transmission with CC. In both AWGN and Rayleigh fading channel scenarios, the HARQ Type-III with IR yields the best throughput performance at low SINR for all the different HARQ options. 

We analyzed the impact of the SINR asymmetry between UL and DL, as is typical in A2G RC links, especially in ISM bands. We also discussed fundamental aspects of processing HARQ indicators in LTE/5G and Wi-Fi to investigate HARQ indicator loss in UL/DL asymmetric SINR scenarios. From the simulation results, the significant performance degradation due to the HARQ indicator loss on the UL side is demonstrated. Lastly, the average latency with different HARQ options is evaluated. The reduction in latency with increasing SINR is characterized for various HARQ options. 

In future work, the scope of our analyses will be extended into multi-input multi-output (MIMO) scenarios to evaluate the performance with higher throughput. Moreover, adaptive UL control strategies such as dynamic power control and advanced interference mitigation schemes will be investigated to tackle the negative impact on the quality of service. Finally, due to its low latency and good cell-edge coverage, burst transmissions can be utilized adaptively only at low SINR scenarios while using one of the other HARQ modes at high SINR, which will also be studied as part of the future work.

\bibliographystyle{IEEEtran}
\bibliography{manuscript}

\end{document}